\documentclass[nofootinbib,twocolumn,showpacs,preprintnumbers,amsmath,amssymb,aps,prd]{revtex4}
\usepackage{hyperref}
\usepackage{url}
\usepackage{graphicx}
\usepackage{dcolumn}
\begin{document}
\title{The implications of Mexican-hat mirrors: calculations of thermoelastic noise and
  interferometer sensitivity to perturbation for the Mexican-hat-mirror proposal for advanced
  LIGO} 
\author{R. O'Shaughnessy}
\email{oshaughn@caltech.edu}
\affiliation{Theoretical Astrophysics, California Institute of
  Technology, Pasadena, CA 91125}
\author{S. Strigin}
\affiliation{Physics Faculty, Moscow State University, Moscow, Russia}
\author{S. Vyatchanin}
\affiliation{Physics Faculty, Moscow State University, Moscow, Russia}
\date{Received ?? Month 2003, printed \today}
\begin{abstract}
Thermoelastic noise will be the most significant noise source in advanced-LIGO interferometers with
sapphire test masses.   The standard plan for advanced-LIGO has optimized the optics, within the
framework of conventional mirrors, to reduce thermoelastic noise.  Recently, we and our
collaborators have proposed 
going beyond the bounds of traditional optics to increase the effective beam spot size and
thus lower thermoelastic noise.  One particular proposal for mirror shapes (``Mexican-hat
mirrors'')  yields the class of
``mesa'' beams. 

In this paper, we outline a general procedure for analyzing light propagating in individual arm
cavities,  
and  the associated thermoelastic noise, in the presence of arbitrary optics.  We
apply these procedures to study the Mexican-hat proposal.  Results obtained by the techniques
of  this paper were presented elsewhere, to demonstrate that the Mexican-hat proposal for
advanced-LIGO both significantly lowers thermoelastic noise and does not significantly
complicate the function of the interferometer.
\end{abstract}
\maketitle

\section{Introduction}
Thermoelastic noise will be the dominant noise source in the advanced LIGO interferometers for
 frequencies at which they are most sensitive,
 should the advanced LIGO upgrade follow the baseline
design \cite{advLIGOmain,advLIGOweb}.  Other noise processes (e.g., conventional bulk thermal noise; coating thermal
noise; unified optical noise) could be important, but are expected to be smaller in noise power, by a factor
$\sim 9$  (at the minimum of the noise curve), than thermoelastic noise.
As a result, 
the sensitivity of the advanced LIGO interferometers can be significantly increased.

In a companion paper---the mesa-beam interferometer paper (MBI) \cite{MBI},  summarized
briefly in a short LIGO technical document (MBI-S) \cite{ShortReport}---we and our
collaborators 
have argued that a significant reduction in thermoelastic noise can be achieved by 
using modified optics that reshape the beam from a conventional (gaussian) profile into a
\emph{mesa-beam} shape.
We furthermore performed calculations which imply (though do not comprehensively demonstrate)
that this proposal would not place
significant burdens on the experimental community (e.g., in the need for slightly improved tilt
control systems and improved mirror figure accuracies).

The companion publications (MBI and MBI-S) survey the physically and practically relevant
features of our proposal---that is, they demonstrate that the proposal should work, and sketch
how to implement it.  Those publications intentionally omitted many supporting details,
to be provided in this paper and a companion paper by Erika D'Ambrosio
\cite{ErikaSimulations}.   In particular, this paper provides the computational details
underlying our evaluations of thermoelastic noise in various interferometer configurations;
both this paper and D'Ambrosio \cite{ErikaSimulations} provide computational details
of our studies of the influence of mirror imperfections (tilts, displacements, figure
errors) on interferometer performance.

Further, this paper considers a broader class of possible improvements than MBI.  The MBI paper
 discussed using mesa beams reflecting off of the standard advanced LIGO cylindrical mirrors.  
In this paper, we analyze mesa beams reflecting off more generically shaped mirrors---not
 merely cylinders of different dimensions, but also frustum-shaped mirrors.\footnote{\label{note:defineFrustum}A frustum
 is a geometric shape arising between two parallel planes that intersect  an axisymmetric cone
 perpendicular to the cone's axis.}

\subsection{Outline of paper}

In this paper, we provide details behind the calculations discussed  in MBI.  For the reader
not  familiar with the principles behind thermoelastic noise, cavity optics, and cavity
perturbation theory, the next
four
sections (Sections \ref{sec:th:Ideal}, \ref{sec:th:Driving}, \ref{sec:th:Defects}, and
\ref{sec:num}) provide a 
brief review of the equations and computational techniques we employ.   
To be more explicit, in Sec.\ \ref{sec:th:Ideal}  we describe how to construct an idealized
interferometer based on mesa beams 
while  remaining within LIGO  design constraints (i.e.\ losses per bounce).  
[In the process, we also briefly review the theory of and introduce notation for optical cavity
eigenfunctions and propagation operators.]
We assume this interferometer, which has mesa beams resonating within the arm cavities, is
driven by conventional (i.e.\ Gaussian) lasers, as described in Sec.\ \ref{sec:th:Driving}.
In  Sec.\ \ref{sec:th:Defects}, we discuss how to  analyze, via perturbation theory, the
 influence of mirror errors on interferometer performance.    [In 
 Section \ref{sec:apply:Defects} we will apply
this general formalism  to mirror transverse position errors (displacement), orientation
 errors (tilt), and figure error (shapes).]  
Finally, in  Sec.\ \ref{sec:num}, we describe how we numerically 
implemented the calculations described in 
Sections \ref{sec:th:Ideal} and \ref{sec:th:Defects}.
With those tools
established, in Sections \ref{sec:apply:Design} and \ref{sec:apply:Defects}  we closely follow
the arguments of MBI, carefully explaining the details behind those arguments.

Specifically, in Sec.\ \ref{sec:apply:Design}, we use explicit forms for the mesa beam 
and its diffraction losses presented in Sections \ref{sec:th:Ideal} and \ref{sec:num} to
determine which mesa beams can operate in a given cavity, when the mesa beams are reflecting 
off of the standard advanced LIGO mirrrors (i.e.\ cylinders of radius $15.7\text{ cm}$ and
thickness $13\text{ cm}$), as well as off mirrors of more generic shape (i.e.\ other cyliinders
and frustums).   We then employ an expression
for the thermoelastic noise associated with a given mirror and beam configuration
(developed generally in Sec.\ \ref{sec:th:Ideal} and discussed practically in
Sec.\ \ref{sec:num}) to determine the thermoelastic noise associated with a given configuration.
Table \ref{tbl:NSRange} summarizes our results for designs which produce lower thermoelastic noise
than the baseline advanced LIGO design.

Section \ref{sec:apply:Defects} describes how sensitive an interferometer using Mexican-hat
mirrors will be to perturbation.  More explicitly, 
Sections \ref{sec:apply:Defects:parasites}, \ref{sec:apply:Defects:displace}, and
\ref{sec:apply:sub:tilt:Cavity} demonstrate by way of explicit comparison that arm cavities
using spherical and mesa-beam mirrors  appropriate to advanced LIGO are roughly equally
sensitive to perturbations.  Section \ref{sec:apply:sub:tilt:Cavity} describes how sensitive
the overall interferometer (as measured by its cavity gain, its dark port power, and its
thermoelastic noise) will be to tilt
perturbations.   Finally, Section \ref{sec:apply:sub:figure} determines how sensitive (as
measured by 
cavity gain, dark port power, and thermoelastic noise) a mesa-beam advanced LIGO design without
signal recycling would
be to mirror figure errors.

Augmented by additional results discussed in MBI, which
demonstrate that the full interferometer including signal recycling will be only moderately
more sensitive, we conclude mesa-beam interferometers could be used
in advanced LIGO without serious concern.

\subsubsection*{Guide to reader}

The MBI paper reports the \emph{results} of calculations developed in this paper and in one by Erika
D'Ambrosio \cite{ErikaSimulations}, emphasizing their practical significance.  On the other
hand, this paper emphasizes the \emph{method} by which those results were obtained.  
As a result, this paper has a dramatically different structure than MBI:  two of the first sections
(Sections \ref{sec:th:Ideal} and \ref{sec:th:Defects}) outline the equations that we have solved;
the next section, Section \ref{sec:num}, describes how we solved them; and  two of the
last sections,
Sections \ref{sec:apply:Design} and \ref{sec:apply:Defects}, briefly summarize  the specific
technical
results quoted in  MBI  and explain how those results were obtained.

The reader who wants only a cursory survey of the techniques used to obtain the MBI
results should skim Sections II-IV, emphasizing the summaries in Sections
 \ref{sec:th:Design:summary} and \ref{sec:th:Defects:summary}; this reader should
completely skip over the implementation section (Sec.\ \ref{sec:num}).   Indeed, we
recommend that a reader interested primarily in better understanding results presented in MBI
should work \emph{backwards}, first establishing a \emph{cross-reference} between a result in
MBI and 
a result in   Sec.\  \ref{sec:apply:Design} or
\ref{sec:apply:Defects} of this paper, then reading in this paper 
the material surrounding the cross-reference.

On the other hand, a reader who wishes to verify or generalize our computations 
should read this paper from beginning to end.

\subsection{Connection with other published work}
The MBI paper \cite{MBI} and the MBI-S LIGO technical document \cite{ShortReport}  survey
our  results, emphasizing  their
practical significance to advanced LIGO interferometer design.  This paper provides details
behind many 
relatively lengthly analytic or numerical calculations which those papers touched upon only
briefly. 

Another of our MBI collaborators, Erkia D'Ambrosio, has performed many of the same calculations
(e.g., the sensitivity of the interferometer to tilt and to defects) using a 
sophisticated, standard, and trusted tool---the so-called ``FFT code''---designed and
developed  specifically to study the behavior of general interferometers
\cite{ErikaSimulations}.  As noted in MBI and in her paper, our two independent approaches
agree well.

\subsection{\label{sec:sub:notation}Notation}
\subsubsection{Symbols}
\begin{itemize}
\item[$b$] The natural diffraction length scale (i.e.\ Fresnel length) associated with the
  previous two parameters: $b\equiv \sqrt{L/k} = 2.6 \text{ cm}$.
\item[$C_V$] Specific heat of test mass, per unit mass, at constant volume
\item[$D$] Mesa beam characteristic scale, Eq.\ (\ref{eq:buildMHcanonical}).
\item[$E$] Young's modulus.  An elastic parameter in the stress
  tensor for an isotropic material [Eq.\ (\ref{eq:elas:Tab})].  (See Appendix
  \ref{ap:sapphire} for a discussion of the explicit value used.) 
\item[ETM] End test mass of an arm cavity (i.e.\ the mass opposite the
  end light enters the cavity).
\item[$F_o$] The net force applied to the front face of the the mirror
  in the thermoelastic model problem of Sec. \ref{sec:th:thermo};
  cf.\ Eqs.\ (\ref{eq:IAraw}) and (\ref{eq:IA}).
\item[$f$] Frequency, units $\text{s}^{-1}$.  Typically, we are interested in values of $f$
  near the peak sensitivity of LIGO (i.e.\ $f\approx 100 \text{Hz}$).
\item[$g_{ab}$] Metric of 3-space.  (See comments on tensor notation, below).
\item[$h_{1,2}$] Height of mirror $1$'s (or $2$'s, respectively) surface.
\item[$I$] Thermoelastic noise integral, Eqs.\ (\ref{eq:IAraw}) and (\ref{eq:IA}).
\item[ITM] Input test mass of an arm cavity.
\item[$k$] The wavevector of light in the cavity, $k=2\pi/\lambda$ with $\lambda = 1064
  \text{nm}$.
\item[$k_b$] Boltzmann's constant [$1.38\times 10^{-16} \text{g}\;  \text{ cm}^2 \; \text{s}^{-2}
  \text{K}^{-1}$]
\item[$L$] The length of the LIGO arm cavity [$L=3.99901 \text{km}$].
\item[${\cal L}$] Some number associated with diffraction losses.  When used alone, denotes the
  total diffraction losses associated with one round trip through the cavity.  When given
  certain subscripts, as with ${\cal L}_1$, denotes the clipping-approximation estimate of
   diffraction
  losses associated with a single reflection off of mirror $1$.
\item[$M$] Mass of the test mass [$40 \text{ kg}$].
\item[MBI] The mesa-beam interferometer paper \cite{MBI}.
\item[$P(r)$] The intensity distribution of  laser light in a cavity, normalized to unity
(i.e.\ $\int P(r) d(\text{area})=1$).
\item[$\vec{r}$] A vector transverse to the optic axis of the cavity (i.e.\ two-dimensional).
\item[$s$] Transverse displacement of an arm cavity's ETM.
\item[$T_{ab}$] Elastic stress tensor for an isotropic medium, Eq.\ (\ref{eq:elas:Tab}).
(See comments on tensor notation, below).
\item[$u$] Electric field on some plane in the arm cavity (typically, the face of the ITM,
  propagating away from   the ITM), renormalized to unit norm (i.e.\ $\int |u|^2
  d(\text{area})=1$).  Subscripts identify the specific system $u$ refers to.
\item[$y^a$] The displacement vector field associated with an elastic distortion.
\item[$x$, $y$] The cartesian components of the 2-vector $\vec{r}$.
\item[$z$] A coordinate measured along the optic axis of the cavity (i.e.\ in a direction
  perpendicular to $\vec{r}$).
\item[$\hat{z}^a$] $\hat{z}$ is a unit vector pointing in the positive
  $z$ direction; $\hat{z}_a=\hat{z}^a$ are the components of
  this vector in our coordinates.  For example, in carteisan coordinates $\hat{z}_a=1$ if $a=z$ and
  $0$ if $a=x,y$.
\item[$\alpha_l$] Coefficient of linear thermal expansion, assumed isotropic.  For an
  isotropic material, the differential change in volume with temperature is $dV/dT/V =
  3\alpha_l$.  (See Appendix \ref{ap:sapphire}.)
\item[$\alpha_{1,2}$] Norm of the first- and second-order corrections to the optical state of
  the cavity when the mirror is tilted
  [i.e.\ in
  Eqs.\ (\ref{eq:tilt:results:gaussians}) and (\ref{eq:tilt:results:MH})].
\item[$\kappa$] Thermal conductivity, assumed isotropic (see Appendix \ref{ap:sapphire}).
\item[$\eta$] An eigenvalue of Eq.\ (\ref{eq:eigenequation}).
\item[$\lambda$] An eigenvalue of Eq.\ (\ref{eq:eigenequationAlt}).
\item[$\rho$] Density of test mass. (See Appendix \ref{ap:sapphire}).
\item[$\sigma$] Poisson ratio.  An elastic parameter in the
  stress tensor for an isotropic material [Eq.\ (\ref{eq:elas:Tab})]
\item[$\theta$] An angle through which the mirror is tilted.
\item[$\Theta$] The expansion associated with a displacement field $\Theta = \nabla_a y^a$ [e.g., in
  Eq.\ (\ref{eq:IA})].
Also denotes a unit step function [e.g., Eq.\ (\ref{eq:propagator:Cut})].
\item[$\omega$] $=2\pi f$.
\item[$\zeta_{1,2}$] Norm of the first- and second-order corrections to the optical state of
  the cavity when the mirror is displaced
  [i.e.\ in
  Eqs.\ (\ref{eq:disp:results:gaussian}) and (\ref{eq:disp:results:MH})].
\item[$\zeta_m$] Used \emph{only} in Eq.\ (\ref{eq:thel:LTp}), $\zeta_m$ denotes the $m$th zero
  of $J_1(x)$.
\end{itemize}

\subsubsection{Tensor notation}
The elasticity calculations (e.g., involving the elasticity tensor $T_{ab}$) involve tensors.
While the authors prefer to interpret these expressions with  an abstract tensor notation
(i.e.\ Wald  \cite{Wald}), the reader may without loss of generality use global cartesian
coordinates and interpret the latin indices $a$ and $b$ as running from $1\ldots 3$, indexing
the coordinate directions; in these coordinates, upper and lower
indicies are entirely equivalent (e.g.\ $y^a = y_a$).   In this case, the reader should interpret 
$\nabla_a$ as the coordinate partial derivative $\partial/\partial x^a$ and $g_{ab}$ as the
$3\times 3$ identity matrix.

\subsubsection{State and operator notation}
As explained briefly in Sec.\ \ref{sec:subsub:paraxial}, because propagation of the state of the
field from one plane to another is merely a unitary transformation, we  substantially
simplify the many integral-equation operations by using a quantum-mechanics-style notation for
the operators.  For example, the integral operation
\begin{equation}
\label{eq:def:notation:integralopFull}
E(\vec{r}', z') = \int d^2 r \; U(\vec{r}',z'; \vec{r},z) E(\vec{r},z) 
\end{equation}
is represented as
\begin{equation}
\label{eq:def:notation:integralopCompact}
E(z') = U(z',z) E(z) \; .
\end{equation}
Further, typically the relevant planes (i.e.\ the $z,z'$) are known, and omitted, in expressions
like the above.

Under a similar philosophy, we also use quantum-mechanics-style notation for the values of the
electric field on a specific plane; in other words, we represent states of the optical cavity
using quantum-mechanical state notation.  For example, the optical state of the cavity may be
equivalently denoted by the explicit functional form $u$ or $u(\vec{r})$, or alternatively as
the state vector $\left| u \right>$.   In a similar spirit, we index the potential solutions to
the eigenequation [Eq.\ (\ref{eq:eigenequationAlt})] by an integer $p$, and denote the set of
basis states as $\left| p \right>$.
 More generally we can represent \emph{any} function  defined on
a two-dimensional plane---not  necessarily normalized; not necessarily a potential state---as a  state
vector. 

We use this notation in particular to simplify notation for inner products between two states
$\left|u\right>$ and $\left| v\right>$:
\begin{equation}
\left<v|u\right> \equiv \int d^2 r \; v^*(r) u(r) \; .
\end{equation}
Finally, to simplify the frequent expressions that involve the norm of a state, we use the
symbol $||\cdot ||$ to denote $L^2$ norms:
\begin{equation}
||u||^2 \equiv \left< u | u \right > = \int d^2 r |u(\vec{r})|^2 
\end{equation}

Many quantum mechanics textbooks written over the past half century employ this notation (cf.,
e.g., \cite{Dirac,Landau,Townsend}).

\section{\label{sec:th:Ideal}Theory of an idealized Mexican-hat arm cavity}
%
%
%
%

In this section, we review the equations needed to understand the design of an arm cavity which
(i)  uses mesa beams, (ii) reduces thermoelastic noise in an advanced LIGO
interferometer, and 
(iii)   satisfies advanced LIGO design constraints (i.e.\ the diffraction losses per bounce
are less than the advanced LIGO design threshold of $10$ppm, and the mirrors have mass
$40\text{ kg}$). 

Specifically, we  introduce the mesa beam, defined for the purposes of this paper by
Eq.\ (\ref{eq:buildMHcanonical}).  We define the mirror surfaces which
confine this beam by Eq.\ (\ref{eq:def:MHheight}).
From Eq.\  (\ref{eq:buildMHcanonical}) we can determine the beam intensity profile.  
We can relate the beam intensity profile
to the power spectrum of thermoelastic noise [Eq.\ (\ref{eq:Sh})] via Eqs.\ (\ref{eq:IA}) and (\ref{eq:elas}).
 Finally,
we describe how to limit attention to only those mesa beams with low diffraction losses, using an
approximate approach [the clipping approximation, Eq.\ (\ref{eq:clip})] which we will test
 against an exact expression [Eq.\ (\ref{eq:diffLossNet})].

\subsection{\label{sec:th:MH}Mexican-hat mirrors and mesa beams}
\begin{figure}
\includegraphics{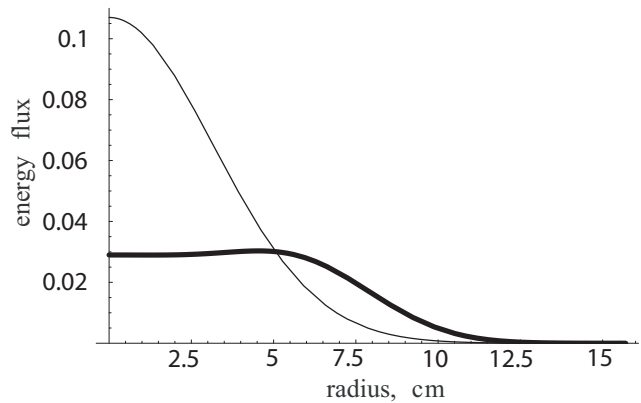}
\caption{\label{fig:fiducialMH}The heavy solid curve is an example of a mesa beam's intensity
  profile; the lighter
  curve is the intensity profile of a Gaussian beam with similar losses.  More precisely, this curve shows the energy fluxes per unit area (normalized to
  unity) for the mesa beam (thick curve) and Gaussian beam (thin curve) that have 10ppm of
  diffraction losses on a mirror with coated radius 15.7 cm.  
}
\end{figure}

The mesa-beam interferometer (MBI) paper \cite{MBI} presented explicit forms for certain specific mesa beams (``canonical mesa
beams'') produced inside a cavity bounded by two \emph{identical} mirrors.  [In MBI, only the
values of the beam fields at the surface of the ITM (specifically, values propagating away from
the ITM surface) were given; Appendix \ref{ap:mh} provides a more comprehensive discussion.]
We summarize MBI Eqs.
 (2.3), (2.6), and (2.7) to provide a compact summary of how to compute  the values
 $u_\text{mesa}$ of a mesa 
beam's normalized electric field on the mirror face, at radius $r$ from the optic axis:
\begin{subequations}
\label{eq:buildMHcanonical}
\begin{eqnarray}
U(\vec{r}, D) &=& 
   \int_{r'<D} d^2 r' \exp\left[ -\frac{(\vec r - \vec {r '})^2 (1+i)}{2 b^2} \right] \\
N^2(D) &\equiv& \int_0^\infty |U(D,r)|^2 2 \pi r dr \\
u_\text{mesa}(D,r) &\equiv& U(D,r)/N(D)
\end{eqnarray}
\end{subequations}
where in the first integral the integration is over all points  with $r'\equiv |\vec{r'}|<D$.
[For readers who wish to numerically explore mesa fields themselves, note that MBI Eq.\ (2.5)
provides a more efficient means of calculating the mesa amplitude function $u(D,r)$.]
Figure \ref{fig:fiducialMH} provides an example of a mesa-beam intensity distribution
$|u(D,r)|^2$. 

\begin{figure}
\includegraphics{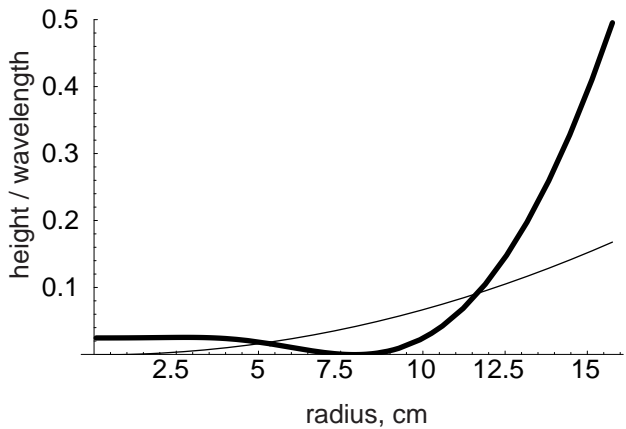}
\caption{\label{fig:fiducialMHmirrors}The heavy solid curve is an example of a Mexican-hat
  mirror; the lighter curve is an example of a spherical (i.e.\ parabolic) mirror.   
 More
  precisely, this curve  shows the spherical and Mexican-hat mirrors
  which produce the beams described in Fig. \ref{fig:fiducialMH}
 }
\end{figure}

As discussed in Appendix \ref{ap:mh} and in MBI Sec.\ II D [cf.\ MBI Eq.\ (2.13)],
the mirrors that reflect mesa beams back into themselves (denoted \emph{Mexican-hat mirrors}) 
necessarily have a continuous height
function $h_\text{MH}$ given by
\begin{equation}
\label{eq:def:MHheight}
k h_\text{MH} = \text{Arg}[u_\text{mesa}(D,r)] \; .
\end{equation}
where $k=2\pi/\lambda$ is the wavenumber of light in the arm cavity.  The solid curve in Figure
\ref{fig:fiducialMHmirrors} provides an example of a Mexican-hat mirror.

\subsection{\label{sec:th:thermo}Thermoelastic noise}
Using the expressions in the previous section, one can design idealized interferometers which
operate with mesa beams and Mexican-hat mirrors.  The thermoelastic noise associated with the
resulting 
interferometer design can be discovered with the techniques of Liu and Thorne (LT)
\cite{LiuThorne}, who employ Levin's direct method \cite{Levin} to derive
the following 
formula for an interferometers' thermoelastic noise in terms of a noise integral $I_A$:
\begin{equation}
\label{eq:Sh}
S_h(f) = 4 \kappa k_B \left( {\alpha_l E T \over C_V (1-2\sigma) \rho \omega L }\right)^2 \sum_{A=1}^4 I_A\;;
\end{equation}
cf.\ LT Eqs.\ (3), (4) and (13); 
the notation used here has been described in 
Sec.\ \ref{sec:sub:notation}. 

The sum on $A=1,2,3,4$ is over the four test masses, and the quantity summed is
the noise integral
\begin{equation}
\label{eq:IAraw}
I_A = {2\over F_o^2} \int_{V_A} \left< (\vec\nabla \Theta)^2 \right> d{\rm volume}\;.
\end{equation}
In Eq.\ (\ref{eq:IAraw}), $\Theta$ is the expansion (i.e.\ fractional volume change) resulting from a 
pressure---sinusoidally oscillating at frequency $f=\omega/2\pi$,
proportional to the light beam intensity, with net force $F_o$---imposed on the face of the
mirror; the integral is over $V_A$, the volume of test mass $A$; and the average
$\left<\ldots\right>$ is over one oscillation period $1/f$.
[Note that the dimensions of $I_A$ are 
length/force$^2 = {\rm s}^4 {\rm g}^{-2} {\rm cm}^{-1}$.]

In the analysis that leads to the above expressions [Eqs.\ (\ref{eq:Sh}) and (\ref{eq:IA})], LT 
\cite{LiuThorne}
have made certain idealizations and approximations, some better than 
others.\footnote{\label{note:idealizeGeneral}For example,  they
idealize the test-mass material as an isotropic material; however, the proposed test mass
material, sapphire, is 
only moderately well approximated as isotropic.
They also assume the elastic response of the solid is \emph{quasistatic} (introducing errors of
order the 
sound crossing time over the gravitational wave period, $\sim 10^{-3}$) and that heat diffuses
slowly so the elastic response of the solid to the imposed pressure profile is \emph{adiabatic}
(introducing errors of order the diffusion length over the beam size, or $\sim
0.3\text{ mm}/10\text{ cm}\sim 3\times 10^{-3}$).
}
We shall employ this expression as it stands, using the precise parameter values provided in Appendix \ref{ap:sapphire}.

\subsubsection{Effective static elasticity model for $I_A$}
For the frequencies at which advanced LIGO will be most sensitive ($f \sim 100 \text{Hz}$),
the interior of the test mass will respond effectively instantaneously to the imposed
pressure profile in the model problem described above.\footnote{\label{note:idealizeSound}The
  mirrors considered in this 
  paper have characteristic dimension of order $H\sim 10\text{ cm}$.   Thus, the sound crossing
  time is of order $H/c_s \sim H/\sqrt{E/\rho}\sim 10^{-5}\text{sec}$.  Since the body responds
  elastically to imposed forces on times of order a few sound-crossing times, the elastic
  response to a force which is imposed at frequency $\sim 100 \text{Hz}$ is effectively
  instantaneous. Equivalently, the same conclusion follows because the gravitational wave
  frequency is far below the frequency of any resonance of the test mass.}
As a result, by going to the
accelerating frame of the test mass and following that accelerating 
frame for one period to simplify the average,  we can 
approximate $I_A$ 
by
\begin{equation}
I_A \approx {1\over F_o^2} \int_{V_A} (\vec\nabla \Theta)^2 d{\rm volume}\;.
\label{eq:IA}
\end{equation}
where the expansion $\Theta = \nabla_a y^a$ arises due to the response (i.e.\ local displacement
field) $y^a$  of the test-mass substrate to the following \emph{static} conditions:
\begin{subequations}
\label{eq:elas}
\begin{enumerate}
\item \emph{Static (accelerating-frame) force}: Some net force $F_o \hat{z}$ acts uniformly throughout
  the test mass (i.e.\ provides a force density $F_o/V_A$).  To be specific, using the equations
  for an isotropic elastic solid (cf.\ Blandford and Thorne \cite{BlandfordThorne}), the
  response field $y^a$ satisfies
\begin{eqnarray}
\label{eq:elas:eom}
\nabla^a  T_{ab} &=& \hat{z}_b  F_o / V_A  \; ,  \\
\label{eq:elas:Tab}
T_{ab} &\equiv& -\frac{E}{(1+\sigma)(1-2\sigma)} \times  \nonumber \\
 & & \left[ \frac{(1-2\sigma)}{2} (\nabla_a y_b +
   \nabla_b y_a ) 
  + \sigma \Theta  g_{ab}     \right] \; .
\end{eqnarray}
\item \emph{Static pressure on mirror face}: An equal and opposite net force $-F_o\hat{z}$ acts
  on the 
  mirror face, with a distribution $P(r)$ proportional to the beam intensity
  profile.  As a result, the displacement field $y^a$ must satisfy
\begin{eqnarray}
\label{eq:elas:bcForce}
T_{ab}(r,z=0) \hat{z}^b &=& - F_o P(r) \hat{z}_a \; ,
\end{eqnarray}
where the mirror's front (reflecting) surface is at $z=0$ and where the mirror lies below
$z=0$.
\item \emph{Break translation symmetry}:  Finally,  translation symmetry must be broken 
 (i.e.\ so   numerical simulations converge to a unique solution).  To
  break translation symmetry, we  fix ``the location'' of the mirror, where that
  location is determined as the average location of points in a set ${\cal R}$.  Therefore,
  to specify a unique solution, we require the displacement field $y^a$ satisfy
\begin{equation}
\label{eq:elas:bcAverage}
\left< y^a \right>_{\cal R}= 0
\end{equation}
 where ${\cal R}$ is some arbitrary nonempty set and $\left < \cdot \right>_{\cal R}$ denotes the
 average over ${\cal R}$.
\end{enumerate}
\end{subequations}
[This quasistatic approximation to $I_A$ and the overall thermoelastic power spectrum
[Eq.\ (\ref{eq:Sh})]  involves errors only of order $10^{-3}$ relative to the general
Liu-Thorne  expression (\ref{eq:IA}) for $I_A$ \cite{LiuThorne};  cf.
footnotes \ref{note:idealizeGeneral} and \ref{note:idealizeSound}, or Appendix
\ref{ap:modelProblem}.]

\subsection{\label{sec:th:propagation}Propagation, the eigencondition, and diffraction losses}


In the previous two sections, we described mesa beams and the technique by
which one can determine, given a  beam and four  mirrors that support
that beam (two identical ITMs and two
identical ETMs), the thermoelastic noise an
interferometer using those  mirrors and beams would experience.  But not all combinations of
mesa beams and mirrors satisfy advanced LIGO design constraints;
the advanced LIGO design specifications require each arm cavity have low diffraction 
losses per bounce ($10$ppm).

In this section, we describe how the diffraction losses can be precisely computed
[Eq.\ (\ref{eq:diffLossNet})] and estimated [the clipping approximation, Eq.\ (\ref{eq:clip})].
Using the diffraction losses, one can then find
combinations of mesa beams and mirrors which produce low diffraction losses per bounce.

Our analysis proceeds by describing in general terms the propagation of light in an arm cavity,
focusing in  particular on eigenmodes and their loss per round trip. 
In the process of justifying our simple estimate of diffraction loss,  the clipping
approximation [Eq.\ (\ref{eq:clip})], we will review the  definitions and properties of
eigenmodes of the arm cavity.   This review  establishes notation and conventions
for our presentation of optical perturbation theory, 
as discussed in Sec.\ \ref{sec:th:Defects}.

\subsubsection{\label{sec:subsub:paraxial}Principles of paraxial optics}
The light in LIGO arm cavities is well-described by the paraxial approximation.  
In this section, primarily to establish notation conventions,  we briefly review the
propagation of light under the paraxial approximation.
A detailed description of the relevant physics can be found in standard references.%
\footnote{
For a more pedagogical and yet brief presentation, we recommend Blandford and Thorne's
treatment \cite{BlandfordThorne}.   Other pedagogical (and technical) treatments can be found
in many laser physics books, e.g., \cite{Siegman,Siegman2,Yariv,LaserHandbook}.   We also recommend the collection of original research articles on laser
and resonator physics compiled by Barnes \cite{Barnes}.
}
Briefly, in this approximation, we can completely describe the state of the wave by the wavelength
$\lambda$  of the light used and the values of the wave amplitude on some fixed plane
$z=$constant (i.e.\ transverse to the optic axis) and at some time $t$  (cf.\ 
\cite{BlandfordThorne}).  
The values at any causally-related later  combination $t',z'$---which must be separated from
the transverse  plane $z$ at 
time $t$   by a light ray
(i.e.\ $t'-t=|z-z'|/c$, mod reflections)---can be deduced by applying the appropriate linear 
functional to these states.

For example, we could characterize the optical state by the value of the electric field
on some plane $z$.\footnote{We limit attention to a single polarization, e.g., the polarization
  excited by LIGO.}   We would then
relate the field at any other plane and at any other time to our initial state via the linear
operation 
\[
E(\vec{r}', z',t') = \int d^2 r \; U(\vec{r}',z'; \vec{r},z) E(\vec{r},z,t) \; ,
\]
cf.\ Eq.\ (\ref{eq:def:notation:integralopFull}).
For compactness and clarity, we employ a quantum-mechanics-motivated notation,\footnote{%
Most quantum mechanics textbooks written over the past half-century (e.g.
\cite{Dirac,Landau,Townsend}) have adopted a similar
operator notation. 
} in which the
integrals are suppressed and the operation above is denoted by [cf.\ Eq.\ (\ref{eq:def:notation:integralopCompact})]
\[
E(z',t') = U(z',z) E(z,t) \; .
\]
By way of example, the following are kernels of integral operators which describe  free propagation down a
length $L$;  reflection off a mirror of height  $h_{1,2}$; and  a
``window'' that cuts out all light outside a radius $r=R_{1,2}$, respectively:
\begin{subequations}
\label{eq:propagatorExamples}
\begin{eqnarray}
\label{eq:propagator:Free}
G_L(\vec{r},\vec{r}\,') &\equiv& -i \frac{k}{2\pi L} \exp i \left[ \frac{(\vec{r}-\vec{r}\,')^2}{2 L/k} + k L \right] \\
G_{1,2}(\vec{r},\vec{r}\,') &\equiv& - \delta(\vec{r}-\vec{r}\,') \exp \left[ -2 i k h_{1,2}(r)
\right] \\
\label{eq:propagator:Cut}
T_{1,2}(\vec{r},\vec{r}\,') &\equiv& \delta(\vec{r}-\vec{r}\,') \Theta(R_{1,2}-r)
\end{eqnarray}
\end{subequations}
where $\Theta(x)$ is a step function equal to $1$ when $x>0$ and $0$ otherwise, and where the negative sign in $G_1$ arises because of boundary conditions on the
electric field at a perfectly reflecting surface;
cf.\ \cite{BlandfordThorne,Siegman,Siegman2}, and other previously noted
references on paraxial optics for further details. 

These propagation operations can be combined to generate more complicated processes. For
example, while $G_1$ describes reflection off an \emph{infinite} mirror of height $h_1$, $G_1
T_1$ describes the reflection of light off a \emph{finite} mirror of height $h_1$ and radius
$R_1$.

\subsubsection{Describing paraxial propagation through an arm cavity}
\begin{figure}
\includegraphics{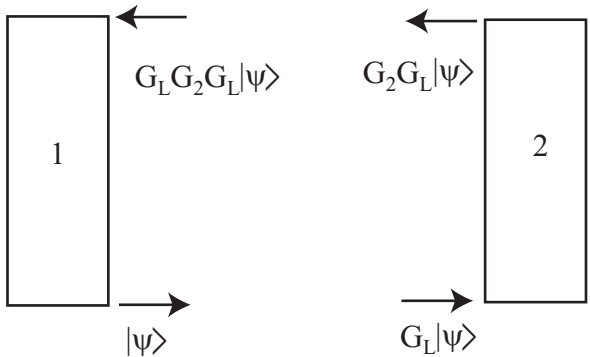}
\caption{\label{fig:propagationDemo}This figure demonstrates how propagation operators described in
  Eq.\ (\ref{eq:propagatorExamples}) are used to evolve the
  field from one plane of constant $z$ to another.  For brevity, the operators $T_1$ and $T_2$,
  which allow for the finite extent of the mirrors' surfaces, have been omitted in this figure.}
\end{figure}

When we have an arm cavity bounded by two cylindrical mirrors separated by a length $L$, of
cylinder radius $R_{1,2}$ and surface heights 
$h_{1,2}$ respectively, we can use the above propagators [Eq.\ (\ref{eq:propagatorExamples})] to describe the free, undriven
propagation of light within the arm cavity\footnote{The description of an arm cavity driven by a
source laser adds only a straightforward inhomogeneous term to our discussion; the eigenmodes
of the homogeneous term are as always in such problems of paramount importance.
}.
For example, as we demonstrate graphically in Fig. \ref{fig:propagationDemo}, we can express the field at the surface of mirror $1$---specifically, the field
for light heading \emph{away} from
the mirror surface---at time $t+2L/c$ in terms of the field at that surface at time $t$ as follows:
\begin{equation}
\label{eq:roundtrip}
u(t+2L/c) = G_1 T_1 G_L G_2 T_2 G_L u(t)
\end{equation}
for the operators on the right side as given in Eq.\ (\ref{eq:propagatorExamples}) and for
$\left| u \right>$ denoting some polarization of the electric field. 

We are in particular interested in eigenmodes---that is, states $u$ so the beam returns
proportional to itself: 
\begin{equation}
\label{eq:eigenequation}
\eta e^{2 i k L} u = G_1 T_1 G_L G_2 T_2 G_L u
\end{equation}
for some $\eta$.  [We factor the phase $e^{2ikL}$ out of $\eta$ to eliminate the effect of the
phase factor $e^{ikL}$ present in the operator $G_L$.]  These eigenmodes are \emph{resonant} when
$\eta e^{2 i k L}$ is \emph{real} and positive.  
The arm 
cavity 
length $L$ and wavevector $k$ are
tuned so one state, the \emph{ground state} $u_o$, is resonant; in this case, $\eta_o$, $k$, and
$L$ satisfy
\begin{equation}
\label{eq:def:makeGroundResonant}
\text{Arg}\left[\eta_o e^{2 i k L}\right] = 0 \; .
\end{equation}

The mesa beam [Eq.\ (\ref{eq:buildMHcanonical})] is designed specifically to be one of the
eigenmodes of the Mexican-hat mirrors [Eq.\ (\ref{eq:def:MHheight})] when the mirrors are infinite
(i.e.\ $T_1=T_2=\textbf{1}$), and when the cavity length is properly tuned to admit it.  When the
mirrors are finite, the true eigenmodes of an arm cavity (be it bounded by spherical or
Mexican-hat mirrors) will have a slightly different shape than the shape supported by infinite
mirrors.  Further, the eigenvalue of the arm cavity $\eta$ will no longer be of unit magnitude (i.e.\ $|\eta|<1$).


\subsubsection{Diffraction losses: Exactly}
From Eqs.\ (\ref{eq:roundtrip}) and (\ref{eq:eigenequation}), we know that when the initial
optical state is an eigenmode, the optical state after $n$  round trips decays as measured
by its norm, which 
evolves as
$||u(t+n 2L/c)|| = |\eta|^n ||u(t)||$.  Thus,  if $|\eta|<1$ the magnitude of
$u$ (i.e.\ the $L^2$ norm of $u$) decreases with each round trip.

 Since our model above only permits losses from
diffraction, and since the power in the cavity is proportional to $\int d^2 r |u|^2$
(i.e.\ the $L^2$ norm of $u$, squared), the following quantity is the diffraction loss
 per round trip:
\begin{equation}
\label{eq:diffLossNet}
{\cal L}_\text{net} = 1 - |\eta|^2 \; .
\end{equation}
When the two mirrors are identical, as is our case, we can subdivide the loss in two, to
obtain a meaningful diffraction loss ``per bounce''.

We can therefore extract from numerical solutions to   Eq.\ (\ref{eq:eigenequation})
(cf.\ Sec.\ \ref{sec:num:modes}) the magnitude of $\eta$  and, by the above expression
[Eq.\ (\ref{eq:diffLossNet})],  deduce the mesa-beam diffraction losses in the
presence of finite mirrors.   We do \emph{not}, however, use this method in this paper.

\subsubsection{Diffraction losses: The clipping approximation}
In practice the exact expression for the diffraction losses given above
[Eq.\ (\ref{eq:diffLossNet})] is slow to evaluate, because we must solve for fine details of
the eigenequation for each mirror size
$R_{1,2}$ of interest.   
To obtain a rough estimate of the diffraction losses, usually accurate to a factor of order
unity, we will often use instead the clipping approximation.

The clipping approximation estimates diffraction losses by
assuming the beam is not significantly changed by those losses; the beam has to a good approximation the same
shape as it would have in the presence of infinite mirrors.  Assuming the beam profile is
known, we then  directly compute the losses by determining the power lost off the edges of finite
mirrors on each reflection.  Specifically, we assert ${\cal L}_\text{net}\approx {\cal
  L}_\text{clip}$, with 
\begin{subequations}
\label{eq:clip}
\begin{eqnarray}
{\cal L}_\text{clip} &=& {\cal L}_1 + {\cal L}_2 \\
{\cal L}_{1} &=& |(1-T_1)u_o|^2 = \int_{r>R_1} |u_o|^2 d^2 r \\
{\cal L}_{2} &=& |(1-T_2)G_L u_o|^2 = \int_{r>R_1} |G_L u_o|^2 d^2 r 
\end{eqnarray}
\end{subequations}
where $u_o$ is an eigensolution to the propagation equation [Eq.\ (\ref{eq:eigenequation})]  computed assuming infinite mirrors.

\subsubsection{\label{sec:sub:diffConstraint}Diffraction losses of the ideal mesa beam,
  computed via the clipping approximation and corrected to estimate the true diffraction losses}
\begin{figure}
\includegraphics{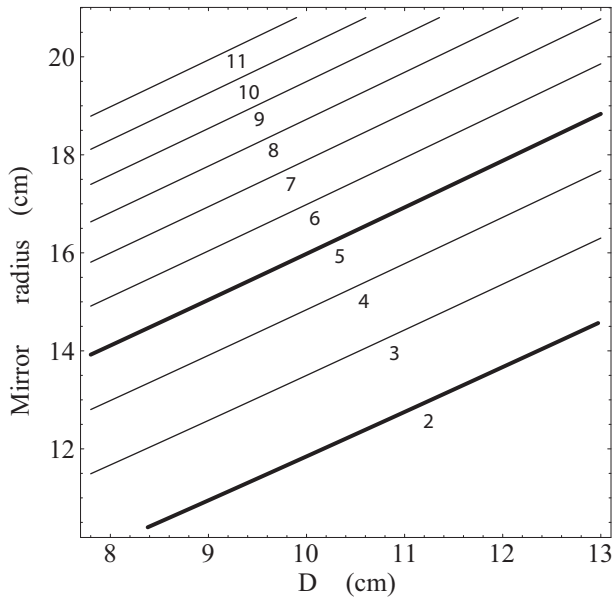}
\caption{\label{fig:mhClippingLosses}A contour plot of the ($\log_{10}$ of the) diffraction
  losses of the mesa beam [Eqs.\ (\ref{eq:buildMH}) with $\bar{z}=L/2$, i.e.
  (\ref{eq:def:mhUgmin})]  
  according to the clipping approximation [Eq.\ (\ref{eq:clip})], as  a function of mirror
  radius $R$ and averaging scale $D$.   Solid curves indicate  losses $10^{-n}$ for
  $n=2,3,\ldots 11$.
  The two heavy solid curves correspond to  $n=2$ and $n=5$; these two curves have special significance to LIGO design (they determine the
  maximum value of losses off the back and front faces of an ITM, per bounce;
  cf.\ Sec.\ \ref{sec:th:Design:constraints}).
}
\end{figure}

In this paper, we  approximate the diffraction losses of the mesa beam resonating between two
identical Mexican-hat 
mirrors by the clipping approximation [Eq.\ (\ref{eq:clip})] applied to the mesa beam
[Eq.\ (\ref{eq:buildMHcanonical})].  
These losses are shown in
Fig. (\ref{fig:mhClippingLosses}).

The true diffraction losses per bounce generally  differ from the estimates of clipping
approximation. 
For example,  for a $D=4\, b$ mesa beam resonating between $R=16\text{ cm}$
cylindrical mirrors, the true diffraction losses are about $18.5 \text{ ppm}$, while the clipping
approximation estimates diffraction loses of $21\text{ ppm}$.  Based on this good agreement, in
this paper we assumed the true diffraction losses were well-approximated by the
clipping-approximation estimate, with relative errors of order a factor $1.2$.

\subsection{\label{sec:th:Design:constraints}Advanced LIGO design constraints} 
The advanced LIGO interferometers are composites of many interrelated systems; each component
of that system has been designed making assumptions about the other components.  In particular,
the advanced LIGO interferometers require (roughly) the following constraints on mirror and beam designs:
\begin{enumerate}
\item \emph{Mirror mass of $40$}kg:  We consider only $40$kg test-mass mirrors, because 
  the present advanced LIGO design for the suspension and seismic isolation system 
  requires $40$kg mirrors. [Also, the advanced LIGO suspension system cannot  support heavier
  mirrors (Phil 
  Willems, private communication)].\footnote{The support system could hold up lighter mirrors.
    However, thermoelastic noise decreases as the mirrors and beams are increased in size (in
    equal proportions).  We therefore limit attention to the most massive possible mirrors.}

\item \emph{Mirror front radius limited by arm-cavity diffraction losses}:
We consider only combinations of mirror front face radius $R$ and beam size (e.g.
if mesa beams are used, $D$) which have diffraction losses equal\footnote{The advanced LIGO 
  design could accept mirrors with lower losses.  However, since thermoelastic noise decreases
  (as a general rule) when the  beam radius increases, we limited attention to the largest
  possible beams.} to  $10$ parts per million
(ppm).%
\footnote{\label{note:diff:explainITM}This constraint is required to keep the diffraction losses bounded by a reasonable portion
 of the total loss.  Specifically, 
for the baseline design there is 125 W of input power to the interferometer and
830 kW of circulating power in each arm cavity.  The 10 ppm of diffraction loss per bounce
results in a diffraction power loss in the arm cavities of $4 \times 10 \hbox{ppm} \times 830
 {\rm kW} = 33$ W, which is 25 per cent of the 125 W of input light, a reasonable value.}

\item \emph{Mirror back radius limited by power-recycling-cavity diffraction losses}:
Finally, we consider test-mass mirrors whose back face is sufficiently large that light
entering the arm cavity through the back face loses at most $1$ percent (i.e.\ $10,000$ppm) of
the input power.\footnote{\label{note:diffloss:explainITMback} As with the constraint involving the front face
  size (cf.\ footnote \ref{note:diff:explainITM}), this constraint is required to keep the
  diffraction losses bounded by a reasonable proportion of the total loss.
Specifically, for the baseline design the power impinging on each ITM is 1.05 kW, so one per
cent diffraction loss (i.e.\ $10,000$ppm) corresponds to 
losing $2 \times 0.01 \times  
1.05 {\rm kW} = 21$ W at the ITM input, which is 17 per cent of the 125 W total laser power.
}
When evaluating this constraint, we assume the input light is in the same state as the arm
cavity light (i.e.\ in a cavity eigenmode).

[This constraint only matters for noncylindrical mirrors (i.e.\ frustums) with a relatively
small back face size; for cylindrical mirrors, it holds automatically.]

\end{enumerate}

\subsection{\label{sec:th:Design:summary}Summary: Exploring mesa-beam arm cavity configurations}
To summarize, we design and evaluate new mesa-beam interferometer configurations by the
following process:
\begin{enumerate}
\item \emph{Pick a configuration}: We (i) select  two axisymmetric mirrors  (i.e.\ cylinders or
  frustums) which have the  \emph{same} front face radius and (ii) select some value $D$.  We
  shape  the
  mirror faces [Eq.\ (\ref{eq:def:MHheight})] and tune the length of the arm cavity 
  [Eq.\ (\ref{eq:def:makeGroundResonant})] so the ground state eigenmode of that 
  cavity---the eigenmode
  which is approximately a mesa beam with length scale $D$---is resonant.

\item \emph{Check that the configuration satisfies the advanced LIGO design constraints}:
  If  (i) the  mirrors masses are not equal to $40\text{ kg}$, if (ii) the physical
  diffraction losses 
  in the arm 
  cavity  are greater than $10\text{ ppm}$ per bounce [Eq.\ (\ref{eq:diffLossNet})], or if
(iii) the diffraction losses for input light are greater than one percent, then
  we stop and try again: the configuration does not satisfy the
  advanced LIGO design constraints described in Sec.\ \ref{sec:th:Design:constraints}.
  [In this paper, we use only the clipping approximation, Eq.\ (\ref{eq:clip}), to estimate
  diffraction losses.]

\item \emph{Evaluate the thermoelastic noise of this configuration}:  We finally evaluate the
  noise integrals $I_1$ and $I_2$ for each of the two arm-cavity mirrors
  [Eqs.\ (\ref{eq:IA}) and (\ref{eq:elas})], using as beam pressure profile the (normalized) beam intensity
  profile [Eq.\ (\ref{eq:buildMHcanonical})]:\footnote{Ideally, we \emph{should} use as beam intensity profile the true eigenstate
    appropriate to two \emph{finite} mirrors, as obtained by solving
    Eq.\ (\ref{eq:eigenequation}).  However, because diffraction losses are low, the 
    eigenstate for finite mirrors is very well approximated by the eigenstate for infinite
    mirrors; and the latter is the mesa-beam amplitude $u_\text{mesa}$.}
\begin{equation}
P(r) = |u_\text{mesa}(r,D)|^2 \; .
\end{equation}
 [Section \ref{sec:num:thermo} will describe in detail precisely how we evaluated the
two  thermoelastic noise integrals.]    

These thermoelastic noise integrals then determine [via Eq.\ (\ref{eq:Sh})] the
overall thermoelastic noise of an interferometer using two identical arm cavities, where
each cavity consists of this particular mirror and beam configuration.
\end{enumerate}

\section{\label{sec:th:Driving}Driving  a mesa-beam arm cavity with a Gaussian beam}
Ideally, a mesa-beam arm cavity would be driven by light already in the mesa-beam state.
But if the optics required to generate mesa-beam inputs are unavailable or too 
inconvenient, Gaussian beams---generated by conventional optics---can be
used to drive the arm cavities.  With the proper choice of Gaussian, the arm cavity behaves
almost as if the ideal mesa beam was used.

In the remainder of this paper, as in the mesa-beam interferometer
paper (MBI) \cite{MBI}, we shall assume properly-chosen Gaussian beams are used to
drive the interferometer. 

In her companion paper \cite{ErikaSimulations}, Erika D'Ambrosio provides a
comprehensive analytic and numerical discussion of Gaussian beams driving mesa-beam cavities
[cf., e.g., ED'A Eq.\ (3)].
In this section, we briefly summarize some of her results on mesa-beam cavities driven by
gaussian beams.

\subsection{Beam and cavity used in this section}
\subsubsection{\label{sec:def:fiducial}Fiducial mesa-beam cavity}
Rather than study all possible interferometer configurations (i.e.\ all possible mirror and beam
sizes), we select  a single \emph{fiducial} mesa-beam arm cavity: the cylindrical mirrors have
front face size $16 \text{ cm}$ and the mesa beams have $D=4b=10.4\text{ cm}$, where 
$b=\sqrt{\lambda L/2\pi}=2.60\text{ cm}$ (and where $\lambda=1.064 \mu\text{m}$ and
  $L=4\text{km}$ are the light wavelength and arm length) [cf.\ MBI Sec.\ IV A 2].
This fiducial arm cavity has diffraction loss $18\text{ ppm}$.\footnote{While not \emph{quite}
appropriate---the diffraction losses are slightly larger than the advanced LIGO design cutoff of
$10\text{ ppm}$ \cite{advLIGOmain}---this fiducial design is sufficiently similar to acceptable
advanced LIGO designs.}

We also assume the mirrors bounding the arm cavity---both 
 the input test mass (ITM) and end test mass (ETM)---have no intrinsic losses,
except for diffraction losses (which we treat separately).  Specifically, we assume the 
ETM is perfectly reflecting, and assume the ITM has power transmissivity $t_\text{I}^2 = 0.005$
\cite{advLIGOmain}.
Thus, our power reflection and transmission coefficients for the ITM and ETM are
\begin{eqnarray}
&&r_\text{I}^2 = 0.995\;, \quad t_\text{I}^2 = 0.005\; ; \nonumber \\
&&r_\text{E}^2 = 1.0 \;, \quad t_\text{E}^2 = 0 . 
\end{eqnarray}

[This same fiducial cavity will appear frequently elsewhere; for example, we explore
perturbations of this particular fiducial cavity when exploring the sensitivity of the
interferometer to tilt.]

\subsubsection{\label{sec:th:drive:overlap}Gaussian beam used to drive the fiducial arm cavity} 
As described in MBI Sec.\ IV B, this fiducial arm cavity will be driven by a Gaussian beam of
form given in MBI Eq.\ (2.8), with Gaussian beam radius $r_{od} = 6.92 cm$.  As in MBI, 
we denote  this    state by $\left| u_d \right>$.  This state has overlap
\begin{equation}
\label{eq:def:overlapValue}
\gamma_o^2 \equiv \left| \left< u_d | u_o \right> \right|^2 = 0.940
\end{equation}
with the arm cavity's ground state $u_o$ [cf.\ MBI Eq.\ (4.2)]. 
[Erika D'Ambrosio has found that this
particular Gaussian beam gives the largest possible overlap with our fiducial beam; cf.\ ED'A 
Eq.\ (3).]

Strictly speaking, we use a  Gaussian state $\left | u_d \right>$ which is shifted by precisely
the phase offset necessary to make $\gamma_o$ real and positive; therefore, 
\begin{equation}
\gamma_o = \sqrt{0.940} = 0.970\; .
\end{equation}

\subsection{Operation of fiducial cavity when driven by a Gaussian beam}

Equation (\ref{eq:def:overlapValue}) implies a
fraction $0.94$ of the  (gaussian) driving-beam light will enter the Mexican-hat-mirrored
cavity; the remaining $6$ percent will be reflected.  The fraction of the light that enters the
cavity is amplified (due to resonant interaction with the bounding cavity mirrors), but
eventually leaks back out the same way it entered (minus that fraction lost through diffraction
losses).   The combination of the $6$ percent 
reflected light and $94$ percent transmitted light (modulo losses) then returns back towards
the beamsplitter.

This small section provides a more quantitative view of this entire process.

\subsubsection{Decomposing light along the cavity eigenstates}
The transmission and reflection processes are best understood in terms of the resonant
eigenstate of the cavity.  We therefore rewrite the driving beam in terms of a projection along
and perpendicular to the ground state:
\begin{eqnarray}
\label{eq:def:inputDecomposition}
\left| u_d \right> &=& {\cal P}_o \left| u_d \right> + (1-{\cal P}_o) \left| u_d \right>  \\
  &=& \gamma_o \left| u_o \right> + (1-{\cal P}_o) \left| u_d \right>  \nonumber
\end{eqnarray}
where the projection operator ${\cal P}_o$ is defined so 
${\cal P}_o v = \left | u_o \right> \left< u_o | v\right>$ for any state $v$, i.e.
\begin{equation}
\label{eq:def:project}
{\cal P}_o \equiv \left | u_o \right> \left< u_o \right|\;,
\end{equation}
and where $\gamma_o$ is defined by (cf.\ Sec.\ \ref{sec:th:drive:overlap})
\begin{equation}
\label{eq:def:gamma0}
\gamma_o \equiv \left< u_o | u_d\right> \; ;
\end{equation}
the phase of $u_d$ has been designed so the unperturbed $\gamma_o$ is real.
Since the input state $u_d$ is normalized, 
\begin{equation}
||{\cal P}_o u_d||^2 + ||(1-{\cal P}_o) u_d||^2 = ||u_d||^2 = 1 \; .
\end{equation}

The portion $\gamma_o u_o$ of the driving field excites the mesa-beam cavity, while the portion
$(1-{\cal P}_o) u_d$ cannot resonate and thus gets fully reflected off the input test mass (ITM).

\subsubsection{Amplification of light entering cavity}
A fraction $\gamma_o^2$  of the light power  enters the arm cavity, resonates, and is
amplified by a factor
\begin{equation}
\label{eq:def:cavityGainRaw}
\frac{1+r_I}{1-r_I} \left( 1- \frac{2 {\cal L}_o}{1-r_I} \right)
\end{equation}
within the arm cavity,\footnote{The second (unnumbered) equation in D'Ambrosio
  \cite{ErikaSimulations} gives the transmitted field through the arm cavity.  Dividing this by
  the second mirror transmissivity, squaring it to get an expression for beam power, and then
  specializing to $r_2=1$ and $r_1=r_I$, we obtain the first factor in 
  Eq.\ (\ref{eq:def:cavityGainRaw}).  The second
  factor arises when losses are properly taken into account on each bounce.}
where ${\cal L}_o$ denotes the diffraction losses per bounce for the
ground state of the arm cavity.
Therefore, relative to the overall input power, the light in the arm cavity is amplified by a factor
\begin{equation}
\label{eq:gain:raw}
{\cal G}_\text{anal} = \gamma_o^2 \frac{1+r_I}{1-r_I} \left( 1- \frac{2 {\cal
  L}_o}{1-r_I}\right)  \approx 
  737 .
\end{equation}
[While the term ${\cal L}_o/(1-r_I)$ is very small in this case and can be ignored, we provide
the entire expression 
now for use in Sections \ref{sec:th:Defects} and \ref{sec:apply:Defects}, when we study the
effect of mirror perturbations on the cavity gain.] 

\subsubsection{Output: Transmission of light out of cavity, and recombination with reflected light}
The cavity's output light is the combination of light reflected at the ITM (which does not enter the cavity)
and light exiting the arm cavity through the ITM.
After some algebra (cf.\ \cite{ErikaSimulations}),\footnote{The portion of the beam
  amplitude proportional to the arm cavity ground state enters the arm
  cavity and resonates.  It therefore exits that
  the arm cavity with the same phase it enters.  On the other hand, the
  rest of the light experiences nearly perfect destructive
  interference.  For the rest of the light, the arm cavity acts like a
  perfectly reflecting mirror, so that portion of the light amplitude
  exits the arm cavity with the opposite phase than it entered with.
    For a more precise demonstration of this effect, see for example
  the first equation in D'Ambrosio
  \cite{ErikaSimulations}
with $r_2=1$, $t_1=t_I$ and $r_1=r_I$, and both exponential factors
  unity.
}
 we find that given a driving beam $u_d$ decomposed according to
  Eq.\ (\ref{eq:def:inputDecomposition}) the reflected light $u_r$ is in the state
\begin{eqnarray}
\label{eq:def:outputDecompositionRaw}
\left| u_r \right> &=& 
    {\cal P}_o \left|u_d\right> - (1-{\cal P}_o)  \left|u_d\right>  \\
   &=& (2 {\cal P}_o -1) \left|u_d\right> \; .  \nonumber
\label{eq:def:outputDecomposition}
\end{eqnarray}
where for simplicity we have ignored the (small) effects of diffraction losses on the light
that resonates in the cavity.

\section{\label{sec:th:Defects}Effect of mirror perturbations on arm cavities and the interferometer}
In Section \ref{sec:th:Ideal}, we described how to design an idealized mesa-beam arm
cavity and to evaluate the thermoelastic noise associated with such a design.

In this section, we shall describe in extremely general terms how one can use perturbation
theory to study the influence of defects on a single ideal (i.e.\ lossless) arm cavity.  We make
no reference to the 
specific details of the cavity used, save that---for technical convenience---we limit
attention in  our perturbation theory to a system with two identical, infinite
mirrors.  [Recall the canonical mesa beams, as defined in Sec.\ \ref{sec:th:MH}, are
designed for 
symmetric cavities.]   Naturally  the methods 
for perturbing arbitrary mirrors bear considerable similarity to methods for perturbing purely
spherical mirrors.   Many explicit tools and concepts carry over with little change to our more
generic approach.

Specifically, in Sec.\ \ref{sec:sub:basisStates} we introduce the  orthonormal basis
set of solutions we shall use to construct the perturbation theory expansion. 
Then, in Sec.\ \ref{sec:th:pt}, we list the explicit second-order
perturbation expansions we use to deduce the effect of mirror defects upon the optical
state of the cavity.  Finally, in Sec.\ \ref{sec:th:pt:imply}, we describe how changes in the
optical state of the cavity influence other quantities that can be deduced from that state,
such as the cavity diffraction losses [Eq.\ (\ref{eq:clip})] and the thermoelastic noise
integrals $I_A$ [Eq.\ (\ref{eq:IA})].

\subsection{\label{sec:sub:basisStates}Basis states for perturbation theory}
We will use perturbation theory to analyze the effect of small changes in the arm cavity
mirrors
on the solutions to the
eigenequation (\ref{eq:eigenequation}).  To construct perturbation theory expansions
by methods similar to those used in quantum mechanics (cf., e.g., \cite{Dirac,Landau,Townsend}), we 
\emph{prefer} to use as a basis for the perturbation expansion the eigenmodes of the initial equation [Eq.\ (\ref{eq:eigenequation})].%
\footnote{Actually, as Erika D'Ambrosio has frequently reminded us, we can construct
  perturbation theory just as well in the more general case, by using a dual basis.  See, for
  example, Appendix A of D'Ambrosio \cite{ErikaSimulations}.
}
Fortunately, as shown in Appendix \ref{ap:basis}, so long as the two mirrors are identical and infinite, the set of eigenmodes of
Eq.\ (\ref{eq:eigenequation}) are a complete, orthogonal set,  independent of the mirror
shape.

At this point, we could introduce basis states (and notation for basis states) which are
defined as eigensolutions to Eq.\ (\ref{eq:eigenequation}).  For technical reasons---our
methods simplify practical evaluation of the perturbation equation%
\footnote{On the one hand, by removing the common factor $\exp(i k L)$
present in $G_L$, we produce a naturally dimensionless eigenvalue problem (as $k$ and $L$ enter
only through the single length parameter $b$).  On the other hand, by focusing on ``half'' of
the propagator rather than a round-trip operator, we can easily deduce the relationship between
the fields at the two ends of the cavity.  The latter proves helpful, because we will study
perturbations of the mirror at one end of the cavity (end $2$), yet represent the state of the cavity field
by its values on the other end (end $1$).
}---we prefer instead to use a basis designed to simplify  ``half'' (i.e.\ a square root of) the round-trip operator, with
a certain phase factor removed.  In other words, we use as basis states the  eigensolutions $\left| p
\right>$, $\lambda_p$ to the equation
\begin{eqnarray}
\label{eq:eigenequationAlt}
\lambda_p \left| p \right> = e^{-i k L} G_1 G_L \left| p \right>
\end{eqnarray}
where $\left| p \right>$ is some state and $p$ denotes some index over all basis states.
[By an argument following that given in Appendix \ref{ap:basis}, this equation
admits a complete set of orthogonal  solutions $\left | p \right>$.]
These states correspond directly to solutions to the full
eigenequation.  Explicitly, 
if we insert a solution $\left| p \right>$ to Eq.\ (\ref{eq:eigenequationAlt}) into the
eigenequation [Eq.\ (\ref{eq:eigenequation})], we immediately conclude that $\left| p \right>$
is also an eigensolution of Eq.\ (\ref{eq:eigenequation}), with eigenvalue
\begin{equation}
\eta_p =  \lambda_p^2 \; .
\end{equation}

\subsection{\label{sec:th:pt}Effect of perturbations on light in the arm cavities}
When the mirror shapes $h_{1,2}$ are deformed, the light propagating in the cavity changes.
Given a basis of states and a specific problem to perturb  [Eq.\ (\ref{eq:eigenequation})],
we employ conventional techniques from quantum mechanics (cf., e.g., \cite{Dirac,Landau,Townsend}) to
compute that change.



\subsubsection{Results of perturbation expansion}
Using conventional quantum-mechanics-style techniques (cf., e.g., \cite{Dirac,Landau,Townsend}, but note the operator we perturb is
unitary rather than hermitian), we can devise a perturbative expansion for the cavity
ground state eigenvalue $\eta_o'$ and  state
$u_o'$ in powers of the height perturbation $\delta_2$.  
The derivations of the sometimes-long expressions noted here are provided in Appendix \ref{ap:pt}.

For simplicity, we provide  perturbation theory expansions only to second order, and only to
describe the effects due to changes in the height  of one mirror (i.e.\ $h_2$ of mirror 2) by an
amount $\delta h$. 
To express these 
changes in height in  dimensionless form, we introduce the variable $\delta_2$:
\begin{equation}
\label{eq:def:deltas}
\delta_2 \equiv  2 k \delta h_2 \; .
\end{equation}

\emph{Change in eigenvalue}: 
The eigenvalue $\eta_o$ of the ground state $\left| u_o \right> = \left| 0\right>$ changes as
\begin{equation}
\label{eq:pt:phase}
\eta_o' =\eta _{o}\left( 
   1
  -i\left\langle 0\left| \delta_2 \right| 0\right\rangle 
 -\frac{\left\langle 0\left| \delta_2 ^{2}\right| 0\right\rangle }{2} 
  -\sum_{k\neq 0}\frac{\eta _{k}\left| \left\langle 0\left| \delta_2 \right| k\right\rangle
   \right| ^{2}}{\eta _{o}-\eta _{k}}
 \right) 
\end{equation}

\emph{Change in eigenstate}:
When we construct the perturbation theory expansion to second order, we find the ground state
changes according to an expression of the form 
\begin{subequations}
\label{eq:pt:state}
\begin{equation}
\left| u'_{o,\text{pt}} \right> \approx \left | 0 \right> 
    + \left| \psi^{(1)} \right> 
    + \left| \psi^{(2)} \right>  \; .
\end{equation}
In this expression, $\psi^{(1)}$ and $\psi^{(2)}$  denote those terms first- and second-order in
  $\delta_2$ in the perturbation
  expansion,  respectively.
To be explicit, when we perform the perturbation theory expansion [details of which are
provided in Appendix \ref{ap:pt}], we find
\begin{eqnarray}
\label{eq:pt:state:psi1}
\left| \psi^{(1)} \right> &=&
   \sum_{k\neq 0}-\frac{i \lambda_o \lambda_k }{\eta _{o}-\eta _{k}}\left|
  k\right\rangle \left\langle k\left| \delta_2 \right| 0\right\rangle  \; ,
\end{eqnarray}
\begin{widetext}
\begin{eqnarray}
\label{eq:pt:state:psi2}
\left| \psi^{(2)} \right> &=&
\sum_{k\neq 0}\left|k\right\rangle \frac{\lambda _{o}\lambda _{k}}{\eta _{o}-\eta _{k}}\left[
  -\frac{1}{2}\left\langle k\left| \delta_2 ^{2}\right| 0\right\rangle 
  +\frac{\eta _{o}}{\eta
    _{o}-\eta _{k}}\left\langle k\left| \delta_2 \right| 0\right\rangle \left\langle 0\left|
      \delta_2 \right| 0\right\rangle
 - \sum_{p\ne 0} \frac{\eta_p}{\eta_o-\eta_p}  
   \left\langle k\left| \delta_2 \right| p\right\rangle
   \left\langle p\left| \delta_2 \right| 0\right\rangle 
  \right]   \; .
\end{eqnarray}
\end{widetext}
\end{subequations}

By construction $\left|\psi^{(1)}\right\rangle$ and $\left|\psi^{(2)}\right\rangle$ are orthogonal to the unperturbed ground state
$\left| 0 \right>$.
As a result, this expression (\ref{eq:pt:state}) for  $\left| \psi' \right\rangle$
 is \emph{not} normalized: we find, working to second order, that  the norm of $\left| u'_{o,\text{pt}} \right\rangle$ is
\begin{equation} 
\left< u'_{o,\text{pt}} | u'_{o,\text{pt}} \right> \approx 
  1+ ||\psi^{(1)}||^2 +  {\cal O}(\delta_2^3) \; .
\end{equation}
where we use the shorthand $|| \psi||^2 \equiv \left< \psi|\psi\right>$. 
Therefore, the physically appropriate normalized perturbed state $\left| u_o' \right\rangle$ is
given by the expression
\begin{eqnarray}
\label{eq:pt:psiexpand}
\left| u_o' \right\rangle
  &\approx & \frac{1}{\sqrt{1 + ||\psi^{(1)}||^2}} \left(
     \left| 0 \right\rangle + \left|\psi^{(1)}\right\rangle + \left|\psi^{(2)}\right\rangle
   \right)  \\ 
  &=&
     \left| 0 \right\rangle + \left|\psi^{(1)}\right\rangle + 
   \left(\left|\psi^{(2)}\right\rangle - \frac{||\psi^{(1)}||^2}{2}  \left| 0 \right\rangle
   \right)  + {\cal O}(\delta_2^3) \; . \nonumber
\end{eqnarray}

\subsubsection{Estimating convergence of perturbation expansion}
Perturbation theory is only effective when higher order terms provide only a small correction
to lower order terms.  To test the convergence of the series, we compare the magnitudes of the
first two perturbative corrections.  When $||\psi^{(2)}||/||\psi^{(1)}||\ll 1$, we believe the
series converges and our expressions should be effective.

\subsection{\label{sec:th:pt:imply}Implications of change in optical state for other quantities}
In the previous section, we described the effect of perturbations $\delta h_{1,2}$ on light
propagating in a single Fabry-Perot arm.  These perturbations cause the beam shape incident on
the two mirrors to change, generally in a different way at each mirror.  Therefore, quantities
that depend on the state of the beam at each mirror, such as the
diffraction losses and thermoelastic noise integrals $I_A$, also change.   In this section, we
loosely describe how the lowest order effect of these changes can be characterized.

\subsubsection{\label{sec:th:pt:imply:diff}Change in diffraction losses under perturbation of
  one mirror}
Unfortunately, a systematic treatment of diffraction losses within the context of perturbation
theory proves very tricky, not the least because we must represent both the height change and
the effect of finite mirror size as perturbations, then use  many states to insure  the perturbation
theory expansion converges properly for the effects of diffraction.  

Rather than perform a truly accurate, well-motivated computation, we will in this paper loosely
estimate diffraction losses by the clipping approximation applied to the perturbed beam state.
By way of example, we can estimate  the diffraction losses produced during a single
reflection off of mirror $1$ or off mirror $2$ in the 
presence of a perturbation of mirror $2$ by expanding the appropriate clipping approximation
estimate 
${\cal L}_1$ or ${\cal L}_2$ [cf.\ Eq.\ (\ref{eq:clip})]:
\begin{subequations}
\begin{eqnarray}
{\cal L}_1 &\approx& \left\langle u | O_1 | u \right\rangle  \; ,\\
{\cal L}_2 &\approx& \left\langle G_L u | O_2 | G_L u \right\rangle  \; , \\
O_1 &\equiv& 1- T_1 \; , \\ 
O_2 &\equiv& 1- T_2 \; .
\end{eqnarray}
\end{subequations}
To be explicit, we can  expand the clipping-approximation estimate for
${\cal L}_1$  to 
second order in  $\delta_2$ as
\begin{eqnarray}
\label{eq:pt:diffloss}
{\cal L}_1   &\approx&  \left\langle 0 | O_1 | 0 \right\rangle  
   +2\text{Re}\left(\left\langle
    \psi^{(1)} |   O_1 | 0 \right\rangle \right)  \\
 &+& \left\langle \psi^{(1)} |   O_1 | \psi^{(1)}  \right\rangle 
    - ||\psi^{(1)}||^2  \left\langle 0 | O_1 | 0 \right\rangle \nonumber\\
 &+& 2 \text{Re}\left(\left\langle \psi^{(2)} |   O_1 | 0 \right\rangle \right) 
  + {\cal O}(\delta_2^3) \; . \nonumber 
\end{eqnarray}
A similar expression is found for ${\cal L}_2$, with $O_1$ replaced by $G_L^\dag O_2
G_L$.  
[Because the beam profile will generally change in different ways at the two ends due to a
perturbation localized at only one end,  generally ${\cal L}_1 \ne {\cal L}_2$.]

\subsubsection{\label{sec:th:pt:imply:gain}Change in cavity gain due to perturbations 
 of one mirror}
The amount of power resonating in the cavity also changes, in part through modified diffraction
losses [Eq.\ (\ref{eq:pt:diffloss})] but also through a change in the resonant ground state of
the cavity from $u_o$ (i.e.\ $\left| 0 \right>$) to $u_o'$ [Eq.\ (\ref{eq:pt:psiexpand})].
Because the resonant state of the cavity changes, the overlap $\gamma_o$ changes to
$\gamma_o'$, which to second order can be approximated by
\begin{subequations}
\label{eq:pt:gamma0}
\begin{equation}
\gamma_o' \equiv \left< u_o'|u_d\right> = \left( 1- \frac{||\psi^{(1)}||^2}{2}\right) \gamma_0
  + \gamma_1 ||\psi^{(1)}|| + \gamma_2 ||\psi^{(2)}|| \; ,
\end{equation}
where we define
\begin{eqnarray}
\gamma_1 \equiv \left< \psi^{(1)}|u_d\right>/||\psi^{(1)}|| \;, \\
\gamma_2 \equiv \left< \psi^{(2)}|u_d\right>/||\psi^{(2)}|| \; .
\end{eqnarray}
\end{subequations}

To be very specific, we can find the perturbed cavity gain ${\cal G}_\text{anal}'$ by
evaluating Eq.\ (\ref{eq:gain:raw}) to second order in $\delta h_2$, 
\begin{equation}
\label{eq:pt:gain}
{\cal G}_\text{anal}' = \gamma_o'^2 \frac{1+r_I}{1-r_I} 
  \left( 1- \frac{ {\cal L}_1' + {\cal L}_2'}{1-r_I}\right)  \; ,
\end{equation}
using Eq.\ (\ref{eq:pt:diffloss}) for the perturbed diffraction losses ${\cal L}_1$ and ${\cal
  L}_2$ and using  Eq.\ (\ref{eq:pt:gamma0}) for the change in overlap between the
perturbed cavity ground state and the driving beam.

[Though we do not write out the resulting second-order expansion in full detail, we will apply these 
methods to compute to second order the effects of mirror tilt on cavity gain in
Sec.\ \ref{sec:apply:Defects}.]

\subsubsection{\label{sec:th:pt:imply:dark}Change in dark port power due to perturbations of
  one mirror}
An interferometer consists of two arms.  When the beamsplitter recombines the two fields
leaving the two arm cavities,   the light going out the dark port of the interferometer
($u_\text{dp}$) is the interference between the light reflected off the two arms (denoted I
and II): 
\begin{equation}
u_\text{dp} = \frac{1}{\sqrt{2}} (u_{r,\text{II}} - u_{r,\text{I}}) \; .
\end{equation}
If both cavities are identical, then $u_{r,\text{II}} = u_{r,\text{I}}=u_r$ for $u_r$ given by
Eq.\ (\ref{eq:def:outputDecomposition}) and no light exits the dark port.
If the cavities are perturbed, however, power will generically go out the dark port.  

For example, if only cavity II is perturbed, then $u_{r,\text{II}}=u'$, $u_{r,\text{I}}=u_r$,
and the dark port beam state is
\begin{subequations}
\label{eq:pt:darkport}
\begin{eqnarray}
u_\text{dp} &=& \frac{1}{\sqrt{2}} (u_r' - u_r)  \nonumber \\
   &=& \sqrt{2} \left({\cal P}_o' - {\cal P}_o\right) u_d \nonumber \\
  & =& \sqrt{2} \left( \gamma_o' u_o' - \gamma_o u_o \right) \nonumber \\
  &=&
 \sqrt{2}\left[
  \left(  - \gamma_o ||\psi^{(1)}||^2 + \gamma_1 ||\psi^{(1)}|| + \gamma_2 ||\psi^{(2)}||\right) \left| u_o\right>  \right. \nonumber\\
 & &    + \left(\gamma_0 + \gamma_1 ||\psi^{(1)}||\right) \left| \psi^{(1)}\right> 
  \nonumber \\
 & & \left. + \gamma_0 \left|\psi^{(2)}\right>
  \right] + O(\delta_2^3) \;  .
\end{eqnarray}
where going from the first line to the second we use 
Eq.\ (\ref{eq:def:outputDecomposition}) [for $u_r$]; from the second to the third we use
Eqs.\ (\ref{eq:def:project}) [for ${\cal P}_o$] and (\ref{eq:def:gamma0}) [for $\gamma_o$]; and
from the third to the fourth we use the perturbation expansions  (\ref{eq:pt:state}) [for
$u_o'$] and (\ref{eq:pt:gamma0}) [for $\gamma_o'$].

The dark port power $P_\text{dp}$ (as a relative fraction of interferometer input power, which
is twice the input power to each arm cavity) can be expressed as
\begin{eqnarray}
P_\text{dp} &=& ||u_\text{dp}||^2 /2\; , 
\end{eqnarray}
\end{subequations}
an expression which we shall not expand here.

\subsubsection{\label{sec:th:pt:imply:thermo}Influence of perturbations of one mirror on  thermoelastic noise}

The thermoelastic noise integrals associated with each mirror ($1$ and $2$) in a given arm
cavity both change because the beam profile at each mirror changes.  To evaluate those changes,
we insert the modified state [Eq.\ (\ref{eq:pt:psiexpand})] into the thermoelastic noise
integral $I_A$ [Eq.\ (\ref{eq:IA})].

For example, to evaluate
the linear-order change in thermoelastic noise at mirror $1$, we first take the new normalized
intensity profile 
$P(r)$ at mirror 1, given by 
\begin{eqnarray}
P_1'(r) &\equiv& \left| u'_0(r) \right|^2 \nonumber\\
\label{eq:pt:dP}
   & \approx& |u(r)|^2 
  + 2 \text{Re}\left(u_0^*(r) \psi^{(1)}(r)\right) + O(\delta_2^2) \; .
\end{eqnarray}
Using this  new intensity profile, we 
solve the thermoelastic noise model problem [Eq.\ (\ref{eq:elas})] for the expansion $\Theta'$
and in particular the first-order change in expansion $\delta \Theta \equiv \Theta' - \Theta$.
Finally, we insert this first order change $\delta \Theta$ into the expression for $I_A$ with $A=1$,
linearized about the background intensity profile:
\begin{equation}
\label{eq:pt:IA}
\delta I_1  = \frac{2}{F_o^2} \int_{V_1} (\nabla_a \Theta) (\nabla^a \delta \Theta) d\text{volume}
\end{equation}
where $\Theta$ is the expansion produced by the unperturbed pressure profile $F_o P$.

The beam shape at mirror $2$ also changes; by a similar construction, we can find its effect on
the thermoelastic noise integral $I_2$.





\subsection{Observations which simplify our computations}
\subsubsection{\label{sec:subsub:firstordersymmetry} Symmetry and the influence of first-order changes in state}
The ground-state beam of physical interest is isotropic.  As a result,  when we perform the
computations outlined above, we find only the 
\emph{axisymmetric} (about the optic axis) part of $\left| \psi^{(1)}\right>$ contributes to
first-order changes in the 
three integral quantities of physical interest (the overlap $\gamma_o$; diffraction losses; 
and thermoelastic noise).

By way of example, consider the first-order change in thermoelastic noise
[Eq.\ (\ref{eq:pt:IA})].  Since the unperturbed pressure 
profile and thus unperturbed expansion ($\Theta$) are isotropic, the above integral couples
only to the axisymmetric part of $\delta \Theta$ and therefore $\delta P$ and $\psi^{(1)}$
[cf.\ Eq.\ (\ref{eq:pt:dP})].

Similarly, to lowest order displacement and tilt perturbations have odd parity; therefore, 
symmetry insures that the first-order change $\gamma_1$ [Eq.\ (\ref{eq:pt:gamma0})] is zero.

\subsubsection{\label{sec:subsub:approximateIdenticalContributions}Assuming (roughly) equal contributions from the two mirrors to changes in $I_A$ and
 diffraction losses}
The change in mirror shape at mirror $2$ produces roughly comparable changes  to the
beam profile at both ends.  
To a rough approximation, then, we can assume that the changes in thermoelastic noise and
diffraction loss at mirror $2$ will be comparable to their changes at mirror $1$.

\subsection{\label{sec:th:Defects:summary}Summary: Using perturbation theory to explore the sensitivity of the arm cavity and
  interferometer to perturbations}
To summarize, we  evaluate how sensitive mesa-beam arm cavities and interferometers are to
mirror perturbations as follows:

\begin{enumerate}
\item \emph{Assume the arm cavity is fiducial}:
The perturbation expansions above can be applied to any arbitrary mirror configuration
(i.e.\ to any specific mesa-beam $D$).\footnote{Our technique assumes the cavity eigenmodes of an
  arm cavity with finite mirrors are
  well-approximated by the modes of a cavity with infinite mirrors.  Therefore, the only
  relevant parameter remaining is the mesa-beam averaging scale $D$.  [This approximation
  ignores quantities of order the diffraction losses,  $10\text{ ppm}\sim 10^{-5}$.]}
In principle, we \emph{could} apply perturbation theory to all possible mirror configurations
generated in Sec.\ \ref{sec:th:Ideal}.  Instead,  to avoid repeating
computations which should yield nearly identical results, we apply perturbation theory
only to the \emph{fiducial} advanced LIGO arm cavity presented in Section
\ref{sec:def:fiducial}. 

\item \emph{Consider only perturbations of one ETM}:
Similarly, we change only one end test mass in one arm cavity, rather than apply
perturbations to each test-mass mirror.

\item \emph{Compute the eigenmodes of the fiducial arm cavity}:
We find the natural eigenmodes of the fiducial arm cavity---and therefore the basis states
$\left | p \right>$ in
our perturbation expansion---by numerically solving the basis-state eigenequation
(\ref{eq:eigenequationAlt}).   [The numerical code  which 
 solved this integral eigenequation is described in Sec.\ \ref{sec:num:modes}.]

\item \emph{Find how  the ground state of arm cavity changes}:  Next, we apply the perturbation expansion
  [Eq.\ (\ref{eq:pt:state})] to find how the ground state of the arm cavity changes when each
  perturbation of physical interest is applied:  tilt,   displacement, and  mirror figure
  error.   In 
  other words, for each of these perturbations $\delta h_2$, we use Eq.\ (\ref{eq:pt:state}) to
  find the first- and second-
  order corrections  $\left| \psi^{(1)}\right>$ and $\left| \psi^{(2)}\right>$ to the state of
  the cavity.  

\item \emph{Determine how the thermoelastic noise integral $I_A$ for each mirror changes}:
Given the changed beam state, we can recompute the thermoelastic noise integral $I_A$ for each mirror 
(cf.\ Sections \ref{sec:th:Ideal} and \ref{sec:num:thermo}), using the perturbed
beam state $u_o'(r)$ and the perturbed beam intensity profile $P'(r)= |u'_o(r)|^2$.  
More directly, the series expansion of thermoelastic noise can be discovered by a series 
expansion  of the thermoelastic noise integral, as sketched (to first order) in
Sec.\ \ref{sec:th:pt:imply:thermo}.

\item \emph{Describe how the cavity gain and dark port power change}:   Finally, given the changed
beam state, we can also recompute the arm cavity gain [Eq.\ 
(\ref{eq:pt:gain})] and interferometer dark port power [Eq.\ (\ref{eq:pt:darkport})] when the
ETM of one cavity is 
perturbed.   These two expressions both depend on how the perturbed arm cavity interacts with the
driving beam [i.e.\  on $\gamma_1$ and $\gamma_2$; cf.\ Eq.\ (\ref{eq:pt:gamma0})] and on
the perturbed arm cavity state itself [i.e.\ on  $||\psi^{(1)}||$ and $||\psi^{(2)}||$;
cf.\ Eq.\ (\ref{eq:pt:state})].  Also, the cavity gain depends on how the diffraction losses of
the arm cavity change [Eq.\ (\ref{eq:pt:diffloss})].

\end{enumerate}
 
\section{\label{sec:num}Numerical implementations of the  thermoelastic noise integral, the optical
  eigenequation, and optical perturbation theory}
In the previous two sections, we described the abstract expressions we must evaluate to design
and analyze an arm cavity bounded by Mexican-hat mirrors.  In this section, we describe how we
implemented and solved those equations.

Where independent methods were used to perform a particular computation, we indicate the
different techniques used.

\subsection{\label{sec:num:thermo}Thermoelastic noise for perfect (undeformed) mirrors}
To evaluate the thermoelastic noise power spectrum [Eq.\ (\ref{eq:Sh})], we need to perform two
tasks.  First, we must  solve the elasticity problem described in
Eq.\ (\ref{eq:elas}); then, using the result, we can evaluate the integral  $I_A$,
which enters directly in Eq.\ (\ref{eq:Sh}).

We employed three methods to address these two tasks.  The first two methods---a numerical
(finite-element) solution and an exact analytic solution---applied only to special
circumstances: axisymmetric beam profiles, and cylindrical mirrors.  The
third method was an approximation based on assuming the mirror to be half-infinite;
we used it only as  a quick,
easy-to-evaluate check on the qualitative behavior of the previous two methods.





\subsubsection{General approach}
We usually solved the elasticity equations
 (\ref{eq:elas:Tab})-(\ref{eq:elas:bcAverage}) using a commercial two-dimensional 
finite-element code  \cite{femlab}.  We chose the region ${\cal  R}$ in
 Eq.\ (\ref{eq:elas:bcAverage})  to be 
one of the mirror faces.\footnote{The code works faster if the region ${\cal R}$ is a point or set of
  points. However, the code is significantly more susceptible (on physical grounds---the
  points act like ``nails'' in the mirror) to small errors in the neighborhood of points,
  errors that contribute significant erroneous expansion.  Therefore, as a practical compromise
  we chose ${\cal R}$ to be a surface. 
}
[To use a two dimensional code, we limited attention to  \emph{axisymmetric} mirrors and pressure
profiles.]

The commercial finite-element code we employed gave us the displacement vector $y^a$.  We then
used postprocessing code discussed in Appendix \ref{ap:postprocess}  to evaluate the
derivatives and  integrals needed in  Eq.\ (\ref{eq:IA}).

\subsubsection{\label{sec:subsub:LT}Special case: cylindrical mirror}
Independently, we employed the analytic elasticity solution Liu and Thorne (LT)
\cite{LiuThorne} developed for cylindrical mirrors with axisymmetric pressure profiles
imposed on them.

Liu and Thorne  constructed a solution to the elasticity equations for a cylinder
[Eqs.\ (\ref{eq:elas:Tab})-(\ref{eq:elas:bcAverage})]
 which they apply to Eq.\ (\ref{eq:IA})
to find an explicit expression for the thermoelastic
noise produced by 
gaussian beams on cylindrical mirrors.
By replacing a single equation in their expressions, we can convert their solution to one
appropriate to \emph{arbitrary} axisymmetric beam intensity profiles.

To be explicit, LT Eq.\ (44) gives an expression that is precisely $1/2$ of $I$ (1/2 because of
 averaging which we have factored out but which LT retain), in terms of quantities defined in
 LT Eqs.\ (35) and  (36).  To generalize to a
generic axisymmetric pressure profile, one need only change LT Eq.\ (36)---the only equation
which involves the specific pressure profile---so that it involves the 
intensity function $P(r)$  defined in Eq.\ (\ref{eq:elas:bcForce}) for $m>0$:\footnote{The LT 
coefficient $p_o$ is
 independent of the pressure profile shape; it is always $1/\pi a^2$.}
\begin{equation}
p_m = \frac{2}{a^2 J_0^2(\zeta_m)} \int_0^a P(r) J_0(\zeta_m r/a) r dr \; .
\label{eq:thel:LTp}
\end{equation}
Here $a$ is the radius of the cylindrical mirror (denoted $R$ elsewhere in this paper); $\zeta_m$ is the $m$th zero of $J_1(x)$; and
the functions $J_0$ and $J_1$ are the zeroth and first order cylindrical Bessel functions.
The sum converges rapidly: typically, only a handful of terms in the infinite sum
[LT Eq.\ (45)] are  required.  

\subsubsection{\label{sec:subsub:halfinf}Approximate technique: half-infinite mirrors}
If the mirror is sufficiently large compared to the imposed pressure profile $P(r)$, the
elasticity problem we must solve [Eq.\ (\ref{eq:elas})] can be well approximated by a solution
to a similar problem with the mirror boundary taken to infinity.  In this case, as the force
density term [Eq.\ (\ref{eq:elas:bcForce})] goes to zero, we need only
solve for the response of a half-infinite (i.e.\ filling the region $z<0$) elastic medium to an
imposed surface pressure.   As described in greater detail in Appendix
\ref{ap:halfInfiniteThermoelastic}, analytic expressions exist for the response, permitting us
to find an compact expression for the thermoelastic noise integral.   We find
\begin{equation}
I =
    \left(\frac{\left( 1+\sigma \right) \left( 1-2\sigma \right) }{2\pi E}\right) ^{2}
  \int d^{2}\vec{K}\,\left| \vec K\right| 
   \left| \tilde{P}\left( \vec{K}\right) \right| ^2 
\end{equation}
where $\tilde{P}(K)$ is the two-dimensional Fourier transform of $P(\vec{r})$:
\begin{equation}
\tilde{P}(\vec{K}) \equiv \int d^2 \vec{r} e^{- i \vec{K}\cdot \vec{r}} P(\vec{r}) \; .
\end{equation}
Recall that $P(\vec{r})$ is normalized to unity (cf.\ Sec.\ \ref{sec:th:thermo}).

\subsection{\label{sec:num:modes}Numerically solving for the resonant optical eigenmodes of a cavity bounded by
  arbitrary axisymmetric mirrors}
To test the validity of the clipping approximation [Eq.\ (\ref{eq:clip})] against the exact  diffraction
losses  [Eq.\ (\ref{eq:diffLossNet})] and to generate the set of basis solutions needed to
construct perturbation theory [Eq.\ (\ref{eq:eigenequationAlt})], we must numerically solve for the cavity eigenmodes.  In this
section, we describe numerically how we converted the cavity eigenmode integral equation into a
numerical eigenproblem, which we then solved with standard numerical tools (e.g., Mathematica).




\subsubsection{Setting up the problem to be solved; preliminary analytic simplifications}
Rather than solve the eigenequation for a full round trip through a symmetric cavity
[Eq.\ (\ref{eq:eigenequation}) with $R_1=R_2$], we exploit symmetry and instead study the
closely related problem of \emph{half} a round trip through a symmetric cavity.  In other
words, we plan to diagonalize the basis state eigenequation [Eq.\ (\ref{eq:eigenequationAlt})]
when that eigenequation is restricted to the space of functions defined on 
$R_1=R_2\equiv b \;  X_\text{max}$:
\begin{equation}
\label{eq:numericallyDiagonalized}
\lambda \left| u \right> = T_1 G_1 G_L T_1 \exp(- i k L) \left| u \right>  
\end{equation}

If we write out the appropriate eigenequation for this ``half-a-round-trip'' operator, making use of the definitions in
Eq.\ (\ref{eq:propagatorExamples}) and using the dimensionless spatial units 
$\vec{X}=\vec{r}/b$, we find the following integral equation:
\begin{subequations}
\begin{eqnarray}
\lambda \psi(\vec{X}) &=& \int d^2 X' {\cal G}(X;X') \psi(\vec{X}') \; , \\
{\cal G}(\vec{X};\vec{X}') &\equiv& 
\frac{-i}{2\pi} \Theta(X_\text{max}- |\vec{X}'|)
   e^{ i \left[ \frac{\left( \vec{X}-\vec{X}'\right) ^{2}}{2}
      -\bar{h}\left( X\right) \right] }  \; ,\\
\bar{h} &\equiv& 2 k h \; .
\end{eqnarray}
\end{subequations}
Here, $\psi(\vec X) = u(\vec r)/b$ denotes a dimensionless representation of the state
$u$. 

Since the mirror surfaces are (ideally) axisymmetric, the operators above all commute with
rotation around to the optic axis.  We can therefore mutually diagonalize this
operator and the generator of rotations.  Therefore, we require $\psi$ be proportional to $\exp
(i m \varphi)$ for some integer $m$:
\begin{equation}
\label{eq:def:Phi}
\psi(X,\varphi) \equiv \Phi(X) e^{i m \varphi} \; .
\end{equation}
Substituting in this form for $\psi$, we reduce the problem to a series of one-dimensional
integral equations, one for each $|m|$:
\begin{subequations}
\label{eq:oneDeigenproblem}
\begin{eqnarray}
\eta \Phi(X) &=& \int_0^{X_\text{max}} \bar{X} d\bar{X} {\cal G}_m(X,\bar{X}) \Phi(\bar{X}) \; , \\
{\cal G}_{m}\left( X,\bar{X}\right) &=&-i^{m+1}J_{m}\left( X\bar{X}\right) 
  e^{i
    \left[ \frac{X^{2}+\bar{X}^{2}}{2}-\bar{h}\left( X\right) \right] } \; .
\end{eqnarray}
\end{subequations}

\subsubsection{Method of numerical solution}
For each $m$ of interest, we represent the  integral operator on the right side of
Eq.\ (\ref{eq:oneDeigenproblem}) as a matrix.
For simplicity, we discretize space in a uniform grid $X_A = A X_\text{max}/(N-1)$ for
$A=0\ldots N$, define 
$\Phi_A = \Phi(X_A)$ and similarly, and  approximate the integral by a simple 
quadrature rule.\footnote{We in fact used the equation Eq.\ (\ref{eq:matrixOneDeigenproblem})
  as stated.  More sophisticated quadrature techniques, such as Gaussian quadrature, offer
  greater accuracy with fewer points (and hence significantly less computation time).
}
 Thus, we approximate the integral equation of
Eq.\ (\ref{eq:oneDeigenproblem}) by the matrix eigenvalue problem
\begin{equation}
\label{eq:matrixOneDeigenproblem}
\lambda \Phi_A  =\sum_{B=0}^N \frac{X_\text{max}^2 B}{(N-1)^2}  {\cal G}_m(X_A,X_B) \Phi_B
\end{equation}
This equation can be solved for $\eta$ and $\Phi$ by any standard eigensolution package.

\subsubsection{Interpreting and applying the numerical solution}
To summarize, to find  numerical approximations to eigensolutions $\psi$ of
Eq.\ (\ref{eq:numericallyDiagonalized})---an eigenequation
similar to the basis-state eigenequation
[Eq.\ (\ref{eq:eigenequationAlt})] but for mirrors of finite radius 
$R = b  X_\text{max}$ ---  
we construct and solve the matrix eigenvalue problem (\ref{eq:matrixOneDeigenproblem}).

The approximate eigensolutions $\eta_p$ and $\Phi$ so obtained provide numerical approximations
to the two cavity-eigenproblems of technical interest:

\begin{itemize}
\item \emph{Solutions for true cavity eigenmodes}: This method provides us with precisely the
  numerical cavity modes needed to understand mesa beams in the presence of finite mirrors.
  [Specifically, 
 eigenmodes  and eigenvalues of
  Eq.\ (\ref{eq:numericallyDiagonalized}) are also eigenmodes of Eq.\ (\ref{eq:eigenequation}).]  In particular, we can
  use the norm $|\eta|$ to determine the round-trip diffraction losses according
  to Eq.\ (\ref{eq:diffLossNet}).

\item \emph{Solutions for the basis states}: Further, by making the mirrors sufficiently large
  that diffraction effects can be ignored\footnote{As a practical matter, we test the quality
    of our solutions by observing the convergence of the norm $|\eta|$ as we increase the
    number of points used to represent the state.
  }, solutions to Eq.\ (\ref{eq:numericallyDiagonalized})
  provide good approximations to solutions to the basis eigenvalue problem
  [Eq.\ (\ref{eq:eigenequationAlt})]. 
\end{itemize}

\subsection{\label{sec:num:pt}Numerical implementation of perturbation theory}
To use perturbation theory exactly [i.e.\ Eqs.\ (\ref{eq:pt:phase}) and (\ref{eq:pt:state})], one
needs an infinite collection of states.  In practice, we limited attention to a handful: the
lowest-lying ten to twenty for each $|m|=0,\ldots 7$; more modes were used when
the rate of convergence of the perturbation theory expansion suggested more were needed.



\subsection{Numerical exploration of changing diffraction losses}
Given the $\delta h_2$-induced perturbed beam state $\left| \psi \right>$ at mirror $1$, we can compute
the (clipping approximation estimate of the) diffraction losses at mirror $1$ for a given
height perturbation $\delta h_2$ at mirror $2$.  Largely, we simply evaluated the integrals
(i.e.\ inner products and norms) required to construct Eq.\ (\ref{eq:pt:diffloss}).   However, to
provide an independent numerical check (i.e.\ to insure we had no typographical or structural
errors), we also evaluated the diffraction losses directly, using the definition
Eq.\ (\ref{eq:clip}), for a sequence of height perturbations $\delta h (\varepsilon) \equiv
\varepsilon \delta h_2$ with $\varepsilon = 0, 0.1, 0.2, \ldots 1$; then fitted a second-degree
polynomial to the resulting data points to extract the series coefficients in
Eq.\ (\ref{eq:pt:diffloss}).

\subsection{Numerical investigation of changes in thermoelastic noise when one mirror shape is
  perturbed} 
Finally, we can use the known form of $\psi^{(1)}(r, \varphi)$ [from Eq.\ (\ref{eq:pt:state:psi1})] in the procedure outlined earlier
[cf.\ Eqs.\ (\ref{eq:pt:dP}) and
(\ref{eq:pt:IA})] to evaluate the first-order change in  thermoelastic noise associated with
mirror $1$ due 
to changes in the shape of mirror $2$.

Specifically, given $\psi^{(1)}(r,\theta)$, we use Eq.\ (\ref{eq:pt:dP}) to find how the
pressure profile changes.  We then use the first-order change $\delta P$ to the pressure
profile in the LT expression for thermoelastic noise (see Sec.\ \ref{sec:subsub:LT}), linearized
about the response to the unperturbed cavity beam intensity $P$.  [We will not provide the
rather lengthy but straightforward linearization of the LT expressions here; any
computer-algebra system can easily reproduce the desired expansion.]

\section{\label{sec:apply:Design}Dependence of thermoelastic noise on mirror and beam shape}



In this section, we explore the 
dependence of the thermoelastic noise integral $I_A$ [Eq.\ (\ref{eq:IA})]  on the various  
arm cavity parameters available to us: 
(i) the  mirror's  dimensions (i.e.\ cylinder height and radius) and shape 
 (i.e.\ frustum or cylinder), 
(ii) the beam size (e.g., the mesa-beam size $D$), and 
(iii) the type of beam   resonating in the arm cavity (i.e.\ Gaussian or mesa).  
The thermoelastic noise integral provides a simple way to characterize how the thermoelastic
noise power spectrum  $S_h$ [Eq.\ (\ref{eq:Sh})] of these configurations compares to $S_h$ for
the baseline advanced LIGO configuration ($S_h^\text{BL}$):
\begin{equation}
\label{eq:apply:Icompare}
S_h/S_h^{\text{BL}} = I/I_\text{BL} \; ,
\end{equation}
where $I_\text{BL}$ (evaluated below) is the value of the thermoelastic noise integral for the
baseline advanced 
LIGO configuration  [Eq.\ (\ref{eq:IBL})], and where $I$ is the value of the thermoelastic noise
integral for all four (identical) mirrors in the interferometer. 

Specifically, in this section we  compare the following types of mirror and beam
configurations, all of which satisfy the advanced LIGO design constraints [Section
\ref{sec:th:Design:constraints}]: (i) the baseline advanced LIGO configuration (which uses
Gaussian beams and cylindrical mirrors; (ii) an improved baseline configuration (which also
uses Gaussian beams and cylindrical mirrors); (iii) configurations with  mesa-beam  light
resonating between identical cylindrical mirrors; and (iv) configurations with mesa-beam light
resonating between identical frustum mirrors.
[In this section, we restrict attention to arm cavities with identical mirrors; in Appendix
\ref{ap:nonidentical} we discuss generalizations to arm cavities bounded by
\emph{nonidentical}  mirrors.]
 In Table \ref{tbl:NSRange} we summarize the optimal (i.e.\ lowest value of $I/I_\text{BL}$) configurations we found for each
class.     
The results of items tabulated in this section are applied, in MBI Sec.\ III (cf.\ MBI Table
I), to produce advanced LIGO designs  with  lower thermoelastic noise than the baseline design.

Our evaluations have been performed independently by all three
co-authors (RO'S, SS, and SV), using multiple methods (both finite-element solutions 
and infinite-sum analytic
solutions) when appropriate.

\subsection{Baseline advanced LIGO configuration}
  The baseline design of an advanced LIGO interferometer \cite{advLIGOmain} has four identical
  cylindrical 
  sapphire test masses (i.e.\ physical radius $R_p = 15.7\text{ cm}$; thickness $H=13\text{ cm}$; mass
  $40 \text{ kg}$) whose  surfaces are coated over most of their surface (i.e.\ out to a radius $R=R_p
  - 8\text{ mm}$).  These mirrors' surfaces are designed so the largest possible Gaussian
  consistent with the $10\text{ ppm}$ diffraction loss constraint
  (Sec.\ \ref{sec:th:Design:constraints}) resonates in the arm cavity  [i.e.\ a Gaussian beam
  with  radius $r_o = 4.23\text{ cm} = 1.63 b$; cf.\ Appendix \ref{ap:spherical} and MBI
  Eq.\ (2.8)] for a discussion of Gaussian beams].

 The thermoelastic noise integral for an arm cavity bounded by  cylindrical mirrors and
  using a Gaussian beam is found by (i) constructing the Gaussian amplitude function
 $u_G(r,r_o)$ [MBI Eq.\ (2.8)] and its associated beam intensity profile 
 $P(r) = |u_G(r,r_o)|^2$; (ii) solving for the elastic expansion $\Theta$ that arises due
  to  $P(r)$ in the elastic model problem of Eq.\ (\ref{eq:elas}) described in
  Sec.\ \ref{sec:th:thermo} (cf.\ Sec.\ \ref{sec:num:thermo} for numerical methods); and (iii)
  inserting the resulting expansion into the definition of 
  the thermoelastic noise integral $I$ [Eq.\ (\ref{eq:IA})].
For the baseline beam and test mass, the resulting value of the noise integral $I$ is 
\begin{equation}
I_{\rm BL} = 2.57 \times
10^{-28} {\rm s}^4 {\rm g}^{-2} {\rm cm}^{-1}\;.
\label{eq:IBL}
\end{equation}

[The advanced LIGO cylindrical mirrors and Gaussian beams are optimal: these
beams 
produce very nearly the lowest thermoelastic noise possible using Gaussian beams reflecting off identical
$40\text{ kg}$ cylindrical mirrors, where those cylinders are coated out to a radius 
     $R=R_p-8\text{mm}$ and where the diffraction losses per bounce are restricted to less than $10\text{ ppm}$.]

\begin{table}
\caption{\label{tbl:CylGauss} The thermoelastic integral $I$ for a cylindrical test mass
and a Gaussian beam, in units
of the value $I_{\rm BL} = 2.57 \times 10^{-28} {\rm s}^4 {\rm g}^{-2} {\rm cm}^{-1}$ for the
advanced LIGO baseline design.  The values of $I/I_{\rm BL}$ are estimated to be accurate to within one
per cent.}
\begin{ruledtabular}
\begin{tabular}{llcccccc}
 & $R$  & $R_p [{\rm cm}]$ & $H [{\rm cm}]$ & $r_o$ & $r_o/b$ & $I/I_{\rm BL}$ & ${\cal L}_0$ [ppm]\footnotemark[1] \\
\hline 
BL\footnotemark[2] & $R_p-8{\rm mm}$ & 15.7 & 13. & 4.23 & 1.63 & 1.000 & 10 \\
 & $R_p$  & 15.7 & 13. & 4.49 & 1.73 & 0.856 & 10  \\
\end{tabular}
\end{ruledtabular}
\footnotetext[1]{Diffraction losses in each bounce off the test mass, in ppm (parts per million),
computed in the clipping approximation}
\footnotetext[2]{Baseline design for Advanced LIGO interferometers}
\end{table}

\subsection{Improved baseline advanced LIGO configuration}
The conventional baseline described above wastes the last $8$mm of mirror face size.  We can
improve upon the thermoelastic noise 
merely by eliminating the uncoated ring in the last $8$mm, i.e.\ by coating the mirror out to the edge and adjusting the
beamspot size to fill in the extra space.
If the mirror coating extends out to the test-mass edge so $R=R_p = 15.7$ cm, and the Gaussian
beam radius is correspondingly increased to $r_o = 4.49$ cm so the diffraction losses are still
10 ppm, then the thermoelastic noise is reduced to $I/I_{\rm BL} = 0.856$; see Table
\ref{tbl:CylGauss}. 

[Again, the \emph{same} advanced LIGO cylindrical mirrors 
(i.e.\ with unchanged phyisical radius and thickness)
produce very nearly the lowest possible thermoelastic 
noise, among all $40\text{ kg}$ cylindrical mirrors coated out to  their physical radius.]

\subsection{Mesa beams reflecting off identical 40kg cylindrical mirrors} 
Arm cavities with \emph{mesa beams} reflecting off cylindrical mirrors admit configurations with
even lower thermoelastic noise   than the improved baseline.    To explore the advantages
of mesa beams, we evaluated the thermoelastic noise integral (via the method described in  Sec.\ \ref{sec:th:Design:summary})
 for two one-parameter\footnote{The arm cavity has three free parameters (mirror radius;
   mirror thickness; and mesa-beam radius parameter $D$) and two constraints (mirror mass and
   diffraction losses per bounce).} families of cylindrical mirrors
and mesa beams: 40 kg mirrors with the largest possible
mesa beams resonating off their front faces (i.e.\ set by $10$ppm diffraction losses; cf.\ 
Sec.\ \ref{sec:sub:diffConstraint}),\footnote{As expected, thermoelastic noise decreased with
  increasing mesa beam radius; we obtain the lowest value of thermoelastic noise integral $I$
  when the mesa beam radius is as large as possible, consistent with the diffraction constraint.} coated either out to (i) their physical radius
$R_p$ or (ii) only out to $R_p - 8\text{ mm}$.
Table \ref{tbl:CylMH} summarizes our results for each one-parameter family.

Whether the mirror is coated out to the full physical mirror radius or not, in both cases
quadratic fits to $I/I_{\rm BL} (R_p)$ give minima rather near the baseline physical radii
$R_p = 15.7$ cm.  Indeed,  
to within our accuracy of computation, the same mirror shape used as the baseline advanced LIGO
design ($R_p= 15.70 {\rm cm}$ and 
$H=13.00 {\rm cm}$) gives the optimal thermoelastic noise for mesa beams with 10 ppm diffraction
loss.  The beam radii $D$ and thermoelastic noise $I/I_{\rm BL}$ for these near-optimal mexican-hat test masses
are shown in Table \ref{tbl:NSRange} below.

\begin{table}
\caption{\label{tbl:CylMH} The thermoelastic integral $I$ for a cylindrical test mass
and a mesa beam, in units
of the value $I_{\rm BL} = 2.57 \times 10^{-28} {\rm s}^4 {\rm g}^{-2} {\rm cm}^{-1}$ for the
Advanced LIGO baseline design.  The values of $I/I_{\rm BL}$ are estimated to be accurate to
within one per cent.  The first four test masses, like the baseline, are mirror coated only out
to $R=R_p - 0.8 \rm{cm}$; the last five are coated all the way out to the test-mass edge,  
$R=R_p$.}
\begin{ruledtabular}
\begin{tabular}{lddddc}
\multicolumn{1}{l}{$R [{\rm cm}]$} & \multicolumn{1}{r}{$R_p [{\rm cm}]$} & \multicolumn{1}{r}{$H [{\rm c
m}]$} & \multicolumn{1}{r}{$D/b$} & \multicolumn{1}{r}{$I/I_{\rm BL}$} & \multicolumn{1}{r}{${\cal L}_0$
[ppm]\footnotemark[1]} \\
\hline
$Rp-8{\rm mm}$ & 14.67 & 14.79 & 3.00 & 0.414 & 10  \\
$Rp-8{\rm mm}$ & 15.70 & 13.00 & 3.43 & 0.364 & 10  \\
$Rp-8{\rm mm}$ &17.11 & 10.87 & 4.00 & 0.442 & 10  \\
$Rp-8{\rm mm}$ & 19.58 & 8.30 & 5.00 & 1.000 & 10  \\
 \\
$R_p$ & 13.94 & 16.38 & 3.00 & 0.373 & 10  \\
$R_p$ & 15.70 & 13.00 & 3.73 & 0.290 & 10  \\
$R_p$ & 16.37 & 11.88 & 4.00 & 0.313 & 10  \\
$R_p$ &18.85 & 8.96 & 5.00 & 0.628 & 10  \\
$R_p$ &21.36 & 6.98 & 6.00 & 1.69 & 10  \\
\end{tabular}
\end{ruledtabular}
\footnotetext[1]{Diffraction losses in each bounce off the test mass, in ppm (parts per million),
computed in the clipping approximation}
\end{table}

\subsection{Mesa beams reflecting off identical 40kg frustum mirrors} 
Roughly speaking, two mirrors with a larger front face radius permit a wider beam to resonate
in the arm cavity and yield even lower thermoelastic noise.  Therefore, we explore arm cavities
bounded by
frustum-shaped 
mirrors (cf.\ footnote \ref{note:defineFrustum}), which expand the front face of the mirror at the
expense of the back face.

More specifically, we considered arm cavities resonating with mesa beams of beam radius
parameter $D$ bounded by 40 kg test-mass mirrors whose front and back faces were as small as
diffraction losses would permit (i.e.\ the front face  produces precisely  $10 \text{ ppm}$
diffraction losses with arm-cavity  mesa beam light of scale $D$; the back face produces precisely one percent
diffraction losses for the same mesa-beam input light; cf.\ Sec.\ \ref{sec:th:Design:constraints}).\footnote{While 
in principle we could consider any combination of front and back face radii $R_1$ and $R_2$
and any mesa beam radius $D$ such that all three satisfy the LIGO design constraints presented
in Sec.\ \ref{sec:th:Design:constraints}, we found that for $R_2$ greater than or equal to the
minimum radius allowed by diffraction losses the thermoelastic noise integral increases with
$R_2$ (for fixed mirror mass and front face size).  Therefore, we limited attention to $R_2$ as
small as possible.  Also, as usual, we limited attention to mesa beams as large as diffraction
losses  on the mirror front face permit. 
}
These arm cavities
satisfy advanced LIGO design constraints (cf.
Sec.\ \ref{sec:th:Design:constraints}).

Table \ref{tbl:ConMH1} summarizes the thermoelastic integrals $I/I_\text{BL}$ for two
one-parameter family of designs: (i) the mirror is coated out to its physical radius $R=R_p$
and (ii) the mirror is only coated out to $R=R_p - 8\text{mm}$.
By fitting a quadratic
to $(D/b, I/I_{\rm BL})$,
we estimate  the optimal mirror dimensions and associated beam radii $D$.  Our optimal results
appear in Table  \ref{tbl:NSRange}.

\begin{table}
\caption{\label{tbl:ConMH1} The thermoelastic integral $I$ for a frustum input test mass (ITM)
and a Mesa beam, in units
of  $I_{\rm BL} = 2.57 \times 10^{-28} {\rm s}^4 {\rm g}^{-2} {\rm cm}^{-1}$.  The values of $I/I_{\rm BL
}$ are estimated to be accurate to within one
per cent.}
\begin{ruledtabular}
\begin{tabular}{ldddddr}
\multicolumn{1}{l}{$R$\footnotemark[1] } & \multicolumn{1}{r}{$R_{p1} [{\rm cm}]$} & \multicolumn{1}{r}{$
R_{p2} [{\rm cm}]$} &\multicolumn{1}{r}{$H [{\rm cm}]$} & \multicolumn{1}{r}{$D/b$} & \multicolumn{1}{r}{
$I/I_{\rm BL}$} & \multicolumn{1}{r}{${\cal L}_0$[ppm]\footnotemark[2] } \\
\hline
$R_p-8{\rm mm}$ & 14.67 & 10.57 & 19.81 & 3.00 & 0.355 & 10  \& $10^4$  \\
$R_p-8{\rm mm}$ & 15.70 & 11.56 & 17.00 & 3.43 & 0.253 & 10  \& $10^4$ \\
$R_p-8{\rm mm}$ & 17.11 & 12.88 & 14.06 & 4.00 & 0.207 & 10  \& $10^4$ \\
$R_p-8{\rm mm}$ & 17.45 & 13.22 & 13.45 & 4.13 & 0.208 & 10  \& $10^4$ \\
$R_p-8{\rm mm}$ & 19.58 & 15.27 & 10.43 & 5.00 & 0.285 & 10 \& $10^4$ \\
& & & & & &  \\
$R_p$ & 13.94 & 9.88 & 22.24 & 3.00 & 0.345 & 10  \& $10^4$  \\
$R_p$ & 15.70 & 11.56 & 17.00 & 3.73 & 0.198 & 10  \& $10^4$ \\
$R_p$ & 16.37 & 12.19 & 15.49 & 4.00 & 0.175 & 10  \& $10^4$ \\
$R_p$ & 17.29 & 13.04 & 13.75 & 4.39 & 0.162 & 10  \& $10^4$ \\
$R_p$ & 18.85 & 14.58 & 11.33 & 5.00 & 0.193 & 10 \& $10^4$ \\
$R_p$ & 21.36 & 17.00 & 8.62 & 6.00 & 0.398 & 10  \& $10^4$ \\
\end{tabular}
\end{ruledtabular}
\footnotetext[1]{Radius $R$ of the mirror-coated portion of both the inner and the outer
faces of the test mass, in units of its physical radius $R_p$.}
\footnotetext[2]{Diffraction losses in each bounce off the test mass, in ppm (parts per million); the
first number is for the light inside the arm cavity, on face 1 of the test mass (radius $R_1$); the
second number is for the light impinging from the beam splitter onto face 2 of the test mass
(radius $R_2$).}
\end{table}






\begingroup
\squeezetable
\begin{table}
\caption{\label{tbl:NSRange} Optimized test-mass and light beam configurations, their
thermoelastic noise compared to the baseline.  [A subset of this table appears as Table I
 in MBI \cite{MBI}.]
}
\begin{ruledtabular}
\begin{tabular}{lld}
Test Masses & Beam Shape & \multicolumn{1}{c}{$\left({S_h\over S_h^{\rm BL}}\right)_{\rm TE}
$}  \\
\{$R_{p1}$, $R_{p2}$; $H$\} & &  \\
\hline
BL: cylinders, $R=R_p-8{\rm mm}$ & BL: Gaussian &  \\
\{15.7, 15.7; 13.0\}  &  $r_o = 4.23{\rm cm}$ & 1.000  \\
BL: cylinders, $R=R_p-8{\rm mm}$  & mesa & \\
\{15.7, 15.7; 13.0\} &  $D/b = 3.73$ & 0.364  \\
identical frustums, $R=R_p-8{\rm mm}$  & mesa & \\
\{17.11, 12.88, 14.06\} & $D/b = 4.00$ & 0.207  \\
\\
BL: cylinders, $R=R_p$ & Gaussian & \\
\{15.7, 15.7; 13.0\}  &  $r_o = 4.49{\rm cm}$ & 0.856  \\
BL: cylinders, $R=R_p$ & mesa  & \\
\{15.7, 15.7; 13.0\} &  $D/b = 3.73$ & 0.290  \\
identical frustums, $R=R_p$ & mesa & \\
\{17.29, 13.04, 13.75\} & $D/b = 4.39$ & 0.162  \\
\end{tabular}
\end{ruledtabular}
\end{table}
\endgroup

\section{\label{sec:apply:Defects}Interferometer sensitivity to mirror perturbations}
In Sec.\ \ref{sec:apply:Design}, we found mirror and beam configurations for the advanced LIGO 
arm cavity which indeed possess lower thermoelastic noise than the  
baseline advanced LIGO design.
But our primary modification---the change to Mexican-hat optics for the cavity arms---involves
employing mirrors which have never before been used in an interferometer.  Naturally, then, we
must make every effort to demonstrate that this radical proposal will not introduce new problems.

For example,  the Mexican-hat mirror has a very flat central region (cf.\ Fig. \ref{fig:fiducialMH}).  In our
early presentations of this proposal, it was suggested to us that such a mirror design might
make the resulting interferometer substantially more susceptible to errors, be they from static
tilts and displacements or mirror figure error.

In this section, we examine this concern by examining the effect of perturbations on both
gaussian and mesa-beam arm cavities.   More explicitly, we by applying the tools described in
Sec.\ \ref{sec:th:Defects} to two fiducial 
beams (cf.\ MBI Sec.\ IV A 1):
\begin{itemize}
\item \emph{Fiducial mesa beam and fiducial Mexican-hat mirrors}:  The fiducial mesa-beam arm cavity, described in
  Sec.\ \ref{sec:def:fiducial},  which has mesa beams with $D=4b=10.4\text{ cm}$.
\item \emph{Fiducial Gaussian beam and fiducial spherical mirrors}: A fiducial Gaussian beam arm cavity, which has
 beam radius $r_o =4.70\text{ cm}$
  (i.e.\  a $g$-value $g=0.952$).\footnote{The value $g=1-L/{\cal R_c}$, for ${\cal R_c}$ the
    radius of curvature of spherical mirrors, can be related to the Gaussian beam radius $r_o$
    using formulae 
    presented in Appendix \ref{ap:spherical}.}
  [This fiducial Gaussian-beam arm cavity \emph{differs} from the baseline
  (cf.\ MBI Sec.\ IV A 1).]  The fiducial Gaussian beam cavity is chosen to have the same
  diffraction losses, on a mirror of the same coated radius, as the mesa-beam arm cavity 
  (i.e.\ so the two fiducial cavities we compare are similar).
\end{itemize}
Using these two fiducial beams, we  demonstrate that mesa-beam
and Gaussian-beam interferometer designs for advanced LIGO will have broadly (i.e.\ within
a factor $\sim$ a few)  similar
sensitivity to perturbations.  In short, we demonstrate that the mesa-beam proposal will not
introduce undue sensitivity of the arm cavities and interferometers to mirror errors.

\subsection{\label{sec:apply:Defects:parasites}Frequency distribution of parasitic modes 
as a  measure of arm cavity sensitivity to perturbations}
Given the denominators present in Eqs.\ (\ref{eq:pt:phase}) and (\ref{eq:pt:state}), a system
will generically be more unstable to perturbations if the eigenphases of excited modes [i.e.\ $\arg
(\eta_k)$] are close to the eigenphase of the resonant state.  We can therefore crudely
characterize the influence of generic perturbations  by the distribution of eigenmodes
nearby the ground state, also called the parasitic mode distribution.
This discussion provides the basis for MBI Sec.\ IV B.

The resonant frequencies $\omega = k c$ are determined when, after one round trip through the
arm cavity, light in a given state interferes constructively with itself.   Therefore, the
light must be in an eigenstate $\left| p \right >$ of the cavity [Eq.\ (\ref{eq:eigenequation})]
and, moreover, the frequency of the light must be chosen so the eigenvalue of the round-trip
eigenequation is \emph{real}, or chosen so [cf.\ Eq.\ (\ref{eq:def:makeGroundResonant})]
\begin{equation}
2 \pi n  = 2 L \omega/c  + \text{Arg}(\eta_p)
\end{equation}
for $n$ some integer.  The same eigenmode $\left | p \right>$ resonates at a specific
frequency, and every other frequency separated from that frequency by the free spectral range
$\omega_{FSR}  = \pi c/L$.   Within each free spectral range, different eigenmodes (i.e.\ $\left
  | p \right>$, $\left| q \right>$) are
occur at different frequencies, separated by an amount uniquely determined by their eigenvalues $\eta$:
\begin{equation}
\label{eq:def:parasiticSeperations}
\Delta \omega_{pq} \equiv \omega_p - \omega_q = \frac{c}{2L}\left[ \text{Arg}(\eta_p) - \text{Arg}(\eta_q)\right] \; .
\end{equation}

For Gaussian beams, the eigenmodes are distributed within each free spectral range regularly;
each nearest neighbor is separated by frequency [cf.\ Eq.\ (\ref{eq:gauss:lambdas})]
\begin{eqnarray}
\Delta \omega &=& \frac{c}{L} \times \cos^{-1} g = \omega_\text{FSR} \times \frac{\cos^{-1}
  g}{\pi}  \\
  &=& 0.099 \times \omega_\text{FSR} \nonumber  \; ,
\end{eqnarray}
using the $g$-value
$g=0.952$ for our Gaussian baseline beams.

For the mesa beams, the frequency distribution of parasitic modes must be obtained numerically,
by (i) solving the eigenequation for the eigenvalues $\eta_p$ [Eq.\ (\ref{eq:eigenequation}), or equivalently
Eq.\ (\ref{eq:eigenequationAlt}),  using the numerical methods of Sec.\ \ref{sec:num:modes}] and then by (ii) using those eigenvalues in
Eq.\ (\ref{eq:def:parasiticSeperations}) to deduce $\Delta \omega_{p0}$ and therefore the
distribution of parasitic modes.  Table \ref{tbl:modeSepMHL} lists the values for 
 $\Delta \omega_{p0}/\omega_\text{FSR}$ for a few states.  Among states with low diffraction losses
 (i.e.\ with ${\cal L} =1-|\eta_p|^2< 10^{-2}$),
 most modes are very well separated from the ground state; the nearest parasitic mode of a mesa
 beam cavity is only a factor $\sim 2.5$ closer to the ground state frequency than the nearest
 parasitic mode of a Gaussian-beam arm cavity.
Therefore, as discussed in MBI Sec.\ IV C, we crudely expect the mesa-beam cavity to be only
marginally more sensitive 
to perturbations than a Gaussian-beam arm cavity.

\begin{table}
\caption{\label{tbl:modeSepMHL} For a LIGO arm cavity with fiducial Mexican hat mirrors
($D/b = 4$, $R=16$ cm), this table gives the
separation $\Delta \omega/\omega_\text{FSR}$ of the eigenfrequencies of parasitic modes
from the eigenfrequency of the fundamental mesa-beam mode.   [The
table is indexed by two numbers, $p$ and $l$, because the
parasitic modes may be written in the form $u(\vec{r})=R_{p,l}(r) \exp(\pm i
l \theta)$; cf.\ Sec.\ \ref{sec:num:modes}.   The index $p$ is the number of radial nodes
in the eigenfunction $R_p(r)$.]
}
\begin{ruledtabular}
\begin{tabular}{ddddd}
 & \multicolumn{1}{c}{$l=0$} & \multicolumn{1}{c}{$l=1$} &\multicolumn{1}{c}{$l=2$} & \multicolumn{1}{c}{$l=3$}  \\
\hline
 \multicolumn{1}{r}{$p=0$} & 0.0 & 0.0404 & 0.1068 & 0.1943 \\
 \multicolumn{1}{r}{$p=1$} & 0.1614 & 0.2816 & 0.4077 & -0.4581\\
 \multicolumn{1}{r}{$p=2$} & 0.4303 & -0.4140 & -0.2570\footnotemark[1] & -0.0812\footnotemark[1] \\
 \multicolumn{1}{r}{$p=3$} & -0.2330\footnotemark[1] & -0.0488\footnotemark[1] & 0.1406\footnotemark[1] &  \multicolumn{1}{c}{---\footnotemark[1]}\\
\end{tabular}
\end{ruledtabular}
\footnotetext[1]{If the cavity length is adjusted so this mode is resonant, then its diffraction losses ${\cal L}$ in each bounce off the test mass exceed 
10000 ppm  (i.e.\ one percent, in contrast with the fundamental mode's 18 ppm), so it cannot 
resonate strongly.} 
\end{table}

\subsection{\label{sec:apply:Defects:displace}Effect of displacement on cavities bounded by spherical and Mexican-hat mirrors}
The discussion of the previous section gives us good reason to believe that mesa-beam and
gaussian-beam arm cavities will display the same sensitivity (i.e.\ within a factor $\sim 2.5$) to
mirror perturbations.   Here, we test this proposition
when  the ETM is displaced through a distance $\vec{s} = s \hat{x}$ --
or, more explicitly, when mirror $2$ is perturbed by $\delta h_2 = \delta h_\text{disp}$,
given by
\begin{eqnarray}
\label{eq:dh:Disp}
  \delta h_\text{disp} &=& (\vec{s}\cdot \vec{r})\frac{1}{r}\frac{dh_2}{dr} \nonumber \\
     & +&\frac{1}{2}\frac{\left|\vec{s}\times \vec{r}\right|^2 }{r^3} \frac{d h_2}{dr}
  + \frac{1}{2}\frac{(\vec{s}\cdot \vec{r})^2}{r^2} \frac{d^2 h_2}{dr^2} 
 \nonumber  \\ &+& O(s^3)  \\
 &=& (x s)\frac{1}{r}\frac{dh_2}{dr}   + \frac{1}{2}\frac{(x s)^2}{r^2} \frac{d^2 h_2}{dr^2}
     \nonumber \\
     & +&\frac{1}{2}\frac{\left(y s\right)^2 }{r^3} \frac{d h_2}{dr}
 + O(s^3) \; . \nonumber
\end{eqnarray}

Section \ref{sec:th:Defects:summary} summarizes the step-by-step process we  follow to
explore 
the influence of perturbations (here, displacement).  However, because we expect---and our
calculations below confirm---that whatever the precise mirror shapes, 
the properties of the interferometer will depend only weakly on displacement,\footnote{The
  natural length parameter for the problem is $b=\sqrt{\lambda L/2\pi}=2.6\text{ cm}$, the diffraction
  length.   When we perform perturbation theory, we find results which vary in powers of
  $s/b$.  Since the LIGO control system will control displacements to much smaller than
  $1\text{ cm}$, displacements have a relatively small effect on the LIGO interferometer.}
we do not complete all the steps that procedure includes (e.g., we do not compute the change in
thermoelastic noise or cavity gain with displacement).

\subsubsection{Displacement of fiducial spherical mirrors}
To evaluate  the perturbation expansion [Eqs.\ (\ref{eq:pt:state}) and (\ref{eq:pt:psiexpand})]
for a cavity bounded by two identical fiducial spherical mirrors subjected to a displacement of its ETM
through a distance $s$ [Eq.\ (\ref{eq:dh:Disp})], we use
analytic techniques specialized to spherical mirrors (e.g., Hermite-Gauss basis functions;
cf.\ Appendix \ref{ap:spherical}).  After some algebra (described in detail in Appendix
\ref{ap:spherical}), we find the ground state of the perturbed cavity,  
to first order, to be [cf.\ MBI Eq.\ (4.12)]
\begin{subequations}
\label{eq:disp:results:gaussian}
\begin{eqnarray}
\left|u \right> &=& \left|0\right> + \zeta^\text{sph}_1 \left|(1,0)\right> 
    + {\cal O}(s^2)\;, \\
\zeta^\text{sph}_1 &=& \frac{(1-g)^{1/4}}{\sqrt{2}(1+g)^{3/4}} \; (s/b) \nonumber \\
  & =&  0.008 (s/1\text{ mm})\; .
\end{eqnarray}
\end{subequations}
[For clarity and for consistency with other work, we have absorbed a phase into the definition
of the $\left|(1,0)\right >$ Hermite-Gauss state.]
Here, $g\equiv 1-L/R_c=0.952$ is the $g$-value for the two fiducial spherical mirrors.    

\subsubsection{Displacement of fiducial Mexican-hat mirrors}
Similarly, if we repeat the above calculation for Mexican-hat mirrors using the methods 
summarized in Sec.\ \ref{sec:th:Defects:summary}, we find the ground state of the cavity changes
to [cf.\ MBI Eq.\ (4.12)]
\begin{subequations}
\label{eq:disp:results:MH}
\begin{eqnarray}
\left|\psi\right\rangle &=& \left|0\right> + \zeta^\text{MH}_1\left| w_1  \right>
    + {\cal O}(d^2) \\
\left|w_1\right> &\equiv&\left|\psi^{(1)}\right>/\zeta^\text{MH}_1 \\
\zeta^\text{MH}_1 &\equiv& 0.263 \; (s/b) = 0.010 \; (s/1\text{ mm}) \; .
\end{eqnarray}
\end{subequations}
Note that unlike the Gaussian-beam case, $\left| w_1\right>$ is a unit-norm
\emph{superposition} of eigenmodes of the unperturbed 
(mesa-beam) cavity, rather than an eigenmode of that cavity itself.

\subsubsection{Power in parasitic modes}
Perturbations cause changes in the resonant ground state.  Equivalently, perturbations couple
the ground state to the parasitic modes, causing power to leak from the resonant ground state
of the unperturbed arm cavity into these other arm cavity eigenmodes.

In the case of displacement, the power in the parasitic modes is easily distinguished from the
carrier light by symmetry: while the carrier
light is axisymmetric,  the lowest-order changes in state are dipolar (i.e.\ $\left|
  \psi^{(1)}\right> \propto \left| (1,0)\right>$ for Gaussian beams).  The fraction of the
total arm cavity light power in the \emph{dipolar} parasitic modes is [cf.\ MBI Eq.\ (4.13)]
\begin{eqnarray}
\label{eq:disp:results:MH:parasiticPower}
P_1 = \zeta_1^2  \simeq \left\{
{100 (s/1.3{\rm mm})^2 {\rm ppm}  \quad \hbox{sph,} \atop
 100 (s/1.0{\rm mm})^2 {\rm ppm} \quad \hbox{MH.}} \right.  
\end{eqnarray}

[In deriving the above expresison  (\ref{eq:disp:results:MH:parasiticPower}), we assume that for
each displacement distance $s$, the input beam has been optimally tuned to match the arm
cavity's perturbed ground state eigenfunction.  In a sense, this expression measures the 
\emph{intrinsic fraction}  of power sent into parasitic modes by the applied perturbation (here,
displacement of one ETM).

In practice, the arm cavity is driven by some fixed driving beam.  As the mirror is displaced,
the fraction $|\gamma_o'(s)|^2$ of input laser power that enters the arm cavity will change [cf.\
Eq.\ (\ref{eq:pt:gamma0})].  All this laser light enters the arm cavity's perturbed ground
state.  Of the power that enters the arm cavity, only a fraction $\approx P_1$ enters
parasitic modes.  Therefore, relative to the total power beamed towards the arm cavity, 
only a fraction $|\gamma_o'|^2
P_1$ enters parasitic modes.]

\subsubsection{Power out the dark port}
If  one arm cavity's ETM is
displaced, then the light leaving the two arm cavities will not interfere destructively at the
dark port.  The precise amount of power $P_\text{dp}$ out the dark port depends on the driving
beam [cf.\ Eq.\  (\ref{eq:pt:darkport}), which depends on $\gamma_0$, $\gamma_1$, and
$\gamma_2$].  If the interferometer is driven by the optimal Gaussian beam
(cf.\ Sec.\ \ref{sec:th:drive:overlap}), then the power out the dark port is approximately
entirely in a dipolar mode, with net power [i.e.\ Eq.\ (\ref{eq:pt:darkport}) to lowest order,
with $\gamma_1=0$; cf.\ MBI Eq.\ (4.14)]
\begin{equation}
P_\text{dp}^{\hbox{total}} =  \gamma_0^2 \zeta_1^2 
\simeq \left\{
{90 (s/1.3{\rm mm})^2 {\rm ppm}  \quad \hbox{sph,} \atop
 90  (s/1.0{\rm mm})^2 {\rm ppm} \quad \hbox{MH.}} \right.
\end{equation}

\subsection{\label{sec:apply:sub:tilt:Cavity}Effect of tilt on the resonant eigenstate of an
  arm cavity bounded by spherical
    and Mexican-hat mirrors}
Because both the spherical and Mexican-hat mirrors planned for advanced LIGO are very flat, the
planned advanced LIGO arm cavities will necessarily be fairly sensitive to tilt.\footnote{For
  example, to
  order of magnitude, a tilt angle  $\theta \sim \sqrt{\lambda/2 \pi L} = 6\times 10^{-6}$ should produce
  a perturbation of order unity in the optical state of the cavity.  Because an order-unity 
 change in state implies fairly dramatic change in the interferometer, tilts much smaller still
 (of order few$\times 10^{-8}$) can cause serious difficulty with the interferometer. }  This section demonstrates
that, though Mexican-hat
mirrors have a very flat central region---much more so than their spherical counterpart
(cf.\ Fig. \ref{fig:fiducialMHmirrors})---an arm cavity bounded by Mexican-hat mirrors will only be
somewhat (i.e.\ a factor $\sim$ few) more sensitive to tilt.

If the ETM of an arm cavity is tilted through an angle $\theta_y$ about its $y$ axis, the mirror
surface is effectively perturbed by $\delta h_2 = \delta h_\text{tilt}$:
\begin{eqnarray}
\label{eq:dh:Tilt}
\delta h_\text{tilt} & =& \theta_y x  + O(\theta_y^3) \; .
\end{eqnarray}

\subsubsection{Tilt-induced changes in the arm cavity ground state}

When this perturbation is inserted into the perturbation expansion (\ref{eq:pt:state}) and
the terms in that expansion are evaluated in the case of \emph{spherical mirrors}
 (using special properties of spherical mirrors and
Hermite-Gauss basis states; cf.\ Appendix \ref{ap:spherical}), we find first- and second-order
corrections to the state [i.e terms in Eqs.\ (\ref{eq:pt:state}) and (\ref{eq:pt:psiexpand})] to
be given by  [cf.\ MBI Eqs.\ (4.3) and (4.4)]
\begin{subequations}
\label{eq:tilt:results:gaussians}
\begin{eqnarray}
\left|\psi_1\right\rangle &=&\alpha^\text{sph}_1 \left|(1,0)\right> \; , \\
\left| \psi_2\right\rangle &=& \alpha^\text{sph}_2 \left|(2,0)\right> \; , \\
\alpha^\text{sph}_1 &\equiv& \frac{1}{\sqrt{2}(1-g^2)^{3/4}} (\theta_y \times \sqrt{k L})
\nonumber \\
  &=& 0.0064 (\theta_y/10^{-8})
  \; , \\
\alpha^\text{sph}_2 &\equiv& \frac{4}{\sqrt{2}(1-g^2)^{1/2}(1-g)} (\theta_y \times \sqrt{k L})^2
  \nonumber \\
 &=& 0.00046 (\theta_y/10^{-8})^2
  \; .
\end{eqnarray}
\end{subequations}
Here $g=0.952$ is the $g$-value for the fiducial cavity,
 the states $\left|(m,n)\right>$ denote states in the Hermite-Gauss basis, and  these states have been adjusted in phase to make $\alpha_1$ and
$\alpha_2$ real.

If this same expansion is evaluated using \emph{Mexican-hat mirrors} via
the numerical procedure outlined in Sec.\ \ref{sec:th:Defects:summary}, we find first- and
second-order 
corrections to the state [i.e terms in Eq.\ (\ref{eq:pt:psiexpand})].  We will not provide an
explicit form for these states here.   These corrections 
have norms given by the Mexican-hat analogues of Eq.\ (\ref{eq:tilt:results:gaussians}) for
$\alpha_1$ and $\alpha_2$ [cf.\ MBI Eq.\ (4.5)]: 
\begin{subequations}
\label{eq:tilt:results:MH}
\begin{eqnarray}
\alpha^\text{MH}_1 &\equiv& ||\psi^{(1)}|| = 14.78 \; (\theta_y  \sqrt{k L}) \nonumber \\
  &=&  0.0227 (\theta/10^{-8})\\
\alpha^\text{MH}_2 &\equiv& ||\psi^{(2)}|| =  74.97 \; (\theta_y  \sqrt{k L})^2 \nonumber \\
  &=&  0.00018 (\theta_y/10^{-8})^2
 \; .
\end{eqnarray}
We can use these expressions to define normalized representations of the first- and second-order
corrections:
\begin{eqnarray}
u_1 &\equiv&\left|\psi^{(1)}\right>/\alpha^\text{MH}_1 \;, \\
u_2 &\equiv&\left| \psi^{(2)}\right>/\alpha^\text{MH}_2 \; .
\end{eqnarray}
\end{subequations}
We display contour maps of these normalized admixtures of modes in
Fig. 5 of MBI \cite{MBI}.


\subsubsection{Parasitic mode power excited by tilt}
The largest correction to the perturbed arm cavity ground state is a \emph{dipolar}
  perturbation (i.e.\ $\left| 
  (1,0)\right>$ for Gaussians; $\left| u_1 \right>$ for mesa beams).  Therefore, interpreting
  this change as an excitation of dipolar parasitic modes, the fractional power in the dipolar
  parasitic modes is [cf.\ MBI Eq.\ (4.6)]
\begin{eqnarray}
P_1 = \alpha_1^2  \simeq \left\{
{0.0005 (\theta_y/3.5\times 10^{-8})^2  \quad \text{sph,}\atop
0.0005 (\theta_y/1.0\times 10^{-8})^2 \quad \hbox{MH,}} \right.  
\end{eqnarray}
when the ETM is tilted through an angle $\theta_y$.

\subsubsection{Tilt-induced changes in the diffraction losses of the ground state of the
  resonant arm cavity (Mexican-hat only)}
The diffraction losses associated with the ground state---which we approximate by the
clipping approximation losses ${\cal L}_1$ and ${\cal L}_2$ [Eq.\ (\ref{eq:clip})]---also
change when the beam state changes.  As described in Sec.\  \ref{sec:th:pt:imply:diff}, we find
the perturbed value for, say, ${\cal L}_1$ merely by expanding the expression for ${\cal L}_1$,
obtaining the general expansion Eq.\ (\ref{eq:pt:diffloss}).

If we evaluate Eq.\ (\ref{eq:pt:diffloss}) for the case of a cavity bounded by two Mexican hat
mirrors with one mirror (the ETM, i.e.\ mirror $2$) tilted through an angle $\theta_y$, we find
we can rewrite Eq.\ (\ref{eq:pt:diffloss}) for the losses at mirror $1$ (the ITM)  in terms of
the expansion mentioned above (i.e.\ in 
terms of 
$\alpha_1$, $\alpha_2$, $u_1$, and $u_2$):
\begin{eqnarray}
\label{eq:pt:apply:tiltedLosses}
{\cal L}_1' &=& {\cal L}_1 + \alpha_1^2 \left({\cal L}_A^{(2)} - {\cal L}_1 \right)
  + \alpha_2 {\cal L}_B^{(2)}  \\
  &\approx& 18\text{ ppm}\; \left[ 1+0.0025 (\theta_y/10^{-8})^2 \right] \nonumber
\end{eqnarray}
[cf.\ MBI Eq.\ (4.7)]
where
\begin{eqnarray}
{\cal L}_1 &=& \left< u_o | O_1 | u_o \right> =18 \text{ ppm} \; , \nonumber \\
{\cal L}_A^{(2)} &=& \left< u_1 | O_1 | u_1 \right> = 96 \text{ ppm} \; , \nonumber \\
{\cal L}_B^{(2)} &=& 2 \text{Re}\left< u_o | O_1 | u_2 \right> = 29 \text{ ppm} \; .
\end{eqnarray}
and where $u_1$ and $u_2$ are defined  by Eq.\ (\ref{eq:tilt:results:MH}).

Using a similar technique we can also evaluate the change in the clipping approximation
diffraction losses at mirror 2.
%
%
  [As discussed
Sec.\ \ref{sec:subsub:approximateIdenticalContributions},  we for simplicity assume the
diffraction losses at 
both mirrors remained the same as the mirror tilted, or ${\cal L}_1={\cal L}_2$.  Since the 
diffraction losses influence physical quantities
like the cavity gain fairly little, we require only a rough estimate of their sensitivity to perturbations.]

\subsection{\label{sec:apply:sub:tilt:Interferometer}Effect of tilt on arm cavities and interferometers using
  Mexican-hat mirrors but driven by Gaussian beams}
Tilt of one ETM causes changes in several important properties of an interferometer 
that uses  mesa beams (and is driven by the optimal Gaussian):
(i) the amount of thermoelastic noise present in the dark port signal; (ii) the amount of
power 
present in the perturbed arm cavity (i.e.\ the arm cavity gain); and (iii) the amount of input
light power leaving the dark port.  In this section we compute the changes in these three
quantities by applying 
 the general techniques 
presented in Sec.\ \ref{sec:th:pt:imply} to the case of tilt.

\subsubsection{Tilt and thermoelastic noise}
To lowest order in  $\theta_y/(b/L)$,
 tilt has 
\emph{no effect} on thermoelastic noise: at first-order, tilt only excites odd-parity modes,
which produce nonaxisymmetric intensity perturbations, and nonaxisymmetric intensity
 perturbations do not contribute to lowest-order changes in the thermoelastic noise integral
 (cf.\ Sections \ref{sec:th:pt:imply:thermo} and   
\ref{sec:subsub:firstordersymmetry}).

\subsubsection{Tilt and cavity gain, when driven by a Gaussian beam}
Equation (\ref{eq:pt:gain}) provides a general expression for the cavity gain for 
a perturbed cavity; this expression depends on (i) the amount of light power $\gamma_o'$ entering
the perturbed arm cavity [Eq.\ (\ref{eq:pt:gamma0}), evaluated using the perturbed mesa-beam state from 
Eq.\ (\ref{eq:tilt:results:MH})], which we can express using 
the perturbation parameters $\gamma_1$ and $\gamma_2$,
\begin{equation}
\gamma_1 = 0  \quad \gamma_2 = -0.070 + 0.013 i   \; ;
\end{equation}
and (ii) on the diffraction losses of the
resonant state of the tilted arm cavity  [Eq.\ (\ref{eq:pt:apply:tiltedLosses})
above].\footnote{When computing the diffraction losses for a tilted cavity, we assume the
same diffraction losses associated with a bounce off mirror $1$ and mirror $2$.}
Combining these expressions, we find the arm cavity power, relative to the input beam power, to
be given by [cf.\ MBI Eq.\ (4.8)]
\begin{eqnarray}
{\cal G}'_\text{anal} &=& {\cal G}_\text{anal} 
 \left[1-\alpha_1^2 + \frac{\alpha_2}{\gamma_0} (\gamma_2 + \gamma_2^*) 
   - 2 \frac{{\cal L}'_1- {\cal L}_1}{1-r_I}
  \right] \nonumber \\
 &=& 737 \left[1- 5.5 \times 10^{-4} (\theta_y/10^{-8})^2 \right] \; .
\end{eqnarray}

\subsubsection{Tilt and dark port power, when driven by a Gaussian beam}
Finally, when one ETM mirror in a mesa-beam interferometer is tilted, the beamsplitter sends
light to the dark port.  Equation  (\ref{eq:pt:darkport}) provides a general expression for the
light $u_\text{dp}$ leaving the interferometer through the dark port; for tilt, this expression
evaluates to 
\begin{equation}
u_\text{dp} = \sqrt{2} \left[
  \left( -\gamma_o \alpha_1^2 + \gamma_2 \alpha_2\right) \left| u_o \right>
 + \alpha_1 \gamma_o \left| u_1 \right> + \alpha_2 \gamma_o \left| u_2 \right>
  \right]
\end{equation}
where we use (i) the definitions   $\alpha_1 = ||\psi^{(1)}||$ and similarly
[cf.\ Eq.\ (\ref{eq:tilt:results:MH})],  and (ii) $\gamma_1
= 0$, to simplify the general equation (\ref{eq:pt:darkport}).  

The corresponding fraction of the interferometer's power that exits the interferometer
through the dark port in the fundamental mode $u_o$ and in the parasitic modes $u_1$ and $u_2$
is [cf.\ MBI Eq.\ (4.9); note the MBI expression will be larger by  a factor 4]\footnote{The
  MBI expression estimates the effect when all four mirrors are tilted about uncorrelated axes;
it therefore is larger than our result, which describes the effect of
tilting only one mirror.}
\begin{subequations}
\begin{eqnarray}
P_\text{dp,0} &=&  
   |-\gamma_o \alpha_1^2+\gamma_2 \alpha_2|^2 \simeq 0.256 (\theta_y/10^{-8})^4 \text{ ppm}
  \\
P_\text{dp,1} &=&  
    \gamma_0^2 (\alpha_1^\text{MH})^2  \simeq 478 (\theta_y/10^{-8})^2 \text{ ppm}\;, \\
P_\text{dp,2} &=&  
    \gamma_0^2 (\alpha_2^\text{MH})^2  \simeq 0.024  (\theta_y/10^{-8})^4 \text{ ppm}\;.
\end{eqnarray}
\end{subequations}

\subsection{\label{sec:apply:sub:figure}Effect of mirror figure error on mesa-beam interferometers}




In this section, we explore the sensitivity of individual arm cavities and the overall
interferometer to mirror figure error, when the beam resonating in the arm cavities is a
fiducial  mesa beam.  More specifically,  in this section we 
(i) distort the ETM by a physically plausible amount (i.e.\ an amount estimated from actual
mirror figure error measurements of LIGO mirrors); (ii) apply perturbation theory (cf.\ 
Sec.\ \ref{sec:th:Defects:summary}) to deduce the change in resonant ground state of the
arm cavities; and then (iii) use the resulting modified beam state to deduce how the
power spectrum of thermoelastic noise will change due to the perturbed beam cross section.

MBI \cite{MBI} uses the computations performed in  this section to place constraints on the
accuracy of machining required  of   the mirrors used in a mesa-beam interferometer; cf.\ 
MBI Sections IV F and IV G.

\subsubsection{GLB's worst-case figure error}
GariLynn Billingsley has provided us with a map of a worst-case figure error, $\delta z_{\rm
wc}(x,y)$ [height error as function of Cartesian coordinates in the transverse plane], produced
by current technologies.  Her map is based on the measured deviation of a LIGO-I beam-splitter
substrate from flatness.  The measured substrate had diameter 25 cm; she stretched its
deviation from flatness (its ``figure map'') to the baseline advanced LIGO mirror diameter of 35.4 cm, fit
Zernike polynomials to the stretched map, and smoothed the map by keeping only the lowest 36
Zernikes. 
MBI Fig. 6 provides a  contour diagram of the resulting figure map (figure ``error'').  In the central region (innermost 10 cm in radius), the peak to valley error
$\Delta z$ is about 30 nm, while in the outer region (10 cm to 16 cm in radius), it is about
110 nm.

\begin{figure}
\includegraphics{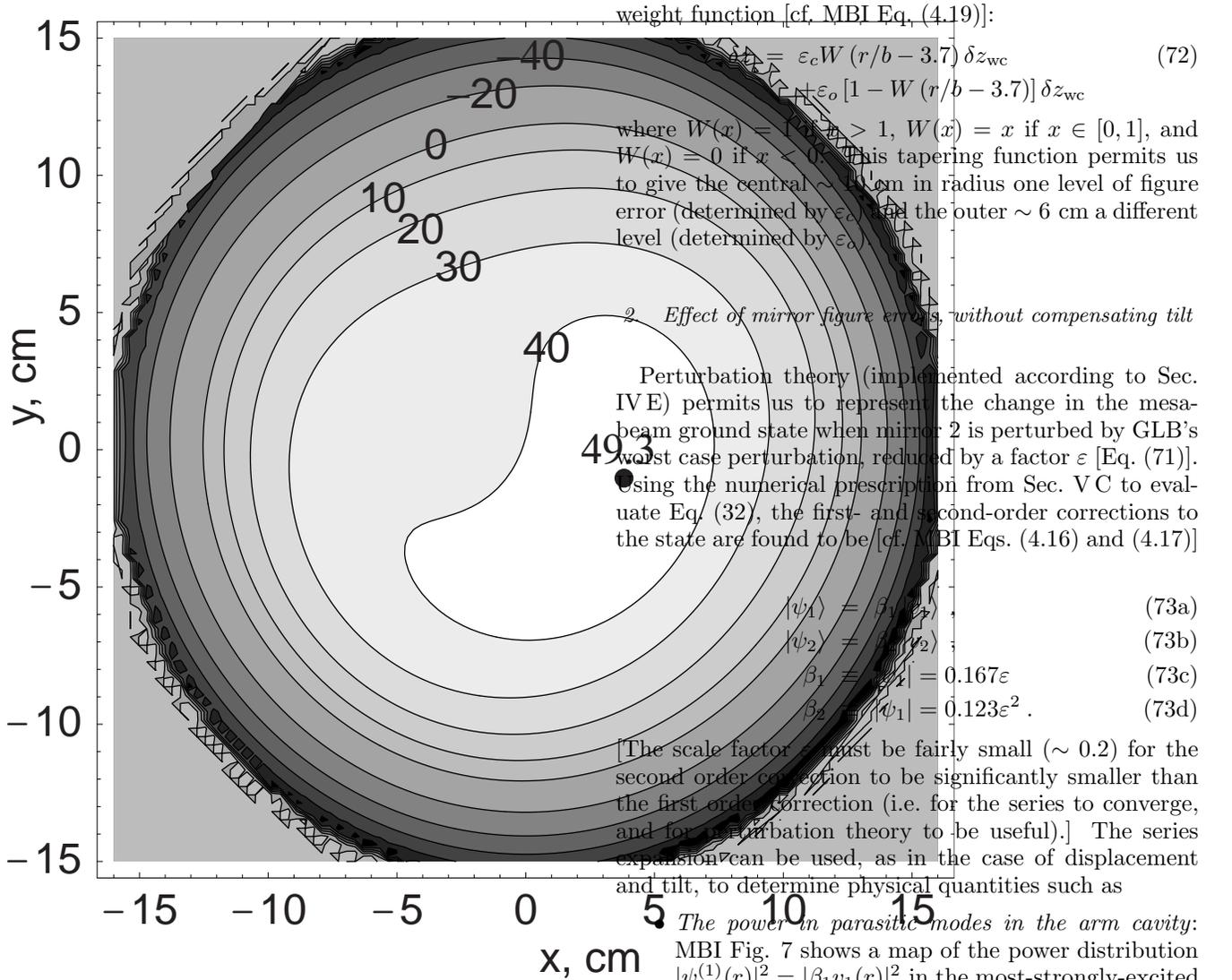}
\caption[The worst-case figure
error.]{\label{fig:mirrorDefects}Contour diagram of GariLynn
  Billingsley's worst-case mirror figure error [height $\delta
  z_\text{wc}$] in nanometers, cf.\ MBI Fig. 6.
}
\end{figure}

 Billingsley (private communications) thinks it likely that in the central region (which dominates our
considerations), peak-to-valley errors of $\Delta z \sim 5$ nm (about 1/5 as large as 
in MBI Fig. 6)  may be achievable; 
Jean Marie Mackowski believes even smaller errors can be obtained with coating methods 
(cf.\ MBI Sec.\ IV F 1).  Accordingly, in
the analyses described below we shall use Billingsley's map, scaled down in height by a factor
$\varepsilon$ [cf.\ MBI Eq.\ (4.15)]:
\begin{equation}
\delta z = \varepsilon \delta z_{\rm wc} (x,y)\;,
\label{eq:def:epsilon}
\end{equation}
and we shall use $\varepsilon= 0.2$ and $\Delta z = 6$ nm as our fiducial values for
$\varepsilon$ and $\Delta z$.  

Also, Billingsley thinks it likely that the outer regions of the mirror will be of
significantly lesser quality than the inner regions.  To study the sensitivity of the
interferometer and thermoelastic noise to errors in the exterior, we divided GLB's
perturbations into two regions by a weight function [cf.\ MBI Eq.\ (4.19)]:
\begin{eqnarray}
\label{eq:defect:dhSplit}
\delta z &=& \varepsilon_c W\left( r/b - 3.7\right)\delta z_\text{wc} \\
  & & + \varepsilon_o \left[1-W\left( r/b - 3.7\right)\right]\delta z_\text{wc}  \nonumber
\label{eq:def:epsilonInOut}
\end{eqnarray}
where $W(x) = 1$ if $x>1$, $W(x)=x$ if $x\in[0,1]$, and $W(x)=0$ if $x<0$.   This tapering
function permits us to give the central $\sim 10\text{ cm}$ in radius one level of figure error
(determined by $\varepsilon_c$)
and the outer  $\sim 6\text{ cm}$ a different level (determined by $\varepsilon_o$).

\subsubsection{Effect of   mirror figure errors, without compensating tilt}
Perturbation theory (implemented according to Sec.\ \ref{sec:th:Defects:summary}) permits us to
represent the change in the mesa-beam ground state when mirror $2$ is perturbed by 
GLB's worst case perturbation, reduced by a factor $\varepsilon$ [Eq.\ (\ref{eq:def:epsilon})].  
Using the numerical prescription from Sec.\ \ref{sec:num:pt} to
evaluate Eq.\ (\ref{eq:pt:state}),
the first- and second-order
corrections  to the state  are found to be 
[cf.\ MBI Eqs.\ (4.16) and (4.17)] 
\begin{subequations}
\label{eq:defect:dstateRaw}
\begin{eqnarray}
\left|\psi_1\right\rangle &=&\beta_1 \left| v_1 \right\rangle \; , \\
\left|\psi_2\right\rangle &=&\beta_2 \left| v_2 \right\rangle \; , \\
\beta_1 &\equiv&|\psi_1| = 0.167 \varepsilon \\  
\beta_2 &\equiv&|\psi_1| = 0.123 \varepsilon^2  \; .
\end{eqnarray}
\end{subequations}
[The scale factor $\varepsilon$ must be fairly small ($\sim 0.2$) for the second order correction
to be significantly smaller than the first order correction (i.e.\ for the series to converge,
and for perturbation theory to be useful).]    The series expansion can be used, as in the case
of displacement and tilt, to determine physical quantities such as
\begin{itemize}
\item \emph{The power in parasitic modes in the arm cavity}:
MBI Fig. 7
shows a map of the power distribution $|\psi^{(1)}(r)|^2 =|\beta_1 v_1(r)|^2$  in the
most-strongly-excited parasitic mode combination $\left| v_1 \right>$.  The net power in
parasitic modes (as a fraction of total power in the arm cavity) is well approximated by the integral of this quantity [cf.\ MBI Eq.\ (4.17)]:
\begin{equation}
P_1 = |\beta_1|^2 = 0.0011 (\varepsilon/0.2)^2 \; .
\end{equation}

\item \emph{The power leaving the dark port}:
Equation (\ref{eq:pt:darkport}) tells us that the fraction of the interferometer's overall
laser light input power going out the dark port is approximately 
[i.e.\ Eq.\ (\ref{eq:pt:darkport}) to lowest order; cf.\ MBI Eq.\ (4.18)]
\begin{eqnarray}
P_{dp} &\approx& \left(|\gamma_o|^2 +|\gamma_1|^2\right)|\beta_1|^2 \nonumber \\
  &\simeq &  \gamma_o^2  |\beta_1|^2  = 0.0010 (\varepsilon/0.2)^2
\end{eqnarray}
where we neglect $\gamma_1$ and compute only the lowest order term in the dark port
  power.\footnote{Unlike displacement and tilt, for mirror figure error $\gamma_1$ is generically nonzero, since the
  perturbation admits an axisymmetric part.}
\end{itemize}

Naturally, the interferometer is more sensitive to errors in the inner $\sim 10\text{ cm}$, where the
beam power is large, than to the outer $\sim 6\text{ cm}$, where the beam power is small.  To
investigate this effect, we considered height perturbations of the form
discussed in Eq.\ (\ref{eq:defect:dhSplit}); to characterize the sensitivity of the beam shape to
these defects,
we examined the norm $\beta_1^2 = ||\psi^{(1)}||^2$ [Eq.\ (\ref{eq:defect:dstateRaw})],
which is necessarily a quadratic form in $\varepsilon_c$ and $\varepsilon_o$ [cf.\ MBI Eq.\
(4.20)].
Using the same methods as for Eq.\ (\ref{eq:defect:dstateRaw}), we found the norm $\beta_1^2$
to be
\begin{eqnarray}
\label{eq:defect:TwoRegions}
\beta_1^2&=& 0.028 \varepsilon_c^2 + 1.06\times 10^{-3} \varepsilon_o (\varepsilon_o - \varepsilon_c) \\
   &\approx& 0.0011 (\varepsilon_c/0.2)^2 
  +  4.3\times 10^{-5} (\varepsilon_o/0.2) (\varepsilon_o - \varepsilon_c)/0.2  \; . \nonumber
\end{eqnarray}
If $\beta_1$ is used as a characteristic example of the sensitivity of physical
quantities (e.g., the power out the dark port) to height perturbations, 
then a mesa-beam interferometer is  around $5$ times more sensitive to 
mirror figure errors in the interior region of each test-mass mirror 
(i.e.\ its inner $\sim 10\text{ cm}$) than it is to perturbations
outside that region.\footnote{In our perturbative computations, the mirrors were
 treated as \emph{infinite}.  The infinite-mirror model proves
 sufficiently accurate for perturbations in the center $10$cm of the
 mirror.  Unfortunately, for mirror defects in the
 outer $6$cm, the resulting perturbed state has  diffraction losses
 roughly comparable to the effect of the perturbation.  Therefore, our
 perturbative simulations of mirror defects in the outer $\sim 6$cm in
 general, and the 
 coefficients of $\epsilon_o$ in Eq.\ (\label{eq:defect:TwoRegions})
 in particular, are not reliable.  
 By contrast, in her simulations Erika
 D'Ambrosio used realistic finite mirrors
 \cite{ErikaSimulations}.   As discussed in MBI [cf.\ MBI Eq.\ (4.20)], she
 finds the inner region to be about 3.5 times more sensitive to mirror
 perturbations than the outer region.}




\subsubsection{Mirror figure errors, with compensating tilt}
The tilt control system automatically and dynamically reorients the mirrors in response to what
it interprets as tilt.  Specifically, the mirror tilt control system
(i) \emph{measures signals containing information about beam asymmetry}, such as the output
  of a quadrant photodiode; 
(ii) \emph{computes the mirror tilt that would generate these asymmetries} using perturbation
  theory expansions  [i.e.\ Eq.\ (\ref{eq:pt:state}), to first order]; and then
(iii) \emph{tilts all four LIGO mirrors to eliminate  the apparent tilts} that the system
  computed in the previous step.
Therefore, since mirror figure error \emph{also} produces beam asymmetries, the tilt 
control  system of the
interferometer will act to partially compensate (the dipolar part of) the static mirror
defect.

 The precise quantity  the tilt control system measures to deduce the tilt angle
is not important: different approaches to tilt control interpret the optical state
of our arm cavity [i.e.\ interpret the state $\left|u_o'\right>$ defined by the corrections in 
Eq.\ (\ref{eq:defect:dstateRaw})] in a manner  fairly independent of the method used (i.e.\ the 
compensating tilt we calculate depends little on the method we use to calculate it).  
In this section, we assume the tilt 
control system acts to minimize the dipolar component of the arm cavity beam power.

Therefore, we find the tilt-compensated state by (i) adding together the results of a tilt 
perturbation [i.e.\ a generalization of Eq.\ (\ref{eq:tilt:results:gaussians}) that depends on
two tilt angles,  
$\theta_x$ and $\theta_y$] 
and the mirror figure error perturbation [Eq.\ (\ref{eq:defect:dstateRaw}), which depends on
$\varepsilon$]; 
(ii) evaluating the norm $||\psi^{(1)}||$ of the first-order perturbation; and then
(iii) finding the tilt angles $\theta_x$ and $\theta_y$ which minimize that norm.
The resulting optimal tilt angle is [cf.\ MBI Sec.\ IV  H 3]
\begin{subequations}
\begin{eqnarray}
\theta_x &=&  +0.98 \times 10^{-8} (\varepsilon/0.2)\; ,\\
\theta_y &=&  +0.69 \times 10^{-8} (\varepsilon/0.2)\; , \\
\theta &=&\sqrt{\theta_x^2 + \theta_y^2 } = 1.2 \times 10^{-8} (\varepsilon/0.2) 
\end{eqnarray}
\end{subequations}
(in the limit of small $\varepsilon$, so linear theory applies).  MBI Figure 8 shows the height of
the surface of the mirror after the compensating tilt is applied.

After the tilt is applied, the first-order correction to the resonant mesa-beam state is a
combination of tilt and mirror-figure perturbations.  For brevity, we denote the net
first-order correction to the mesa-beam state when tilt is applied by 
\begin{subequations}
\begin{eqnarray}
\left|\psi^{(1)}\right\rangle &=&\beta_{1,c} \left| v_{1,c} \right\rangle \;, \\
\beta_{1,c} &\equiv& ||\psi^{(1)} || 0.02 (\varepsilon/0.2) \; .
\end{eqnarray}
\end{subequations}
The square of the norm (i.e., $\beta_{1,c}^2$), as usual, is the fraction of arm cavity power
which is in parasitic modes [cf.\ MBI (4.23)]:
\begin{equation}
P_{1c}^\text{arm} = |\beta_{1,c}|^2 = 0.0004 (\varepsilon/0.2)^2 \; .
\end{equation}
MBI Figures  9 shows a map of the power in parasitic modes, $|\beta_{1,c} v_{1,c}(r)|^2$.
Also, the norm of this first-order correction provides an estimate of the power leaving the
dark port of this  interferometer (i.e.\ an interferometer with one tilted, defective mirror)
when the interferometer is driven  with Gaussian beams [i.e.\ Eq.\ (\ref{eq:pt:darkport} to
lowest order; cf.\ MBI (4.24)]:\footnote{As in the
  untilted case, we neglect the (nonzero) term $|\gamma_1|^2$.}
\begin{eqnarray}
P_{dp} &\approx& \left(|\gamma_o|^2 +|\gamma_1|^2\right)|\beta_{1,c}|^2 \nonumber \\
  &\simeq &  \gamma_o^2  |\beta_1|^2  = 0.00038 (\varepsilon/0.2)^2
\end{eqnarray}

\subsubsection{Influence of mirror figure errors on thermoelastic noise}
Because the ETM's figure error distorts the  beam resonating in the arm cavity, the
thermoelastic noise produced by each mirror bounding that arm cavity changes by some
small amount.   Given the change in state deduced above, we know how the beam profile at the
ITM changes.  We can therefore evaluate, using the discussion of Sec.\ \ref{sec:th:pt:imply},
the change in  thermoelastic noise associated with the beam reflecting off the ITM's face.

To be explicit, to compute the
first-order effects of a perturbation $\delta P$ to the thermoelastic noise integral $I_1$ of
mirror $1$ in this fiducial case, we first
select the 
axisymmetric portion $\delta P_o(r) =\int d\varphi P(\vec{r})/2 \pi$ of $\delta P$; we then linearize the analytic expressions derived by
Liu and Thorne for cylindrical mirrors
(cf.\ Sec.\ \ref{sec:subsub:LT}), using $\delta P_o$ as the magnitude of the
perturbation\footnote{When linearizing the LT equations, the natural relationship between
     $\delta P$ and 
      $\delta p_m$ follows from
     Eq.\ (\ref{eq:thel:LTp}). 
Note, however, that  $p_o$ is
     independent of $P(r)$ [because $P(r)$ are all normalized] and thus $\delta p_o = 0$.}; and
finally we extract from this linearization the first-order change in $I_1$.

When we apply this technique to the fiducial beam reflecting off a cylindrical mirror of radius
$16 \text{ cm}$ and thickness $13 \text{ cm}$ which is deformed by GariLynn Billingsley's
mirror distortion [Eq.\ (\ref{eq:def:epsilon})], we find the thermoelastic noise integral 
for that mirror changes to
\begin{equation}
I_1/(2\times 10^{-28}) = 0.632\left[1 +  0.035 (\varepsilon/0.2) \right]
\end{equation}
where $I_1$ denotes the total thermoelastic noise integral for mirror $1$ when mirror $2$ is
deformed by our scaled height perturbation (\ref{eq:def:epsilon}).
[Roughly speaking, we expect the beam and hence the thermoelastic noise to change in a qualitatively
similar fashion at mirror $2$ (cf.\ Sec.\ \ref{sec:subsub:approximateIdenticalContributions})].

MBI Section IV G applies this result to deduce how sensitive the power spectrum of
thermoelastic noise is to uncorrelated mirror figure errors on all four mirrors 
[cf.\ MBI Eq.\ (4.27)].

\section{Conclusions}
In this paper, we have described both the theory and practice needed to obtain the 
results summarized in MBI.  Specifically, we have developed analytic and numerical tools to
evaluate the following:
\begin{itemize}
\item \emph{Thermoelastic noise integrals}:  We developed practical techniques for finding the
  thermoelastic noise for nonstandard optical systems 
(i.e.\ noncylindrical finite mirrors and unusual beam shapes) [Sections \ref{sec:th:thermo}
 and \ref{sec:num:thermo}].    We tabulated the thermoelastic
 noise, relative to the current advanced LIGO baseline, for many alternative mirror and
 beam configurations [Tables \ref{tbl:CylGauss} - \ref{tbl:NSRange} in Section
 \ref{sec:apply:Design}].  We found many configurations with lower thermoelastic noise than
 the baseline advanced LIGO configuration.
\item \emph{Eigenmodes of an arm cavity bounded by Mexican-hat mirrors}: We also wrote numerical code 
to find the eigenmodes of an optical cavity  [Sections \ref{sec:th:propagation} and 
\ref{sec:num:modes}].   We computed and tabulated  many of the eigenmodes of an arm cavity
bounded by Mexican-hat mirrors (none of which appear explicitly in this paper).  
\item \emph{Second-order optical perturbation theory}:  Finally, we developed expressions for
  second-order optical perturbation theory [Sec.\ \ref{sec:th:Defects}].  We applied
  perturbation theory extensively to study the sensitivity of mesa-beam interferometers to
  perturbations (i.e.\ to mirror figure error, mirror tilt, and mirror displacement).
\end{itemize}
The results found here are used in MBI to  conclude
 that mesa-beams interferometer designs offer clear advantages over the baseline
advanced LIGO design,  without being substantially more sensitive to mirror figure error,
tilt, or displacement perturbations.

In MBI \cite{MBI}, in Erika D'Ambrosio's paper \cite{ErikaSimulations}, and this paper, we and
our collaborators  have only taken the first steps towards the design of a practical
mesa-beam advanced LIGO proposal.  For example, more perturbative calculations---this time,
applied to the final  design, rather than to a fiducial case---are needed, for the design of
the control system (i.e.\ so the relationship between light on various photodiodes and the
correcting tilt applied to the LIGO mirrors can be established).   Further, in this paper we
have only begun to explore the space of all possible mirrors.  For simplicity, we chose to fix
the mirror mass to $40\text{ kg}$.  In practice, however, with sapphire, mirror designs are
limited by fabrication limits (i.e.\ the radius of the mirror is limited by the radius of the
sapphire boule one can grow) rather than by weight limits.  Therefore, before a final
design is chosen, 
more mirror designs (including cylinders with $m>40\text{ kg}$) should be examined.
Finally, in this paper, our Mexican-hat designs were limited to symmetric cavities (i.e.\ using
\emph{identical} mirrors).   In practice, asymmetric designs offer the possibility of lower
thermoelastic noise and greater practical convenience\footnote{For example, rather than design
  all four mirrors to be identical MH, one may want to operate with some spherical mirrors
  first, then replace a few mirrors (i.e.\ the ETMs) with MH-like mirrors later. 
  Bill Kells has proposed using a flat ITM and first a 
  spherical and then an MH-like ETM.
}.  Further work is necessary on the design, construction, and operation  of asymmetric 
cavities with mexican-hat-like mirrors.

\begin{acknowledgements}
Without the assistance of our collaborators on our extended project to study LIGO thermoelastic
noise in the presence of Mexican-hat mirrors (Erika D'Ambrosio and Kip Thorne), this paper would
never have appeared; we thank them for their support and their assistance with the text.  
We thank GaryLynn Billingsley for providing and explaining plausible
deformations for LIGO mirrors.   

ROS thanks: Bob Spero for helpful discussion about practical computations of diffraction loss;
Erika D'Ambrosio, for invaluable discussions on perturbation theory for optics, on assistance
with references, and for helpful suggestions regarding thermoelastic noise simulations
(i.e.\ on the choice of the set ${\cal R}$); and Lee Lindblom, for helpful suggestions on
the postprocessing numerical methods.  He also thanks Guido Muller
(our kind LSC reviewer)  and
Pavlin Savov for locating  typographical errors in early drafts.  ROS has been supported in part
by NSF Grant PHY-0099568.

SS and SV thank: Vladimir Braginsky and Farid Khalili for stimulating
discussion.  This research was supported by NSF grant PHY-0098715, the Russian Foundation for
Fundamental Research, Russian Ministry of Industry and Science, and (for SPV) by the NSF
through Caltech's Institute for Quantum Information.

\end{acknowledgements}

\appendix

\section{\label{ap:sapphire}Sapphire material parameters: Notation and values}
While sapphire is not an isotropic material, it can be reasonably approximated by isotropic
elastic and thermodynamic properties.  In this paper, we therefore treat sapphire as an
isotropic medium, with the following specific values for physical parameters:%
\footnote{These numbers are taken directly from the  advanced LIGO project book \cite{advLIGOmain},
  and the advanced LIGO summary web page \cite{advLIGOwebOld}.
}
\begin{eqnarray}
E &=& 4\times 10^{12} \text{erg/cm}^3\\
\rho &=& 4.0 \; \text{g/cm}^3 \\
\sigma &=& 0.28 \\
\alpha &=& 5.5\times 10^{-6} K^{-1} \\
\kappa &=& 3.3\times 10^{6} \text{erg}\, \text{cm}^{-1}\, \text{K}^{-1}\, \text{s}^{-1}\\
C_V &=& 7.7 \times 10^6 \text{erg}\, \text{g}^{-1}\,   \text{K}^{-1}
\end{eqnarray}

\section{\label{ap:mh}General description of mesa beams}
In Section \ref{sec:th:MH} we only briefly describe the form of mesa beams evaluated at the
surfaces of an arm cavity's two mirrors.  In this appendix, we provide a much more thorough
treatment of 
mesa beams.

\subsection{Constructing mesa beams}

Mesa beams are
constructed by averaging some gaussian beams (with waist size $A_o$ and waist location $z_w$)
over some disc of size $D$ and then normalizing the result.  
Specifically, we construct 
mesa beams that propagate towards positive $z$ by first averaging a (non-normalized) gaussian
beam over a disc of radius $D$: 
\begin{subequations}
\label{eq:buildMH}
\begin{eqnarray}
\label{eq:buildMH:average}
U(\vec{r}, z, D) &=& e^{ikz} \nonumber \\
 &\times & \int_{r'<D} d^2 r' U_g(\vec{r}-\vec{r'}, z; \bar{z}, z_w) 
\\
\label{eq:buildMH:gaussian}
U_g(\vec{r}, z; \bar{z}, z_w) &=& 
\exp {\left[ -\frac{r^2 \left[1-i\frac{(z-z_w)}{\bar{z}}\right]}{2  A_o \sqrt{1+\frac{(z-z_w)^2}{\bar{z}^2}}}
  \right]}\\
 &\times&  e^{-i \tan^{-1}(z/\bar{z})} \nonumber
\end{eqnarray}
[where $A_o$ is a function of $\bar{z}$, given by $\bar{z}\equiv k A_o^2$, and where $z_w$ is
the location of the beam waist].  The result of the 
average is necessarily axisymmetric.  Since $U(\vec{r},z,D)$ is
axisymmetric (because of the symmetric integral which defines it), we may normalize the result with a one-dimensional integral:
\begin{eqnarray}
\label{eq:buildMH:norm}
N^2(D,z) &\equiv& \int_0^\infty |U(r, z, D)|^2 2 \pi r dr \\
\label{eq:buildMH:normalize}
u_\text{mesa}(\vec{r}, z, D) &\equiv& U(r,z,D)/N(D,z) e^{i k z}
\end{eqnarray}
\end{subequations}

To construct the associated Mexican-hat mirror which will reflect this mesa beam back into itself
(propagating in the opposite direction) at location
$z_r$, we require the Mexican-hat mirror height function $h_\text{MH}$ to be continuous and to satisfy
\begin{equation}
\label{eq:defHeight}
u_\text{mesa}(r,z_r,D) \exp(-2 i k h_\text{MH}) = u_\text{mesa}(r,z_r,D)^*
\end{equation}
for $=u(\vec{r},z_r,D)$  the values of a mesa beam at the mirror plane $z=z_r$.  This expression
is equivalent to MBI Eq.\ (2.13)  [cf.\  Eq.\ (\ref{eq:def:MHheight}) in this paper].
These
requirement uniquely 
specify the Mexican-hat mirror shape $h_\text{MH}$.

\subsubsection{\label{sec:subsub:MHspecial}Canonical mesa beams: mesa beams for symmetric
  cavities}
The mesa beams  presented in MBI and discussed henceforth in this paper --- denoted
\emph{canonical mesa beams} ---  are assumed to have particularly
special form.  First, it is assumed that  the cavity confining the mesa beam
is symmetric, with mirror $1$ placed at $z=-L/2$ and mirror 2 placed at $z=L/2$; also it is 
assumed that the waist location, $z_w$, is placed precisely between them (i.e $z_w = 0$).  
Second, it is
assumed that the gaussian used to construct the mesa beam is the \emph{minimal-diffraction}
gaussian for the symmetric cavity.  The minimal diffraction gaussian has  $\bar{z}=L/2$ and $z_w=0$,
implying that the beam width is $b =\sqrt{L/k}$ at the mirror surfaces and $b/\sqrt{2}$ at
the beam waist at $z=0$.

\subsection{\label{sec:subsub:MHsurfaces}Canonical mesa beams at the mirror surfaces}
In this paper and in MBI, we only need to know the form of canonical mesa beams at the mirror
surfaces.   In this case, since the beam is canonical (i.e.\ has symmetric cavities and minimal
diffraction gaussians, so $z_w = 0$ and $\bar{z}=L/2$, implying $A_o = b/\sqrt{2}$) and since
we evaluate it at the mirror surfaces $z=\mp L/2$, we find that Eq.(\ref{eq:buildMH:gaussian})
simplifies to 
\begin{eqnarray}
\label{eq:def:mhUgmin}
U_{g,\text{min}}(\vec{r}, z=\mp L/2 ) &=& 
\exp {\left[ -\frac{r^2 \left[1 \pm i \right]}{2  b^2}
  \right]  } e^{\pm i \pi/4}  \; .
\end{eqnarray}
Inserting this expression into the definition (\ref{eq:buildMH}) of general mesa beams, we find that  the
mesa beam form at mirror $1$ (i.e.\ the mirror at $z=-L/2$) is given in terms of 
the construction described in Eq.\ (\ref{eq:buildMHcanonical}).
[For simplicity, in Eq.\ (\ref{eq:buildMHcanonical}) we omit the overall phase factor 
$\exp (i \pi/4)$.]

\section{\label{ap:modelProblem}Converting the fluctuation-dissipation model problem to a
  static model problem at low frequencies}

In Sec.\ \ref{sec:th:thermo}, we claim that at low frequencies, the elastic response of a cylinder  to an oscillating
pressure profile can be reconstructed to a good approximation using the static response of a
cylinder to the same pressure profile, in an appropriate accelerating frame.

Briefly speaking, a relationship between the dynamic and static problems exists because, 
when the oscillations are sufficiently slow, the effect of
dynamical terms in the elastic equations of motions can be neglected.  These physical 
considerations
have been discussed at greater length elsewhere, cf.\ Sec.\ II of LT \cite{LiuThorne}.  

In this appendix, we demonstrate in more technical detail precisely how to establish the desired
(approximate)  relationship.  We assume only basic familiarity with elasticity, on the level of
Blandford and Thorne 
\cite{BlandfordThorne}. 

\subsection{General quasistatic approach for the elastic response to an oscillating surface stress}
We wish to solve the elasticity equations, which can be expressed as
\begin{equation}
\rho \partial_t^2 y_a + \nabla^b T_{ab} = 0 \; ,
\end{equation}
for a test-mass mirror subject to an oscillating  surface stress on its inner face (i.e.\ on the
surface 
$z=0$, where the optic axis of the advanced LIGO arm cavity is  the $z$ axis),
\begin{equation}
T_{az}(t,r,z=0)\equiv \bar T_{az}(r,z=0) \cos(\omega t) \; ,
\end{equation}
and otherwise subject to no other stresses.
[Here, $T_{ab}$ is given in terms of $y_a$ by Eq.\ (\ref{eq:elas:Tab}).]

\subsubsection{Step 1: Express the problem in the accelerated frame}
We can
better understand the response of the mirror substrate if we go into a frame comoving with the 
center of mass of the test mass mirror.
The test mass experiences a net force $F_a \cos(\omega t)$, determined by the surface stress:
\begin{equation}
F_a \equiv - \cos (\omega t )\int d^2 r \bar T_{az}(\vec{r},z=0)  \; .
\end{equation}
In response to this net force, the mirror center  of mass  $R_\text{ cm}$ accelerates:
\begin{equation}
\vec{R}_\text{ cm} = - \frac{F_a}{M \omega^2} \cos(\omega t) \; .
\end{equation}

Therefore, to go to the comoving frame, we perform the following transformation:
(i) define the comoving displacement field $y'_a = y_a - R_\text{cm}$:
\begin{equation}
y'_a \equiv y_a +   \frac{F_a}{M \omega^2} \cos(\omega t) \; ;
\end{equation}
(ii) define the comoving stress-energy tensor as $T'_{ab} = T_{ab}(y')$.  In terms of these two
new quantities, the elasticity equations in the accelerated (primed) frame are
\begin{equation}
\label{eq:elasmodel:accel:general}
\rho \partial_t^2 y'_a + \nabla^b T'_{ab} = - \rho \frac{\vec{F}_a}{M}\cos(\omega t) \; .
\end{equation}
The right hand side is simply the inertial force associated with working in an accelerated frame.

The boundary conditions remain unchanged.

\subsubsection{Step 2: Factor out all sinusoidal dependence}
In the accelerated frame, we now assume all quantities oscillate sinusoidally in response to
the sinusoidally-oscillating pressure profile and inertial force:
\begin{eqnarray*}
y'_a &=& \bar y_a \cos(\omega t) \; , \\
T'_{ab} &=& \bar T_{ab} \cos(\omega t) \; .
\end{eqnarray*}
Substituting these expressions into Eq.\ (\ref{eq:elasmodel:accel:general}), we find
 we can reconstruct a solution
to the dynamic solution by solving the static partial differential equation
\begin{equation}
\label{eq:elasmodel:accel:sinusoidal}
-\rho \omega^2 \bar{y}_a + \nabla^b \bar T_{ab} = 
   - \rho \frac{\vec{F}_a}{M} 
\end{equation}
subject to force-free boundary conditions on all surfaces except the top surface, which is
subjected to a \emph{constant} pressure profile $\bar T_{az}(\vec{r}, z=0)$.


\subsubsection{Step 3: Approximate the problem as static in the accelerated frame}
Finally, at sufficiently low frequencies (i.e.\ frequencies so low that sound crosses the
cylinder many times within 
one period, as is the case for advanced LIGO; cf.\ notes \ref{note:idealizeGeneral} and
\ref{note:idealizeSound}), the first term in the accelerated-frame
elasticity 
equations [i.e.\ the term $\propto y'_a \omega^2$ in 
Eq.\ (\ref{eq:elasmodel:accel:general})]
can be neglected (cf., e.g., LT Sec.\ II).  The remaining problem [i.e.\ Eq.\ (\ref{eq:elas})] can be interpreted precisely
as equations for the static elastic response of a solid to an imposed pressure profile.

\subsubsection{Summary}
To summarize, then, we find an approximate solution for the elastic response of a solid to an
imposed surface stress $T_{az}$  at $z=0$ by
\begin{subequations}
\label{eq:ap:quasistatic}
\begin{equation}
y_a = \cos(\omega t) \left[- \rho \frac{\vec{F}_a}{M\omega^2}  + \bar{y}_a \right]
\end{equation}
where $\bar{y}_a$ is obtained as a solution to the static elastic equations in an accelerated frame
\begin{equation}
 \nabla^b \bar T_{ab} = 
   - \rho \frac{\vec{F}_a}{M} 
\end{equation}
subject to the effective static surface stress $\bar T_{az}$ and bulk acceleration $\vec{F}_a/M$ given by
\begin{eqnarray}
T_{az}    &=& \bar{T}_{az} \cos(\omega t) \\
\vec{F}_a &\equiv& - \int d^2 r \; \bar T_{az}(\vec{r}, z=0)
\end{eqnarray}
\end{subequations}

To summarize our conclusions on physical grounds, because the elastic response occurs much more
rapidly than the surface pressure profile changes
(i.e.\ because the sound crossing time is much shorter than the oscillation period of the
imposed force), we can effectively treat the elastic response as instantaneous.   The elastic
solid moves slowly through a sequence of static configurations. 

This quasistatic approximation, however, must be performed with care.  If one neglects the
dynamical terms entirely, as other authors working on this and related subjects have done
\footnote{%
Because of the fluctuation-dissipation theorem, many other authors working on thermal and
thermoelastic noise (e.g., Cerdonio and Conti \cite{CerdonioConti}; Liu and 
Thorne \cite{LiuThorne}; ...) must solve a similar or identical elastic problem to deduce the effects of noise.
And most choose to approach it using a similar quasistatic approximation, ignoring any
dynamical effects.    However, most make the quasistatic approximation before they go to the
accelerated frame, rather than after.  Only Liu and Thorne have correctly accounted for the
effects of acceleration, though they add those effects in by hand later.
}, then one finds
elasticity equations without the bulk acceleration term.  That equation is inconsistent with
the static boundary conditions we impose.

\subsection{Using the quasistatic elastic solution to simplify our thermoelastic noise integral}
We can apply the quasistatic elastic solution we just developed
[Eq.\ (\ref{eq:ap:quasistatic})] 
to find the thermoelastic
integral $I_A$ [Eq.\ (\ref{eq:IA})] associated with our specific elastic model problem, where the
surface stress imposed has (i) $T_{az}=0$ unless $a=z$  and (ii) 
 $T_{zz} = - \cos(\omega t) F_o P(r)$ [Eq.\ (\ref{eq:elas:bcForce})].  
Since the accelerated-frame transformation does not change $\Theta$
   (i.e.\ $\Theta = \Theta'$) and since  $\left < \cos^2(\omega t) \right> = 1/2$, we conclude that
\begin{equation}
I_A = \frac{2}{F_o^2} \int d^3 r \; \left< |\nabla \Theta|^2 \right>
    \approx  \frac{1}{F_o^2} \int d^3 r \;  |\nabla \bar\Theta|^2   \; ,
\end{equation}
where the approximation neglects terms which are small in the quasistatic limit (cf.
notes \ref{note:idealizeSound} and \ref{note:idealizeGeneral}).

\section{\label{ap:basis}On the completeness of basis states for an arm cavity with two
  identical, infinite mirrors}
In this appendix, we demonstrate that the eigenproblem for two identical infinite mirrors 
[Eq.\ (\ref{eq:eigenequation})] admits a
complete set of orthogonal eigensolutions.  In the text, we use the resulting eigensolutions
as a basis for building perturbative expansions.

Simply, the eigenequation admits a complete set of orthonormal states
because it is an eigenvalue problem for a \emph{unitary} operator. 
Unitary operators always admit a complete set of orthogonal
eigenvectors.  
The technical details for the infinite-dimensional
($L^2$) proof we need here are far beyond the scope of this paper.
However, the reader can understand the result by two simple arguments:

\emph{Orthogonality of eigenvectors for unitary matricies I: Via
  hermetian intermediary}: From any unitary operator we can define a
  hermetian operator $H=
  i \ln U$ (e.g., by a series expansion).  The hermetian operator
  admits a complete set of orthogonal states.  Hence so does
  $U=\exp(-i H)$.

\emph{Orthogonality of eigenvectors for unitary matricies II:
  Directly}: Alternatively, a familiar argument can be applied to demonstrate that, if $U$ is a
unitary matrix, the eigenspaces associated with distinct eigenvalues
 are mutually orthogonal.  

Let $\lambda_a$ and
$\left| a \right>$ denote all the eigenvalues and (normalized)
eigenvectors of a unitary operator $U$.  Since
$U^\dag U = 1$, we know the eigenvalues are pure phase: $\left<a|U^\dag U|a\right> =
|\lambda_a|^2 = 1$.
Further, Since $U^{-1}$ has $\left| a \right>$ an eigenvector with
eigenvalue $1/\lambda_a=\lambda_a^*$, we can show
$\left< b\right| U = \lambda_b \left< b \right| $ and thus
\begin{equation}
0=(\lambda_a - \lambda_b) \left< b| a \right > \; .
\end{equation}
Therefore, eigenvectors associated with \emph{distinct} eigenvalues are
mutually orthogonal.

Furthermore, in each degenerate subspace, we can diagonalize to find
an orthogonal basis.  Therefore, for unitary matricies, all
eigenvectors can be assumed orthonormal.

\section{\label{ap:pt}Second-order perturbation theory for eigenstates of an individual arm cavity}
In this appendix, we develop second-order perturbation expansions which relate changes in shape
of one specific mirror of a symmetric cavity (the ETM) to changes in the resonant states in
general
  (and to changes of the the ground state in particular) of
that cavity.  

Since the precise form of the eigenequation depends on how one chooses to represent the state
(i.e.\ at what plane, in what direction), many alternative and equivalent perturbation
expansions can be derived\footnote{The relationship between these expressions need not be
transparent.  For example, one can represent the eigenequation using states represented
``halfway'' through a reflection of one of the mirrors (a choice which conveniently renders the
resulting 
equation always perfectly symmetric).  The transformation between our representation
and this one depends on the mirror height.  Thus, when the mirror heights are perturbed, the
relationship between these two representations involves a unitary transformation that depends
on the perturbation.
}.  The form presented in this appendix is that developed by O'Shaughnessy; Sergey Strigin and
Sergey  Vyatchanin performed their calculations  using an independently-derived (but provably equivalent) approach.

As in the text, in this section we make heavy use of standard quantum-mechanics operator
notation for states and inner products, a notation described briefly in Section
\ref{sec:sub:notation} (cf., e.g.,  \cite{Dirac,Landau,Townsend}). 

\subsection{Preliminaries: Setting up notation for the expansion}
We study the effect of  changes in height of mirror $2$ on the solutions to the eigenequation
 (\ref{eq:eigenequation}) in the case of a \emph{symmetric} cavity with \emph{infinite}
 mirrors. 
   For clarity and simplicity of notation,  we redefine
the eigenequation problem we perturb still further, into the following expression:
\begin{subequations}
\label{eq:ap:eigenequationPT}
\begin{eqnarray}
\eta \left| \psi \right> &=& G T G \left| \psi \right> \;. \\
G         &\equiv& e^{-i k L} G_1 G_+ \; ,\\
T         &\equiv& e^{-  i \delta_2 }  \equiv 1+ \delta T\; , \\
\delta_2         &\equiv&  2  k \delta h_2  \; .
\end{eqnarray}
\end{subequations}
Here, $\delta h_2$ is the change in the (inward-pointing) height of the far mirror (mirror $2$)
and $\delta T \equiv T-1$.
[The operator $T$, used only in this section, should not be confused with the truncation
operators defined in Eq.\ (\ref{eq:propagatorExamples}).]

\subsection{Perturbation theory expansion, expressed using operators}
We construct a perturbation theory expansion by expanding both sides of the eigenequation
Eq.\ (\ref{eq:ap:eigenequationPT}) in series, giving
\begin{eqnarray}
\left( \eta _{o}+\varepsilon \eta _{1}+\varepsilon ^{2}\eta _{2}+\ldots
\right) \left( \psi _{o}+\varepsilon \psi _{1}+\ldots \right)  \\
=G\left(
1+\delta T\right) G\left( \psi _{o}+\varepsilon \psi _{1}+\varepsilon
^{2}\psi _{2}\right) \; ,  \nonumber
\end{eqnarray}
and then matching orders on both sides; in this expression, $\varepsilon$ is a formal
perturbation parameter, added to rescale the change in the reflection operator: $\delta
T(\epsilon) = \epsilon \delta T$.
When we work out order matching, we find (for $P$ a projection orthogonal to
the ground state $\psi_o$, also denoted $\left| 0 \right>$): 
\begin{subequations}
\label{eq:ap:pt:Tform}
\begin{eqnarray}
\eta &=&\eta _{o}
   +\left\langle 0\left| \delta \tilde{T}\right|0\right\rangle \\ 
 &+&\left\langle 0\left| 
   \delta \tilde{T}P\frac{1}{\eta_{o}-G^{2}}P\delta \tilde{T}
  \right| 0\right\rangle  \; , \nonumber \\  
\psi &=&\left| 0\right\rangle 
 +\frac{1}{\eta _{o}-G^{2}}P\delta \tilde{T} \left| 0\right\rangle  \\ 
 &+&\frac{1}{\eta _{o}-G^{2}}P\left[ \delta \tilde{T}
 -\left\langle 0\left| \delta \tilde{T}\right| 0\right\rangle \right] \frac{1%
}{\eta _{o}-G^{2}}P\delta \tilde{T}\left| 0\right\rangle \; , \nonumber
\end{eqnarray}
\end{subequations}
where the operation $\tilde O$ on an operator $O$ is defined by
\begin{equation}
\tilde{O}\equiv G O G \; .
\end{equation}
We can furthermore substitute into the above expression the expansion
\begin{eqnarray*}
\delta \tilde T &\approx& G\left(- i \delta_2 - \frac{1}{2} \delta_2^2\right)G 
  = -i \tilde \delta_2 - \frac{1}{2} G \delta_2^2 G
\end{eqnarray*}
to give us the final form of the second-order expansion of the state:
\begin{widetext}
\begin{subequations}
\label{eq:ap:pt:OperatorExpansion}
\begin{eqnarray*}
\eta &=&\eta _{o}
  -i\left\langle 0\left| \tilde\delta_2 \right| 0\right\rangle 
  - \frac{1}{2} \left\langle 0\left| G \delta_2^2 G \right| 0\right\rangle 
  - \left\langle 0\left| 
   \tilde \delta_2 P\frac{1}{\eta_{o}-G^{2}}P \tilde \delta_2 
   \right| 0\right\rangle +O(\delta_2^3) \; .\\
\psi &=&\left| 0\right\rangle 
   -i\frac{1}{\eta _{o}-G^{2}}P\tilde \delta_2 \left| 0\right\rangle 
   -\frac{1}{2}\frac{1}{\eta _{o}-G^{2}}P G \delta_2^2 G \left| 0\right\rangle 
  -\frac{1}{\eta _{o}-G^{2}}P\left[ 
       \tilde \delta_2 
       -\left\langle 0\left| \tilde \delta_2 \right| 0\right\rangle 
   \right] 
   \frac{1}{\eta _{o}-G^{2}}P\tilde \delta_2 \left| 0\right\rangle  + O(\delta_2^3)\; .
\end{eqnarray*}
\end{subequations}
\end{widetext}

\subsection{Perturbation theory expansion, expressed using basis states}
As a practical matter, we compute the perturbation series expansion using not the operators
themselves, but rather through a finite collection of matrix elements of the relevant operators
relative to basis states.  Therefore, we insert the identity operator, represented as a sum
over all basis states (i.e.\ $\textbf{1} = \sum_k \left| k \right> \left< k \right|$), at
several points in the above expression.  Since the basis states are eigenvectors of the
propagation operator $G$ and since
the operator $G$ is unitary, we know
\begin{subequations}
\begin{eqnarray}
 G \left | k \right > &=& \lambda_k \left| k \right > \; , \\ 
  \left < k \right | G&=& \lambda_k \left< k \right | \; .
\end{eqnarray}
\end{subequations}
Therefore, we conclude that the eigenvalue changes as
\begin{subequations}
\begin{eqnarray}
\eta &=&\eta _{o} \bigg( 1
 -i\left\langle 0\left| \delta_2 \right|
0\right\rangle   \\
 & &  
  -\frac{1}{2}\left\langle 0\left| \delta_2^{2}\right|
0\right\rangle -\sum_{k\neq 0}\frac{\eta _{k}\left| \left\langle 0\left|
\delta_2 \right| k\right\rangle \right| ^{2}}{\eta _{o}-\eta _{k}} \bigg)
+O\left( \delta_2 ^{3} \right) \nonumber
\end{eqnarray}
while the state changes according to the expansion
\begin{widetext}
\begin{eqnarray}
\label{eq:ap:pt:state}
\left| \psi \right\rangle &=&
  \left | 0 \right>
  -i \sum_{k\neq 0}\frac{\lambda_o \lambda_k}{\eta _{o}-\eta _{k}} 
   \left| k\right\rangle 
   \left\langle k\left| \delta_2 \right| 0\right\rangle  \\
&+&\sum_{k\neq 0}\left|k\right\rangle
    \frac{\lambda _{o}\lambda _{k}}{\eta _{o}-\eta _{k}}\left[
  -\frac{1}{2}\left\langle k\left| \delta_2^{2}\right| 0\right\rangle 
  +\frac{\eta _{o}}{\eta
    _{o}-\eta _{k}}\left\langle k\left| \delta_2 \right| 0\right\rangle \left\langle 0\left|
      \delta_2 \right| 0\right\rangle
 - \sum_{p\ne 0} \frac{\eta_p}{\eta_o-\eta_p}  
   \left\langle k\left| \delta_2 \right| p\right\rangle
   \left\langle p\left| \delta_2 \right| 0\right\rangle 
  \right] + O(\delta_2^3) \; .   \nonumber
\end{eqnarray}
\end{widetext}
\end{subequations}


\section{\label{ap:spherical}Perturbation theory for cavities bounded by two identical
  spherical mirrors}
For cavities bounded by spherical mirrors---that is, mirrors with height function $h_{1,2} =
r^2/2{\cal R}_{1,2}$---the eigenfunctions are known and of simple,
tractable Hermite-Gauss form.  We can therefore perform perturbation-theory calculations
analytically.  In this appendix, we describe these basis states and their application to
perturbation theory in greater detail.

\subsection{Background: Notation and definitions}
A spherical mirror of height  $h\left( r\right) $ is
uniquely characterized by its radius of curvature ${\cal R}$: $h=r^{2}/2{\cal R}$. We
consider two identical such mirrors, placed symmetrically at $\pm L/2$. We
drive this cavity so it has $kL/2\pi \gg 1$ wavelengths between the mirrors (so $kL$ is
to an excellent approximation independent of mirror shape).
\subsubsection{Coordinate representation of eigenfunctions}
The eigenfunctions for a cavity bounded by two identical spherical mirrors are known 
\cite{Siegman,Siegman2,Yariv,LaserHandbook,Barnes}:
\begin{eqnarray}
\psi _{m,n}  &=&\sqrt{\frac{1}{2^{m+n}m!n!}}H_{m}\left( \frac{x}{A\left(
z\right) }\right) H_{n}\left( \frac{y}{A\left( z\right) }\right) \nonumber \\
&\times& \frac{1}{\sqrt{\pi }A\left( z\right) }\exp \left[ -\frac{r^{2}}{%
2A\left( z\right) ^{2}}\left( 1-i\frac{z}{\bar{z}}\right) \right] \nonumber \\
&\times& \exp \left[ {-i\left( N+1\right) \tan ^{-1}\frac{z}{\bar{z}}} + ik z \right] \; .
\label{eq:def:gaussianStates}
\end{eqnarray}
Here $H_m$ are Hermite polynomials,
 $A_o \equiv \sqrt{\bar{z}/k}$,
 $N\equiv m+n$,  $g\equiv 1-L/{\cal R}$,  $\theta = \cos^{-1} g$,  $k = (q\pi +
\theta)/L$ for $q$ a large integer, and 
\begin{subequations}
\begin{eqnarray} 
A\left( z\right) ^{2}&\equiv&A_{o}^{2}\left( 1+\left( z/\bar{z}\right)^{2}\right)  \; , \\
\bar{z} & \equiv&\frac{L}{2}\sqrt{\frac{1+g}{1-g}} \; .
\end{eqnarray}
\end{subequations}

In particular, at the mirror faces at $z_\pm=\pm L/{2}$, we have 
\begin{subequations}
\begin{eqnarray}
z_\pm/\bar{z} &=& \pm \sqrt\frac{1-g}{1+g} \; ,\\
\label{eq:gauss:theta}
\tan^{-1}(z_\pm/\bar z) &=& \pm \frac{1}{2} \cos^{-1} g = \pm \theta/2 \; ,\\
\label{eq:gauss:AatMirrors}
A(z_\pm )  &=& \sqrt{\frac{L}{k}} \frac{1}{(1-g^2)^{1/4}} \; .
\end{eqnarray}
\end{subequations}

\subsubsection{Basis as eigensolutions of cavity}
After some algebra, one can verify these states $\psi_{m,n}$ are solutions to the round-trip
eigenequation Eq.\ (\ref{eq:eigenequation}) [or equivalently Eq.\ (\ref{eq:eigenequationAlt})] 
with $h_1(r)=h_2(r)=r^2/2{\cal R}$, and
\begin{equation}
\label{eq:gauss:lambdas}
\lambda_{mn} =\sqrt{\eta_{mn}} =  e^{-i \theta(m+n+1)} \; .
\end{equation}

\subsubsection{Establishing a quantum-mechanical correspondence}
For technical reasons not discussed further here, the eigenfunctions of a cavity bounded by
spherical mirrors correspond directly to the states of a 2-dimensional quantum-mechanical
scalar nonrelativistic particle in a quadratic potential (i.e.\ a 2-d quantum simple harmonic
oscillator, or SHO).  Like the SHO, the states are highly degenerate.  If we index the states
using their cartesian symmetry properties [i.e.\ express them  as a Hermite-Gauss basis, as in
Eq.\ (\ref{eq:def:gaussianStates})], these quantum states of a SHO
have the form $\left|m,n\right> = \Psi_m(x) \Psi_n(y)$ for 
\begin{equation}
\Psi _{n}\left( x\right) \equiv \frac{1}{\left[ \sqrt{\pi }A\right] ^{1/2}}\frac{1%
}{\sqrt{2^{n}n!}}H_{n}\left( \frac{x}{A}\right) e^{-x^{2}/2A^{2}} 
\end{equation}
for $A$ some length scale.  Therefore, if we evaluate the fields at
the mirror surface $z=-L/2$ and define $A=A(-L/2)$, we find
\begin{subequations}
\begin{equation}
\psi_{m,n}(x,y,z=-L/2) = e^{i \Phi_N} \Psi_m(x) \Psi_n(y)
\end{equation}
where $\Phi_N$ is given by
\begin{equation}
\Phi_N = (N+1) \theta/2  
       + (\text{independent of } N)\; .
\end{equation}
\end{subequations}

Since the terms in $\Phi_N$ which vary with $N$ do not vary with
position, we can establish a \emph{correspondance} between matrix
elements of SHO states and optical states for those operators which do
not involve derivatives.  Explicitly, if ${\cal O}$ is an operator
(e.g., some function of $x$), then
\begin{equation}
\left< m, n \right | {\cal O} \left | \bar{m}
  \bar{n}\right>_\text{opt} 
 = e^{i(\bar{N}-N)\theta/2}\left< m, n \right | {\cal O} \left | \bar{m}
  \bar{n}\right>_\text{sho}  \; .
\end{equation}

\subsubsection{Creation and annihilation operators for SHO states}
In quantum mechanics, many computations involving simple harmonic oscillator states can be
rewritten in a more readable form by the use of creation and annihilation operators
\cite{Dirac,Landau,Townsend}.   These simplifications occur because we can relate
any excited state $\left| 0 \right> $ to the ground state via the action of these operators:
\begin{eqnarray}
\left| m \right>_\text{sho} &=& \frac{\left(a_x^\dag\right)^m}{\sqrt{m!}} \left| 0\right>_\text{sho} \\
a_{x}^{\dag }&\equiv&\frac{1}{\sqrt{2}}\left( \frac{x}{A}-A\frac{d}{dx}\right) 
\end{eqnarray}

\subsection{Tabulating useful matrix elements}
When performing perturbation-theory calculations, we need the matrix elements of various
operators (i.e.\ $x$ and $x^2$) relative to the Hermite-Gauss basis.    These elements can be
expressed as follows:
\begin{subequations}
\label{eq:ap:pt:matrixelements}
\begin{eqnarray}
\left\langle m,n \right| &x&\left| \bar{m},\bar{n}\right\rangle_{opt} \\
  &=&e^{i\left( \bar N- N \right) \theta /2}
    \left\langle m,n\left| \frac{A  \left( a_{x}+a_{x}^{\dag }\right) }{\sqrt{2}}
   \right| \bar{m},\bar{n}\right\rangle _{sho} \nonumber\\
 &=&\frac{A}{\sqrt{2}}\delta _{n\bar{n}}\left[ 
    e^{i\theta/2}\sqrt{\bar m}\delta _{m,\bar{m}-1} 
   +e^{-i\theta /2}\sqrt{ m} \delta _{\bar m,m-1}\right] \nonumber
\end{eqnarray}
\begin{eqnarray}
\left\langle m,n \right| &x^2&\left| \bar{m},\bar{n}\right\rangle_{opt} \\
 &=&e^{i\left( \bar N- N \right) \theta /2}
   \left\langle m,n\left| \frac{A^2  \left( a_{x}+a_{x}^{\dag }\right)^2 }{2}
  \right| \bar{m},\bar{n}\right\rangle _{sho} \nonumber\\
  &=&\frac{A^2}{2}\delta _{n\bar{n}}\left[
     e^{ i\theta}\sqrt{m(m+1)}\delta _{m,\bar{m}-2}   \right . \nonumber \\
  & &     + (2m+1)\delta_{m,\bar m} \nonumber \\
  & &
    +\left . e^{- i\theta }\sqrt{\bar{m}(\bar{m}+1)}\delta _{\bar{m},m-2}\right] \nonumber
\end{eqnarray}
\end{subequations}

In the above,  we
use the subscript $opt$ to denote an optical state and $sho$ to denote a SHO state.


\subsection{Effects of tilt on beam state}
If we tilt  mirror $2$ about the $y$ axis, the mirror's height changes according to $\delta_2 =
2 k x \phi$ [cf.\ Eqs.\ (\ref{eq:dh:Tilt}) and (\ref{eq:def:deltas})].  Using that expression in
the two perturbative expansion expressions, 
Eqs.\ (\ref{eq:pt:state:psi1}) and (\ref{eq:pt:state:psi2}), and taking account of the above tools, we
can find explicit expressions for the first and second order changes in the beam state
evaluated at mirror $1$.

\begin{eqnarray}
\left| \psi^{(1)} \right> &=&
-\left| 1,0\right>
 (\phi \sqrt{Lk}) \frac{e^{-i\theta/2}}{\sqrt{2}} \frac{1}{(1-g^2)^{3/4}}
 \\
\left| \psi^{(2)} \right> &=& 
 - \left| 2,0\right> (  \phi\sqrt{k L})^2 \frac{e^{i \theta}}{(1-g)\sqrt{2(1-g^2)}} 
\end{eqnarray}

In performing the above calculation, we have explicitly substituted in $A=A(z_\pm)$ from
Eq.\ (\ref{eq:gauss:AatMirrors}) and $e^{i\theta} = g + i \sqrt{1-g^2}$  [derived from
Eq.\ (\ref{eq:gauss:theta})]  when appropriate.

\subsection{Effect of displacement on beam state}
When the mirror is displaced through a distance $d$ in the $x$ direction, the mirror height is
perturbed from $h(r) = r^2/2 {\cal R}$ into 
\begin{equation}
h'(r)=h(\vec{r}+s\hat{x}) = h(r) + \frac{xs}{\cal R} + \frac{s^2}{2\cal{R}} \; .
\end{equation}
We can in a straightforward manner insert $\delta h_\text{disp} = h'-h$ into the perturbative
expansion for the state [Eq.\ (\ref{eq:ap:pt:state})]; note that $\delta h$ has terms both first
and second order in $\psi$; and thereby deduce the appropriate first and second order changes
in the optical state of the cavity.   

We will not provide a comprehensive calculation
here.  Instead, we note that the first-order changes in height $\delta h$ due to tilt and
displacement are identical, involving merely a change in factor:
\begin{equation}
\phi \sqrt{Lk}\rightarrow \left( s/b\right) \left( 1-g\right) \; .
\end{equation}
Therefore, we deduce the first-order effect of a displacement of mirror $2$ is
\begin{eqnarray}
\left| \psi^{(1)} \right> &=&
-\left| 1,0\right>
 (d/b) \frac{e^{-i\theta/2}}{\sqrt{2}} \frac{(1-g)}{(1-g^2)^{3/4}}
\end{eqnarray}

\section{\label{ap:postprocess} Postprocessing performed on finite-element computations used in
  thermoelastic noise integrals}
We used a commercial finite-element code, FEMLAB \cite{femlab}, to solve the
fluctuation-dissipation-motivated axisymmetric elastic model problem presented in
Eq.\ (\ref{eq:elas}).   This code expresses its solution for $y^a$ in terms of a piecewise
linear function.  Unfortunately, the thermoelastic noise integral depends on second derivatives
of $y_a$.  Therefore, even though the FEMLAB program includes many useful tools to evaluate
physically interesting quantities, we had to evaluate the thermoelastic noise integral
ourselves.   We therefore
sent the FEMLAB output for $y^a$ to a two-dimensional MATLAB array, then used our own MATLAB
routines to generate expressions for $|\nabla \Theta|^2$ and thus $I_A$.

The two groups performing this computation (Caltech and Moscow) each developed relatively
independent postprocessing codes.  In this appendix, we describe 
the Caltech postprocessing code.

\subsection{Caltech postprocessing code}
The Caltech group cobbled together a rough set of MATLAB routines to convert the
finite-element representation of the response $y^a$ to useful values for $I_A$.  While this
code contains more than a few kludges used to get it to work, we found the combination of the
finite-element solution and this postprocessing routine gave answers for $I_A$ that agreed well with and
converged to the values of $I_A$ for known, analytic solutions (i.e.\ the LT cylindrical
solutions). 
\subsubsection{Postprocessing technique}

\begin{enumerate}
\item
\emph{Preliminaries}: We used the FEMLAB finite-element code, with some number $N$ elements, to find the
elastic response of a frustum of front radius  
$R_1$, back radius $R_2$, and thickness $H$.  We increased $N$ until we felt comfortable with
the resulting response curve.  The FEMLAB code worked in cylindrical coordinates (i.e.\ $r$,$z$,
$\phi$), and assumed axisymmetry.

\item
\emph{Output to rectangular grid}: The code allowed us to extract the values of the
displacement vector (components $U=y_r$ and $V=y_z$) at any point.  We used this ability to
obtain values of $U$ and $V$ on a rectangular grid.  The code naturally provided smoothly
extrapolated values of $U$ and $V$ when we asked for a point outside the volume simulated.

The total number of points on the rectangular grid was chosen to be comparable to $N$.
Specifically, the numbers  of points in the $r$ and $z$ directions were chosen as $\sqrt{N
  R_1/H}$ and $\sqrt{N H/R_1}$, respectively.

\item
\emph{Select cutoff region for grid}: Because we used a grid containing too large a region, we
must provide a cutoff filter to select only those gridpoints which contain physical values.
Furthermore, as a practical matter, the FEMLAB code gave odd results\footnote{In other words, we
  found NaN (i.e.\ ``not a number'') answers from our code.} when we evaluated
the response at the edges of the computational domain.  We therefore chose a cutoff filter
which eliminated all exterior and ``near-to-the-boundary''  points.

This filter was applied repeatedly (i.e.\ during each derivative process).  Because the filter
was such a universal ingredient to each subsequent action, we will not mention each occurrence
on which it is used in the following postprocessing.

\item
\emph{Compute expansion}: Using the relationship between $\Theta$ and $y^a$, namely
\begin{equation}
\Theta = \partial_z V + \partial_r  U + U/r
\end{equation}
we compute $\Theta$.  Derivatives are formed as centered differences, which are then
interpolated (or, for the endpoints, extrapolated) back to the gridpoints.  The values of $U/r$ 
at $r=0$ are found by extrapolating the values of $U/r$ for $r\ne 0$.  

[To circumvent  problems that arose due to dividing by small numbers near $r=0$, we
typically erased the values for $\Theta$ we obtained on four gridlines near $r=0$ and 
replaced them by extrapolated values of the region immediately outside.]

\item
\emph{Compute $|\nabla \Theta|^2$}: Next, we computed the two derivatives $\partial_r U$ and
$\partial_z V$ and used them to form
\begin{equation}
|\nabla \Theta|^2 = (\partial_r U)^2 + (\partial_z V)^2 \; .
\end{equation}
As before, derivatives were evaluated with centered differences which were interpolated to
gridpoints. 

\item
\emph{Compute $I_A$}: Finally, we used our own two-dimensional (Simpson's rule) integrator to
evaluate  the two-dimensional integral 
\begin{equation}
I_A = \int r dr dz \; |\nabla \Theta|^2 \;,
\end{equation}
which provides an explicit form for Eq.\ (\ref{eq:IA}).

\end{enumerate}

\subsubsection{Testing the result}
Our postprocessing code is far from polished.  To insure the cutoff filter is operating
properly and to otherwise guarantee that the results of the postprocessing code seem physically
plausible, we usually plotted $|\nabla \Theta|^2$ to verify that the integrand is indeed a
well-behaved (i.e.\ smooth-looking) function.

Because we had a serious limitation on the number of points we could practically employ in a
reasonable amount of desktop computing time ($\sim$ few $\times 10^4$), we did not perform
systematic convergence testing.  However, what testing we did, corroborated by comparisons
between our numerical method and exact analytic solutions, suggests our results are relatively
accurate.  More critically, our computations agreed well with independent computations 
performed by Sergey Strigin and Sergey Vyatchanin.








\section{\label{ap:halfInfiniteThermoelastic}Thermoelastic noise of half-infinite mirrors }
To evaluate the thermoelastic noise associated with a given beam shape $P(r)$, we must evaluate
the 
integral $I_A$ [Eq.\ (\ref{eq:IA})] given the solution $y^a$ to a model elasticity problem
[Eq.\ (\ref{eq:elas})].  As discussed in Sec.\ \ref{sec:subsub:halfinf}, if the mirror is
sufficiently large compared to the beam shape $P(r)$, we can effectively treat the mirror as
half-infinite (i.e.\ filling the whole volume $z<0$) in the elasticity problem.  In this case,
the bulk acceleration term in the elasticity problem drops out [i.e.\ $V_A \rightarrow \infty$
in Eq.\ (\ref{eq:elas:eom})] and
we seek only the elastic response of a half-infinite medium to an imposed surface stress.  This
last problem has been discussed extensively in the literature --- cf., e.g.,
\cite{LandauElasticity,CoatingNoise} --- and there exist simple fourier-based
computational techniques to generate and manage solutions.   We apply these known solutions from
the literature  to evaluate the thermoelastic integral $I_A$.

\subsection{Elastic solutions for the expansion ($\Theta$)}
In the case of half-infinite mirrors, the  response $y^a$ to the imposed
pressure profile $P(r)$ can be found in the literature [cf.\ Eqs.\ (8.18) and
(8.19)  of Landau and Lifshitz's book on elasticity \cite{LandauElasticity}, where, however,
the half-infinite volume is chosen \emph{above} the $z=0$ plane rather than below; see also
Nakagawa et al. \cite{CoatingNoise}, especially their Appendix A].  These expressions
allow us to explicitly relate the expansion $\Theta$ to the imposed pressure profile $P(r)$:
\begin{subequations}
\begin{eqnarray}
\label{eq:ap:thetaConvolution}
\Theta \left( \vec{r},z\right) 
  &=&\int G^{\left( \Theta \right) }\left( \vec{r},z;r'\right) P\left( r'\right) d^{2}\vec{r'}
  \\
G^{\left( \Theta \right) }(\vec{r},z;\vec{r}_o) 
  &=& -\frac{\left( 1+\sigma \right) \left( 1-2\sigma \right)  z H(-z)}{2\pi E
   \left| (\vec{r}-\vec{r}_o)^{2}+z^{2}\right| ^{3/2}}
\end{eqnarray}
\end{subequations}
where $H(x)$ is a step function which is $1$ when $x>0$ and $0$ otherwise.

Because we have complete transverse translation symmetry, we can make our results more
tractable by fourier-transforming in the transverse dimensions:
\begin{eqnarray}
\tilde{\Theta}\left( K,z\right) &\equiv&\int e^{-i\vec{K}\cdot \vec{R}}\Theta(R,z) d^2 R \; , \\
\tilde P(\vec K) &\equiv& \int e^{-i\vec{K}\cdot \vec{R}}P(R)  d^2 R \; .
\end{eqnarray}
For example, the convolution relating light intensity profile to the associated elastic
response, Eq.\ (\ref{eq:ap:thetaConvolution}), can be re-expressed as
\begin{eqnarray}
\tilde \Theta(K,z) &=& G^\Theta(z,\vec K) \tilde P(\vec K)  \\
\tilde{G}^{(\Theta )}\left( z,\vec{K}\right) 
&=&-\frac{\left( 1+\sigma \right) \left( 1-2\sigma \right) }{2\pi E}e^{-\left|
Kz\right| } \; .
\end{eqnarray}


\subsection{Thermoelastic integral $I_A$}
Inserting the solution discussed above into Eq.\ (\ref{eq:IA}) and using fourier-transform
techniques to 
simplify the resulting integral, we find

\begin{equation}
I_A =\left(  \frac{\left( 1+\sigma \right) \left( 1-2\sigma \right) }{2\pi E}\right)^2
\int d^{2}\vec{K}\,\left| K\right| \left| \tilde{P}\left(
K\right)\right|
^{2}  \; .
\end{equation}

\section{\label{ap:nonidentical}Configurations with nonidentical frustum
  mirrors and mesa-like beams} 
Physically the ETM and ITM need not be identical: while the ITM back face size is constrained
to be above a certain radius (for a given mesa-beam $D$) by diffraction losses, the ETM back
face has no size restriction.  Therefore, if we  allow different mirrors, we introduce more
parameters and therefore more possibilities for finding a mirror and beam configuration with
very low noise.

Unfortunately, strictly speaking the mesa beams presented in this paper are designed
specifically for identical cavities.  Therefore, in the main text, when we explored
 how thermoelastic noise varied with mirror dimensions and shape,
we  restricted attention to cavities bounded by \emph{identical} mirrors.  

Nonetheless, we expect that \emph{generic} flat-topped beams will always closely resemble the
mesa-beam profile, with roughly the mesa-beam diffraction losses.  Therefore,  this appendix
generalizes the  computations of Sec.\ \ref{sec:apply:Design} to include distinct ETM and ITM
mirrors. 

More specifically, we (i) assume that, for any $D_1$ and $D_2$, there is some flat-topped beam
whose intensity very closely mimics the mesa-beam intensity profile $|u_\text{mesa}(r,D_1)|^2$
at mirror 1 and  $|u_\text{mesa}(r,D_2)|^2$ at mirror 2; (ii) select $D_1$,  $D_2$, $R_1$, and
$R_2$ so the clipping approximation to the diffraction losses is $10\text{ ppm}$ at each mirror
face (Sec. \ref{sec:th:Design:constraints}); and  (iii) use the resulting beam intensity
profiles to compute the thermoelastic noise integrals $I_1$ and $I_2$ (Sec.\
\ref{sec:num:thermo}).
The power spectrum of thermoelastic noise for an interferometer with two identical arm
cavities, each of which has an ITM like mirror 1  and an ETM mirror 2 can be expressed relative
to the baseline  [cf.\ Eq.\ (\ref{eq:Sh})] by
\begin{equation}
S_h/S_h^{\text{BL}} = \frac{I_\text{ITM} + I_\text{ETM}}{2I_\text{BL}} \; .
\label{eq:apply:IcompareNonidentical}
\end{equation}

\subsection{Thermoelastic noise integral for ETM designs}
The ETM mirror has only two design constraints: (i) its mass must be $40\text{ kg}$ and (ii) the
diffraction losses off its front face must be less than or equal to $10\text{ ppm}$.
Table \ref{tbl:ConMH2} shows the value of the
thermoelastic noise integral for various values of $R_{p1}$, $R_{p2}$, and $D$, chosen to
satisfy the various constraints (mirror mass $40\text{ kg}$; diffraction losses off the front
face of $10\text{ ppm})$.   

While we have complete freedom to adjust the back face size of the ETM, from the table,
we see the back face size matters little so long as it is less than  $R_{p2} \lesssim 9$
cm.  By fitting a  quadratic
to $(D/b, I/I_{\rm BL})$ [for only those ETM  entries with $R_{p2}\lesssim 9$cm],
we estimate  the optimal ETM dimensions;  our results are tabulated in Table \ref{tbl:NSRangeLarger}.

In Table \ref{tbl:NSRangeLarger} we also combine the optimal ETM with the optimal ITM and use 
Eq. (\ref{eq:apply:IcompareNonidentical}) to express the thermoelastic noise of a
nonidentical-mirror configuration relative to the baseline.   When the ETM can be different
from the ITM, the optimal thermoelastic noise is significantly reduced (because the ETM inner
radius can be larger).

\begin{table}
\caption{\label{tbl:ConMH2} The thermoelastic integral $I$ for a frustum end test mass (ETM)
and a Mexican-Hat Beam, in units
of  $I_{\rm BL} = 2.57 \times 10^{-28} {\rm s}^4 {\rm g}^{-2} {\rm cm}^{-1}$.  The values of $I/I_{\rm BL
}$ are estimated to be accurate to within one
per cent.}
\begin{ruledtabular}
\begin{tabular}{ldddddd}
\multicolumn{1}{l}{$R_1$} & \multicolumn{1}{r}{$R_{p1} [{\rm cm}]$} & \multicolumn{1}{r}{$R_{p2} [{\rm cm
}]$} &\multicolumn{1}{r}{$H [{\rm cm}]$} & \multicolumn{1}{r}{$D/b$} & \multicolumn{1}{r}{$I/I_{\rm BL}$}
 & \multicolumn{1}{r}{${\cal L}_0$[ppm]\footnotemark[1] } \\
\hline
$R_{p1}-8{\rm mm}$ & 17.11 & 2.00 & 28.85 & 4.00 & 0.175 & 10 \\
$R_{p1}-8{\rm mm}$ & 17.11 & 8.00 & 19.34 & 4.00 & 0.175 & 10 \\
$R_{p1}-8{\rm mm}$ & 17.11 & 12.00 & 14.87 & 4.00 & 0.192 & 10 \\
\\
$R_{p1}-8{\rm mm}$ & 19.58 & 3.00 & 21.17 & 5.00 & 0.134 & 10 \\
$R_{p1}-8{\rm mm}$ & 19.58 & 9.00 & 14.91 & 5.00 & 0.133 & 10 \\
$R_{p1}-8{\rm mm}$ & 19.58 & 13.00 & 11.83 & 5.00 & 0.180 & 10 \\
\\
$R_{p1}-8{\rm mm}$ & 19.98 & 5.00 & 18.22 & 5.15 & 0.133 & 10 \\
\\
$R_{p1}-8{\rm mm}$ & 22.08 & 3.00 & 16.97 & 6.00 & 0.156 & 10 \\
$R_{p1}-8{\rm mm}$ & 22.08 & 9.00 & 12.45 & 6.00 & 0.157 & 10 \\
$R_{p1}-8{\rm mm}$ & 22.08 & 15.00 & 9.15 & 6.00 & 0.310 & 10 \\
\\
$R_{p1}$ & 13.94 & 5.00 & 32.93 & 3.00 & 0.333 & 10 \\
$R_{p1}$ & 13.94 & 7.50 & 26.80 & 3.00 & 0.340 & 10 \\
$R_{p1}$ & 13.94 & 10.00 & 21.96 & 3.00 & 0.345 & 10 \\
\\
$R_{p1}$ & 16.37 & 2.00 & 31.34 & 4.00 & 0.161 & 10 \\
$R_{p1}$ & 16.37 & 4.00 & 27.33 & 4.00 & 0.160 & 10 \\
$R_{p1}$ & 16.37 & 6.00 & 23.74 & 4.00 & 0.160 & 10 \\
$R_{p1}$& 16.37 & 8.00 & 20.63 & 4.00 & 0.160 & 10 \\
$R_{p1}$ & 16.37 & 10.00 & 17.96 & 4.00 & 0.162 & 10 \\
$R_{p1}$ & 16.37 & 12.00 & 15.70 & 4.00 & 0.173 & 10 \\
$R_{p1}$ & 16.37 & 14.00 & 13.78 & 4.00 & 0.207 & 10 \\
\\
$R_{p1}$ & 18.85 & 1.00 & 25.45 & 5.00 & 0.112 & 10 \\
$R_{p1}$& 18.85 & 3.00 & 22.69 & 5.00 & 0.112 & 10 \\
$R_{p1}$ & 18.85 & 5.00 & 20.12 & 5.00 & 0.112 & 10 \\
$R_{p1}$ & 18.85 & 7.00 & 17.81 & 5.00 & 0.111 & 10 \\
$R_{p1}$ & 18.85 & 9.00 & 15.56 & 5.00 & 0.113 & 10 \\
$R_{p1}$ & 18.85 & 11.00 & 13.97 & 5.00 & 0.119 & 10 \\
$R_{p1}$ & 18.85 & 13.00 & 12.41 & 5.00 & 0.144 & 10 \\
\\
$R_{p1}$ & 19.86 & 5.00 & 18.42 & 5.42 & 0.107 & 10 \\
\\

$R_{p1}$ & 21.36 & 3.00 & 18.04 & 6.00 & 0.116 & 10 \\
$R_{p1}$ & 21.36 & 5.00 & 16.24 & 6.00 & 0.115 & 10 \\
$R_{p1}$ & 21.36 & 7.00 & 15.39 & 6.00 & 0.115 & 10 \\
$R_{p1}$ & 21.36 & 9.00 & 14.58 & 6.00 & 0.116 & 10 \\
$R_{p1}$ & 21.36 & 12.00 & 11.15 & 6.00 & 0.133 & 10 \\
$R_{p1}$ & 21.36 & 15.00 & 9.55 & 6.00 & 0.224 & 10 \\
\end{tabular}
\end{ruledtabular}
\footnotetext[1]{Diffraction losses in each bounce off inner face of the test mass (radius
$R_1$), in ppm (parts per million).}
\end{table}

\subsection{Aside: Constructing mesa-like beams appropriate to nonidentical mirrors}
In this appendix, we assume that for any $D_1$ and $D_2$, there is some flat-topped beam whose
intensity  closely mimics the mesa-beam intensity profile $|u_\text{mesa}(r,D_1)|^2$
at mirror 1 and  $|u_\text{mesa}(r,D_2)|^2$ at mirror 2.  In fact, we may explicitly construct
such a beam through a generalization of the mesa-beam proposal (though the authors have not
numerically explored this possibility).

The mesa beams constructed in this paper were built by averaging minimal gaussian beams which all 
travelled \emph{parallel} to the optic axis [cf.\ Eq.\ (\ref{eq:buildMHcanonical})].   
More generally, we can construct mesa-like beams by averaging gaussians which travel at 
a position-dependent angle to the optic axis; for example, at mirror $1$, we could choose the 
beams to have angle distribution
\begin{equation}
\phi(r) = \frac{R_2 - R_1}{L}\frac{r}{R_1} \; .
\end{equation}

The  resulting mesa-type beams will have power distributions at the two mirrors which closely 
resemble the power distributions of mesa beams: (i) they will be flat over a large central region,
of size $D_1$ at mirror $1$ and size $D_2 \approx D_1 (R_2/R_1)$ at mirror $2$; and (ii) they
will fall of rapidly outside this region, with a falloff rate nearly the same as that of the 
minimal gaussian.

\begingroup
\squeezetable
\begin{table}
\caption{\label{tbl:NSRangeLarger} Optimized test-mass and light beam configurations, their
thermoelastic noise compared to the baseline.  [A subset of this table appears as Table I
 in MBI \cite{MBI}, and as Table \ref{tbl:NSRange} in this paper.]  
}
\begin{ruledtabular}
\begin{tabular}{lld}
Test Masses & Beam Shape & \multicolumn{1}{c}{$\left({S_h\over S_h^{\rm BL}}\right)_{\rm TE}
$}  \\
\{$R_{p1}$, $R_{p2}$; $H$\} & &  \\
\hline
BL: cylinders, $R=R_p-8{\rm mm}$ & BL: Gaussian &  \\
\{15.7, 15.7; 13.0\}  &  $r_o = 4.23{\rm cm}$ & 1.000  \\
BL: cylinders, $R=R_p-8{\rm mm}$  & mesa & \\
\{15.7, 15.7; 13.0\} &  $D/b = 3.73$ & 0.364  \\
identical frustums, $R=R_p-8{\rm mm}$  & mesa & \\
\{17.11, 12.88, 14.06\} & $D/b = 4.00$ & 0.207  \\
 different cones, $R=R_p-8{\rm mm}$  & MH &  \\
 ITM\{17.42, 13.18, 13.51\}  & $D/b = 4.00$ &  \\
 ETM\{19.96, $\lesssim 9.$, $\gtrsim14.$\} & $D/b = 5.15$ & 0.170 \\
\\
BL: cylinders, $R=R_p$ & Gaussian & \\
\{15.7, 15.7; 13.0\}  &  $r_o = 4.49{\rm cm}$ & 0.856  \\
BL: cylinders, $R=R_p$ & mesa  & \\
\{15.7, 15.7; 13.0\} &  $D/b = 3.73$ & 0.290  \\
identical frustums, $R=R_p$ & mesa & \\
\{17.29, 13.04, 13.75\} & $D/b = 4.39$ & 0.162  \\
 different cones, $R=R_p$ & MH &  \\
 ITM\{17.29, 13.04, 13.75\} & $D/b = 4.39$ &  \\
 ETM\{19.91, $\lesssim 9.$, $\gtrsim14.$\} & $D/b = 5.42$ & 0.135  \\
\end{tabular}
\end{ruledtabular}
\end{table}
\endgroup


\end{document}